\documentclass[prl,superscriptaddress,twocolumn,longbibliography]{revtex4-2}

\usepackage{amsmath,amssymb}
\usepackage{physics}
\usepackage{graphicx}
\usepackage{bm}
\usepackage{bbm}
\usepackage[colorlinks,bookmarks=false,citecolor=blue,linkcolor=black,urlcolor=red]{hyperref}
\usepackage{color} 
\usepackage{newtxtext,newtxmath}
\usepackage{multirow}
\usepackage{tikz}
\usetikzlibrary{positioning}
\usetikzlibrary{backgrounds}
\usetikzlibrary{calc}
\usepackage{booktabs,multirow}
\usetikzlibrary{arrows.meta}
\allowdisplaybreaks

\begin{document}

\title{Quantized pumping of boundary Hilbert space in chiral Floquet phases}

\date{\today}

\author{Ryota Nakai}
\affiliation{RIKEN Center for Quantum Computing (RQC), Wako, Saitama, 351-0198, Japan}

\author{Taozhi Guo}
\affiliation{Department of Physics, Princeton University, Princeton, New Jersey 08544, USA}
\affiliation{Pritzker School of Molecular Engineering, University of Chicago, Chicago, Illinois 60637, USA}

\author{Shinsei Ryu}
\affiliation{Department of Physics, Princeton University, Princeton, New Jersey 08544, USA}
\affiliation{Institute for Advanced Study, Princeton, New Jersey 08540, USA}

\begin{abstract}
    Chiral Floquet phases support quantized chiral flow along their boundaries even when the bulk is fully localized.
    In this Letter, we establish a complementary phenomenon, which we call Hilbert-space pumping: under a boundary condition twisted by time translation, a part of the boundary Hilbert space, of quantized dimension, is transferred between the opposite boundaries.
    The twist is the temporal analogue of Laughlin's flux insertion in the quantum Hall effect.
    For noninteracting fermions, the number of transferred boundary modes is determined by the winding number, as we show analytically for the boundary dynamics and numerically for the anomalous Floquet Anderson insulator.
    For interacting systems whose boundary dynamics is a quantum cellular automaton, the transferred dimension is set by the Gross--Nesme--Vogts--Werner (GNVW) index: the twisted boundary Floquet operator is the Floquet operator on a tilted time slice, whose Hilbert-space dimension is multiplied by the index for each period of twist.
    We derive this relation for matrix product unitaries and test it numerically in random unitary circuits.
    Applied to the matrix product operators of non-invertible symmetries, the same twisted rank returns the quantum dimension of the symmetry operator in addition to its flow index.
\end{abstract}

\maketitle

Quantized transport is one of the most striking phenomena in topological phases \cite{RevModPhys.82.3045,RevModPhys.83.1057}.
In the integer quantum Hall effect, chiral boundary channels carry charge along the boundaries with quantized conductance, while the adiabatic insertion of a flux quantum through a cylinder pumps a quantized charge from one boundary to the other \cite{PhysRevB.23.5632,PhysRevB.25.2185,PhysRevB.38.9375}.
These two forms of transport are complementary manifestations of the same bulk topology.

Periodically driven (Floquet) systems provide a distinct route to topological transport \cite{
PhysRevB.79.081406,
PhysRevB.82.235114,
Lindner2011,
PhysRevLett.106.220402,
annurev:/content/journals/10.1146/annurev-conmatphys-031218-013423,
annurev:/content/journals/10.1146/annurev-conmatphys-031218-013721,
Rudner2020}.
Two-dimensional chiral Floquet phases exhibit robust chiral boundary dynamics even when the bulk is fully localized \cite{
PhysRevX.3.031005,
PhysRevLett.114.056801,
PhysRevB.96.245116,
PhysRevX.6.021013,
PhysRevX.6.041070,
PhysRevLett.118.115301,
PhysRevB.95.195128}, reflecting the topology of the full time evolution over a period rather than that of the Floquet bands.
In the anomalous Floquet Anderson insulator (AFAI), a noninteracting fermionic realization \cite{
PhysRevX.6.021013,
PhysRevB.99.235408,
PhysRevB.101.041403,
PhysRevResearch.3.013117,
PhysRevB.109.L121114,
Wintersperger2020,PhysRevLett.130.043201,
Maczewsky2017,Mukherjee2017,Li2023,
Peng2016,
PhysRevX.5.011012}, the boundary dynamics is characterized by a winding number that determines the quantized charge pumped along the boundary per period.
More generally, even without U(1) charge conservation, the boundary of an interacting chiral Floquet phase supports a quantized chiral flow of quantum information, characterized by the Gross--Nesme--Vogts--Werner (GNVW) index \cite{Gross2012,PhysRevX.6.041070,PhysRevLett.118.115301,PhysRevB.95.195128}, which classifies one-dimensional quantum cellular automata and has recently been used to characterize lattice anomalies \cite{m188-w1ct,zrs1-j2xd}.

\begin{figure}
    \centering
    \includegraphics[width=0.47\textwidth]{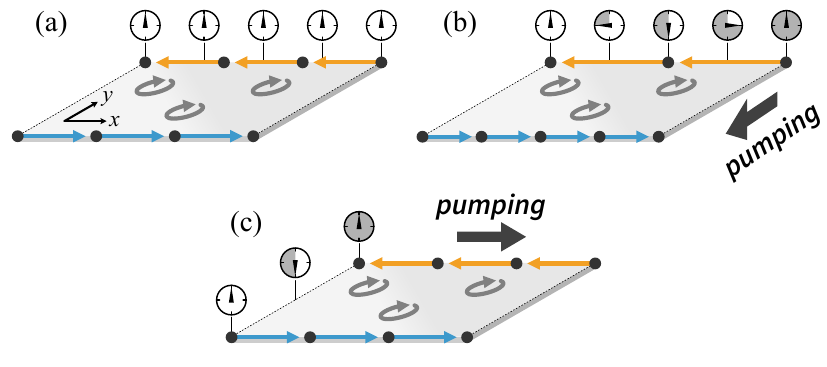}
    \caption{
        Time-translation twist and Hilbert-space pumping in a chiral Floquet phase on a cylinder, periodic in $x$ and open in $y$.
        Clocks indicate the local phase of the drive, i.e., the time shift of the Floquet Hamiltonian at each position.
        Circular arrows represent the localized bulk, and blue (orange) arrows the chiral motion of the bottom (top) boundary modes.
        (a) Without the twist, the drive is uniform and each boundary hosts three modes.
        (b) Under the time-translation twist, the drive is delayed in proportion to $x$ [Eq.~(\ref{eq:afaibulk_twist})], so that a boundary mode is delayed by $aT$ when it goes once around the cylinder.
        A mode is pumped from the top to the bottom boundary.
        (c) When the delay depends on $y$ instead, the boundary modes are pumped along the boundary (see text).
        \label{fig:tttwist}
    }
\end{figure}

What flows along the boundary of a chiral Floquet phase is not a conserved charge but the local degrees of freedom themselves.
This raises the question of what the analogue of Laughlin's flux insertion \cite{PhysRevB.23.5632} is in chiral Floquet phases, and what is pumped by it.
In Laughlin's argument, the flux is a background U(1) gauge field whose holonomy $e^{i\alpha}$ is encoded in the twisted boundary condition $\psi(x+L)=e^{i\alpha}\psi(x)$, where $\psi$ is the electron wavefunction or annihilation operator, $x$ is the coordinate along the periodic direction, and $L$ is the circumference.
The symmetry underlying chiral Floquet phases is the discrete time translation by one period $T$, and the analogue of the flux insertion is the boundary condition twisted by time translation (time-translation twist) \cite{PhysRevB.95.165405,PhysRevB.106.155128}: a local operator $O(x,t)$ in the Heisenberg picture that goes once around the periodic direction comes back translated in time by $a$ periods,
\begin{align}
    O(x+L,t)=O(x,t+aT),
    \label{eq:tttwist}
\end{align}
which for free fermions is a condition on the field operator, $\psi(x+L,t)=\psi(x,t+aT)$ [Fig.~\ref{fig:tttwist} (b)].
Here, $t$ is time, and the twist parameter $a$ is real, while the pumping is quantized at integer $a$.
In a bulk Floquet system, we impose (\ref{eq:tttwist}) by delaying the drive in proportion to $x$ [Eq.~(\ref{eq:afaibulk_twist}), Fig.~\ref{fig:tttwist} (b)], so that a boundary mode is delayed by $aT$ when it goes once around the cylinder; for a boundary Floquet operator written as a tensor network, the legs that cross the seam are displaced by $a$ rows [Eq.~(\ref{eq:mpu_withtwist})].
Similar twists have been used in the studies of thermal transport \cite{PhysRevB.95.165405,PhysRevB.106.155128,PhysRevB.108.075108,y88d-tx4l,hsyt-nv2n}, of time crystals \cite{1dc6-y1x8}, and of time vortices \cite{Kishony2025}.

\textit{\color{blue}Main results.---}
In this Letter, we show that the twist (\ref{eq:tttwist}) transfers a part of the boundary Hilbert space between the opposite boundaries, a phenomenon we call Hilbert-space pumping, and that the dimension of the transferred part is quantized, while the localized bulk states are insensitive to the twist.
For noninteracting fermions,
\begin{align}
    \frac{1}{2}\left(
    M_\text{bottom}(a)-M_\text{top}(a)
    \right)
    =
    \mathcal{W}a 
    \label{eq:afairelation}
\end{align}
for integer twists $a\in\mathbb{Z}$, where $M_\text{bottom(top)}(a)$ is the number of boundary modes on the bottom (top) boundary and $\mathcal{W}$ is the winding number of the bottom-boundary dynamics: $\mathcal{W}a$ boundary modes are transferred from the top to the bottom boundary.
We show this analytically for the boundary dynamics and confirm it numerically in the anomalous Floquet Anderson insulator (AFAI).
For interacting systems whose boundary dynamics is a spatially translation-invariant matrix product unitary (MPU) on qudits of dimension $d$,
\begin{align}
    \frac{1}{2}\log_d\frac{\rank \hat{U}_\text{bottom}^t(a)}{\rank \hat{U}_\text{top}^t(a)}
    =
    \text{ind}\cdot a
    \label{eq:mpurelation}
\end{align}
for $a\in\mathbb{Z}$ and $t\ge |a|$, where $\hat{U}_\text{bottom(top)}(a)$ is the bottom-(top-)boundary Floquet MPU under the twist, $\hat{U}^t(a)=[\hat{U}(a)]^t$ is the evolution over $t$ driving periods, and $\text{ind}=\log_d I_\text{GNVW}$ is the logarithm of the GNVW index of $\hat{U}_\text{bottom}(0)$.
Since the rank of a unitary operator is the dimension of the Hilbert space on which it acts, the left-hand side of (\ref{eq:mpurelation}) is the transferred dimension counted in qudits; for free fermions, each boundary mode carries a two-dimensional Fock space ($d=2$), and (\ref{eq:mpurelation}) reduces to (\ref{eq:afairelation}).
We derive (\ref{eq:mpurelation}) for $t\le 3$ from the MPU structure and for $t\ge |a|$ from a tilted-slice argument, which identifies the nonzero spectrum of $\hat{U}^t(a)$ with that of a unitary on $d^{\,N+\text{ind}\cdot a}$ states for $N$ boundary qudits [Eq.~(\ref{eq:tilted})], and verify it exactly for the shift operator and numerically in random unitary circuits (RUCs).
Finally, applied to the matrix product operator (MPO) of a non-invertible symmetry, the twisted rank returns the quantum dimension in addition to the flow index.

\begin{figure}
    \centering
    \includegraphics[width=0.47\textwidth]{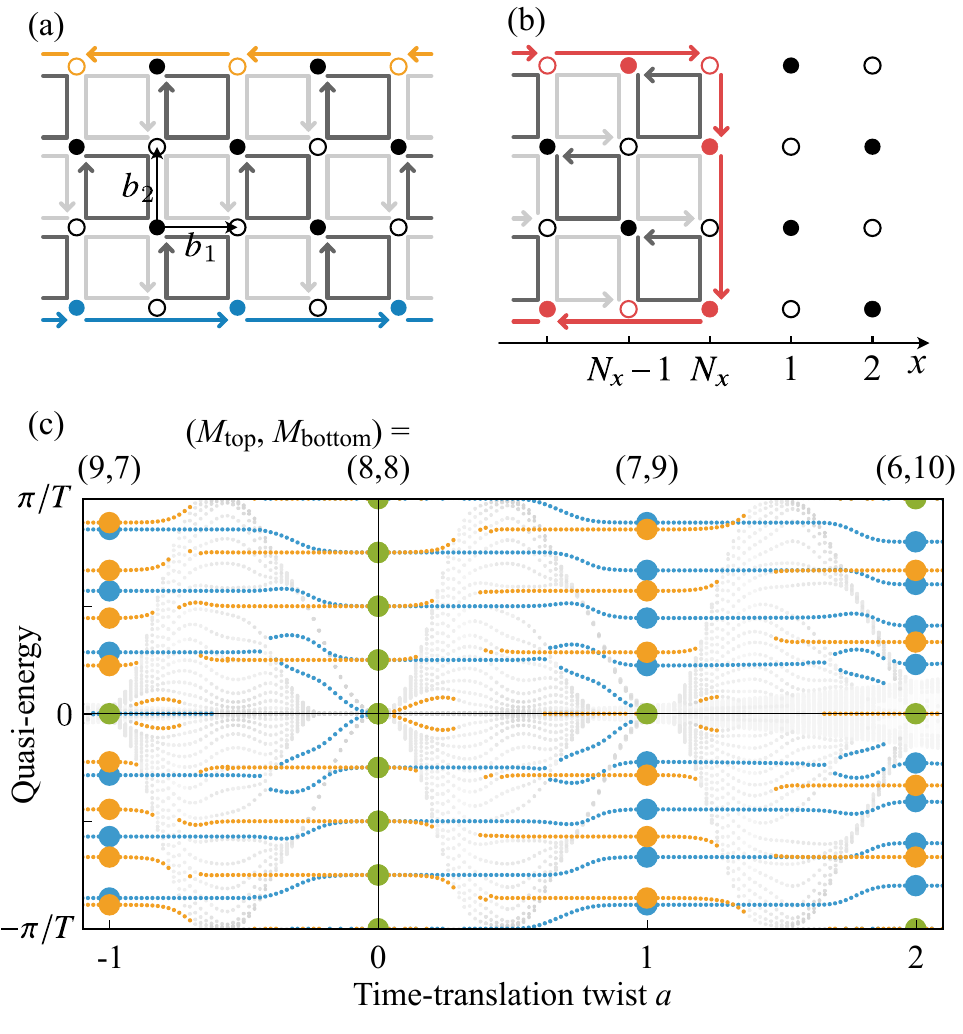}
    \caption{
    (a) The Floquet evolution in the anomalous Floquet Anderson insulator in the cylinder geometry.
    Filled (open) circles are sites on the A (B) sublattice.
    (b) Boundary modes are transferred from the top to bottom boundary during a continuous change of the twist parameter from $a=0$ to $1$.
    (c) The quasi-energies of the AFAI without disorder as a function of the time-translation twist $a$. 
    The system size is $16\times 16$.
    Boundary modes are highlighted in blue (bottom) and orange (top), and in green when overlapped.
    In particular, the boundary-mode quasi-energies at integer $a$ are shown by large circles.
    The other modes are plotted in grayscale as determined by the inverse participation ratio (larger when darker).
        \label{fig:afai_numerical}
    }
\end{figure}

\textit{\color{blue}Anomalous Floquet Anderson insulator.---}
We consider the AFAI \cite{PhysRevX.3.031005,PhysRevX.6.021013}, whose Hamiltonian $H(t)=H_n+H_\text{disorder}$ for $t\in[(n-1)T/5,nT/5]$, $n=1,\ldots,5$, consists of five sectors,
\begin{align}
    &H_n=\sum_{r\in A} H_{n,r},\quad 
    H_{n,r}=-J\left(c_r^\dagger c_{r+b_n}^{\ }+\text{H.c.}\right),
    \label{eq:hamiltonian_afai_drive}
\end{align}
for $n\in[1,4]$ and $H_5=\sum_ru_r n_r$, where $r=(x,y)$ labels the sites of an $N_x\times N_y$ square lattice, $ n_r=c_r^\dagger c_r^{\ }$, $b_1=-b_3=(1,0)$ and $b_2=-b_4=(0,1)$ are the bond vectors, $J$ is the hopping amplitude, $u_r$ is an on-site disorder potential, and $H_\text{disorder}$ is a time-independent disorder Hamiltonian that localizes the bulk.
At $J=5\pi/2T$ each sector swaps the fermions on the corresponding bonds, the single-particle Floquet operator $U_F=\mathcal{T}\exp[-i\int_0^Tdt\,h(t)]$ of the bulk is the identity under periodic boundary conditions, and on a cylinder the fermions on the two boundaries flow in opposite directions, decoupled from the bulk even for $u_r\neq0$ \cite{PhysRevX.3.031005} [Fig.~\ref{fig:afai_numerical} (a)].

On the bottom boundary, $U_F$ is the shift $c_j\to c_{j+1}$ on the $N=N_x/2$ boundary sites, with the phase $e^{-i\theta}$, $\theta=\alpha+\theta_0$, on the bond that crosses the seam, where $\alpha$ is an imposed U(1) twist $c_{N+1}=e^{i\alpha}c_1$ and $\theta_0$ is the disorder phase [Eq.~(\ref{eq:shifttwist_transfermatrix})].
The time-translation twist by $a$ periods replaces $\theta\to\theta-a\epsilon T$ for a mode of quasi-energy $\epsilon$, and the self-consistent solution of the resulting eigenvalue problem is (End Matter)
\begin{align}
    \epsilon_j=\frac{2\pi j+\alpha+\theta_0}{(N+a)T},
     \quad 
    j\in[1,N+a].
    \label{eq:quasienergy_afai}
\end{align}
The bottom boundary thus hosts $N+a$ modes: a boundary fermion is translated by one site per period and is delayed by $aT$ when it crosses the seam, so that the boundary behaves as a chain of $N+a$ effective sites [Fig.~\ref{fig:tttwist} (b)].
The top boundary, related to the bottom one by inversion, hosts $N-a$ modes, and the twist transfers $a$ modes from the top to the bottom.
For a general winding number, the spectral flow under the U(1) twist shifts the quasi-energies to their $\mathcal{W}$-th neighbors per $2\pi$, and the same argument gives
\begin{align}
    M(\mathcal{W},a)=N+\mathcal{W}a.
    \label{eq:boundarymodenumber_afai}
\end{align}
(End Matter and Sec.~\ref{sec:windingnumber} in \cite{SM}).
The two boundaries have opposite winding numbers, $M_\text{bottom(top)}(a)=M(\pm\mathcal{W},a)$, which is (\ref{eq:afairelation}).

We confirm (\ref{eq:boundarymodenumber_afai}) including the bulk by delaying the drive in proportion to $x$,
\begin{align}
    H_{n,r}(t)\to H_{n,r}\left(t-\frac{(r+b_n/2)_xaT}{N_x}\right),
    \label{eq:afaibulk_twist}
\end{align}
where $H_{n,r}(t)$ coincides with $H_{n,r}$ for $t\in[(n-1)T/5,nT/5]$ and vanishes otherwise, and $(r+b_n/2)_x$ is the $x$ coordinate of the bond center; a boundary mode is then delayed by $aT$ when it goes once around the cylinder [Fig.~\ref{fig:tttwist} (b)].
Figure~\ref{fig:afai_numerical} (c) shows the quasi-energies of a $16\times16$ cylinder as functions of $a$ (End Matter and Sec.~\ref{sec:afai_details} in \cite{SM}): the number of bottom (top) boundary modes is $8+a$ ($8-a$) for integer $a$. %, and the spectral interval of the bottom boundary under the U(1) twist is accordingly smaller than that of the top.

The transfer is a genuine pumping between the boundaries rather than a redistribution along each boundary.
A continuous change of the twist from $a=0$ to $1$ amounts, by (\ref{eq:afaibulk_twist}), to a backward evolution by $xaT/N_x$: for a long circumference the electrons near $x=N_x$ are evolved backward by almost $aT$ while those near $x=1$ are not, so that the line between them acts as a virtual boundary, absent in the Floquet dynamics, along which a mode is carried from the top to the bottom [Fig.~\ref{fig:afai_numerical} (b)].
This is the analogue of Laughlin's argument, in which the inserted flux drives the charge across the cylinder.
Conversely, a time translation that depends on $y$ instead of $x$ [Fig.~\ref{fig:tttwist} (c)] evolves the top boundary backward by $aT$ relative to the bottom one and translates the boundary modes along the boundary by $\mathcal{W}a$ sites; the pumping between the boundaries and the quantized charge pumping along the boundary \cite{PhysRevX.6.021013} are thus related by a $\pi/2$ rotation, in the same way as Laughlin's and Halperin's pictures of the quantum Hall effect are \cite{PhysRevB.23.5632,PhysRevB.25.2185}.
The relation 
(\ref{eq:boundarymodenumber_afai})
is also consistent with the $\pi$ Majorana mode bound to a time vortex in the Floquet Kitaev honeycomb model
\cite{Kishony2025}, which is a time-translation twist contracted to a plaquette: without the vortex, a plaquette binds two modes at $\epsilon= \pm \pi / 2 T$, i.e., (\ref{eq:quasienergy_afai}) with $N=2$ and $\alpha+\theta_0=\pi$, and the vortex removes one of them $(N+a=1)$, leaving the $\pi$ Majorana mode at $\epsilon=\pi / T$.

\textit{\color{blue}Matrix product unitaries.---}
The relation (\ref{eq:boundarymodenumber_afai}) extends to interacting systems with $\mathcal{W}$ replaced by $\text{ind}=\log_dI_\text{GNVW}$.
The boundary dynamics of a chiral Floquet phase is a one-dimensional locality-preserving unitary, i.e., a quantum cellular automaton, characterized by the GNVW index \cite{Gross2012}.
We represent it by a spatially translation-invariant MPU on a ring of $N$ qudits of dimension $d$, $\langle\bm{j}|\hat{U}_F|\bm{k}\rangle=\text{Tr}[\mathcal{U}_{j_1k_1}\cdots\mathcal{U}_{j_Nk_N}]=\text{Tr}[(\mathcal{U}_N)_{\bm{j}\bm{k}}]$, where $\mathcal{U}_{jk}$ are $D\times D$ matrices, $\mathcal{U}_N$ is the tensor obtained by blocking $N$ sites, and the trace is over the $D$-dimensional virtual space \cite{Cirac_2017,PhysRevB.98.245122}.
For a sufficiently large $N$, $\mathcal{U}_N$ is \textit{simple} \cite{Cirac_2017}, and the GNVW index is given by the rank-ratio index $r_N/l_N=I_\text{GNVW}^2$, where the right (left) rank $r_N$ ($l_N$) is the rank of $\mathcal{U}_N$ as a map between the legs separated by a diagonal cut, with $r_Nl_N=d^{2N}$ [Eqs.~(\ref{eq:isometryseparation}) and (\ref{eq:rankratioindex})].

In the spacetime tensor network of $\hat{U}_F^t(a)$, the time-translation twist displaces the horizontal legs that cross the seam by $a$ rows [Eq.~(\ref{eq:twistnetwork})].
Extracting a single row, $\hat{U}_F(a)$ is the operator on the $N$ physical qudits and $|a|$ auxiliary qudits of dimension $D$ obtained by attaching $|a|$ translation tensors to $\mathcal{U}_N$ [Eq.~(\ref{eq:mpu_withtwist})].
The twisted leg carries the right (left) virtual index of $\mathcal{U}_N$ to the output, so that $\rank\hat{U}_F(\pm1)=r_N\,(l_N)$, and each further twist multiplies the rank by $D$; hence $\rank\hat{U}_F(a)/\rank\hat{U}_F(-a)=I_\text{GNVW}^{2\,\text{sign}[a]}$.
For $t>1$ we block $N/t$ sites and contract the twisted legs of the $t$ rows using the isometry and separation properties of simple tensors.
Carrying this out for $t=2$ and $3$ (End Matter and Sec.~\ref{sec:shift} in \cite{SM}), we find
\begin{align}
    \rank \hat{U}_F^t(a)
    =
    d^N\cdot
    \left\{
    \begin{array}{ll}
        I_\text{GNVW}^{-t}D^{|a|-t}\quad &(a\le -t)\\[+3pt]
        I_\text{GNVW}^{a}\quad &(|a|\le t)\\[+3pt]
        I_\text{GNVW}^tD^{a-t}\quad&(a\ge t)
    \end{array}
    \right.
    \label{eq:mpu_floquetrank}
\end{align}
for $a\in\mathbb{Z}$, which we conjecture to hold for arbitrary $t\in\mathbb{N}$.
Since the rank of a unitary operator on $N$ qudits is $d^N$, Eq.~(\ref{eq:mpu_floquetrank}) states that for $|a|\le t$ the twisted boundary Floquet operator acts on $N+\text{ind}\cdot a$ effective qudits, in parallel with (\ref{eq:boundarymodenumber_afai}).
The opposite boundary has the index $-\text{ind}$, and the ratio of the two ranks, from which $D$ drops out, is (\ref{eq:mpurelation}).
We verified (\ref{eq:mpu_floquetrank}) exactly for the shift operator, $\rank\hat{S}^t(a)=d^{N-a-2\min(N,t,-a)}$, and numerically for brickwork RUCs on qubits with ($I_\text{GNVW}=2$) and without ($I_\text{GNVW}=1$) a shift layer, where it holds once $\min(t,|a|)\lesssim N/2$ (Secs.~\ref{sec:shiftrank} and \ref{sec:ruc} in \cite{SM}); the relation thus extends to spatially non-uniform circuits.

\begin{figure}
    \centering
    \includegraphics[width=0.47\textwidth]{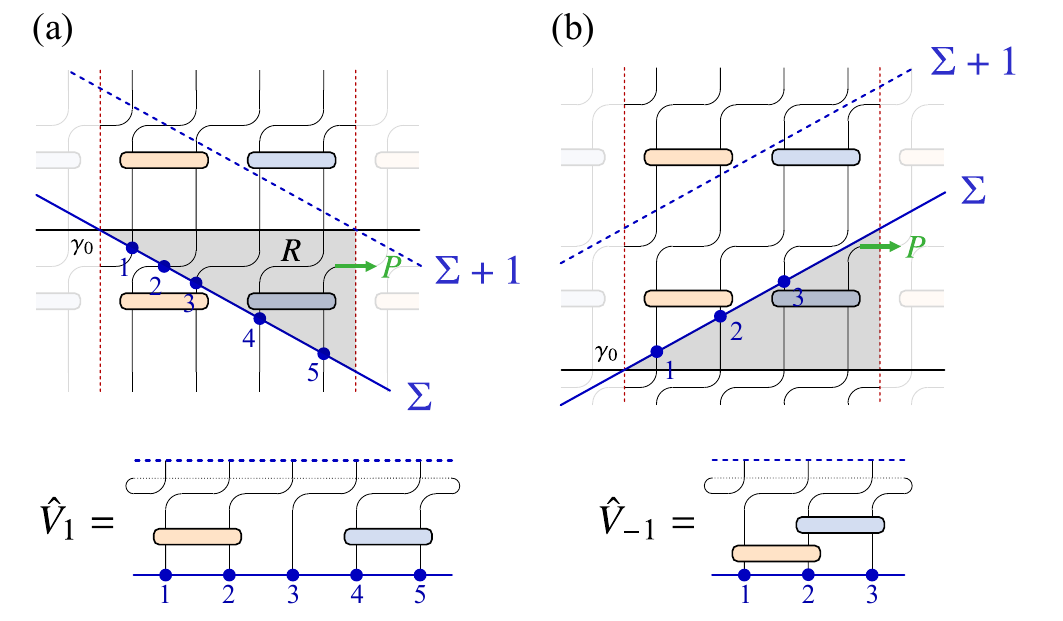}
    \caption{
        Tilted time slices for the circuit $\hat{U}_F=\hat{S}\hat{U}_4$ on $N=4$ qubits (two-site gates on $(1,2)$ and $(3,4)$ followed by a right shift, $I_\text{GNVW}=2$); time runs upward and the dashed vertical lines are the seam $x=L$.
        (a) $a=+1$: the slice $\Sigma$ (blue) descends by one period across the ring and crosses $N+1=5$ wires, against $4$ for the flat slice; the extra wire is the shift wire. % that leaves the region $R$ through the seam ($P$).
        The slab between $\Sigma$ and $\Sigma+1$ is $\hat{V}_1$ on 5 sites shown below.
        (b) $a=-1$: $\Sigma$ crosses $N-1=3$ wires, and the corresponding $\hat{V}_{-1}$ on 3 sites is shown below.
        %(c), (d) The slabs redrawn as circuits: the same gates followed by a shift on $N\pm1$ wires.
        \label{fig:tilted}
    }
\end{figure}

\textit{\color{blue}Tilted time slice.---}
The rank formula has a geometric origin that also fixes the spectrum of $\hat{U}_F(a)$ and does not rely on translation invariance.
We can deform a tensor network with the time-translation twist to a position-dependent delay as was done for the AFAI (\ref{eq:afaibulk_twist}), which defines a tilted
time slice $\Sigma$ that descends by $a$ periods as it goes once around the ring (Fig.~\ref{fig:tilted}).
The time-translation twist on flat slices is the periodic boundary condition on tilted slices, the lattice counterpart of the relation between an energy twist and a boost \cite{PhysRevB.106.155128}.
For circuits composed of local unitary gates and wires, the deformation consists of reordering of gates, that is, moving gates connected to input legs to those connected to output legs or vice versa, and removing or adding caps and cups.

Let $\Sigma+1$ be the time slice translated by one period from $\Sigma$. 
There are several options of choosing $\Sigma$ depending on which gates we place between $\Sigma$ and $\Sigma+1$ and which legs the time slices cut.
Properly choosing the time slice, we can make the evolution operator $\hat{V}_a$ from $\Sigma$ to $\Sigma+1$ a unitary operator on the Hilbert space $\mathcal{H}_\Sigma$ of the wires crossing $\Sigma$, where the Hilbert space dimension is minimized by (Fig.~\ref{fig:tilted}, Sec.~\ref{sec:tilted} in \cite{SM})
\begin{align}
    \dim\mathcal{H}_\Sigma=d^{\,N+\text{ind}\cdot a}.
    \label{eq:tilted_dimension}
\end{align}
The dimension follows from counting wires: deforming the flat slice into $\Sigma$ sweeps $a$ periods of traffic through the seam, and in each period the wires that cross the seam rightward are gained while those crossing it leftward are lost; the ratio of the two dimensions is $I_\text{GNVW}$ \cite{Gross2012,PhysRevX.6.041070}.
During the deformation, the nonzero spectra are preserved since reorderings are unitary transformations, and caps and cups are null spaces.
Equation (\ref{eq:tilted_dimension}) states that $\hat{U}_F(a)$ has exactly $d^{\,N+\text{ind}\cdot a}$ nonzero eigenvalues, all of modulus one, the remainder being nilpotent (for the shift operator, Jordan blocks of size at most $|a|$, Sec.~\ref{sec:shiftrank} in \cite{SM}).
As is also evident from Fig.~\ref{fig:tilted},
\begin{align}
    \text{Tr}\,\hat{U}_F^t(a)=\text{Tr}\,\hat{V}_a^t,
    \label{eq:tilted}
\end{align}
for all $t\ge 1$ (see Sec.~\ref{sec:tilted} in \cite{SM}).
Unlike the trace, the rank depends on the slicing: $\rank \hat{V}_a^t=\dim\mathcal{H}_\Sigma$ for all $t$, whereas $\rank\hat{U}_F^t(a)$ contains the not-yet-decayed nilpotent part of dimension $(I_\text{GNVW}^{-\text{sign}[a]}D)^{|a|-t}$ for $|a|>t$, and equals $\dim\mathcal{H}_\Sigma$ only for $|a|\le t$, which is the condition in (\ref{eq:mpurelation}).
For the shift by $v$ sites per period, $\hat{V}_a$ is the shift on $N+av$ sites and $\text{Tr}\,\hat{S}^t(a)=d^{\gcd(N+av,\,vt)}$ exactly; for the brickwork circuit of Fig.~\ref{fig:tilted}, $V_{\pm1}$ is the same circuit on $N\pm1$ sites.
The counting does not require the index to be an integer power of a single qudit dimension and applies, e.g., to the circuit of Gong, Nahum, and Piroli \cite{Gong2022} with $I_\text{GNVW}=p/q$ \footnote{In this circuit, sites of dimensions $p$ and $q$ alternate and every gate maps $\mathbb{C}^p\otimes\mathbb{C}^q\to\mathbb{C}^q\otimes\mathbb{C}^p$, so that in each period a $p$-dimensional degree of freedom moves to the right and a $q$-dimensional one to the left by one unit cell, $I_\text{GNVW}=p/q$ \cite{PhysRevX.6.041070}. With $N/2$ unit cells, $\dim\mathcal{H}_\Sigma=(pq)^{N/2}(p/q)^a=p^{N/2+a}q^{N/2-a}$: each unit of twist replaces a $q$-dimensional site of the slice by a $p$-dimensional one.}.

\textit{\color{blue}Twisted spectral form factor.---}
Hilbert-space pumping can be detected spectrally, without identifying individual boundary modes.
The twisted spectral form factor (SFF) $K(t,a)=|\mathrm{Tr}\,\hat U_F^t(a)|^2=|\mathrm{Tr}\,\hat{V}_a^t|^2$ probes the unitary evolution on the tilted slice and is insensitive to the nilpotent sector.
For nondegenerate eigenphases, its infinite-time average equals $\dim\mathcal H_\Sigma$, so that $\overline K(a)/\overline K(0)=I_\mathrm{GNVW}^a$.
We verify this response in random circuits with integer and rational indices; the averaged SFF exhibits a ramp and a plateau whose height measures the effective boundary dimension (End Matter).

%\textit{\color{blue}Twisted spectral form factor.---}
%Since $V_a$ is unitary, Hilbert-space pumping has a spectral signature that does not require identifying individual boundary modes.
%The SFF $K(t,a)=|\text{Tr}\,\hat{U}_F^t(a)|^2=|\text{Tr}\,V_a^t|^2$ is insensitive to the nilpotent part of $\hat{U}_F(a)$ and, for a chaotic boundary circuit, follows the ramp $\langle K\rangle=t$ of the circular unitary ensemble (CUE) \cite{PhysRevX.8.041019,PhysRevLett.121.264101} up to the Heisenberg time $t_H=d^{\,N+\text{ind}\cdot a}$ and then saturates at the plateau $d^{\,N+\text{ind}\cdot a}$.
%We confirmed this for brickwork RUCs on qubits with ($I_\text{GNVW}=2$) and without ($I_\text{GNVW}=1$) a shift layer, and for the circuit of \cite{Gong2022} with $(p,q)=(3,2)$ (Sec.~\ref{sec:sff} in \cite{SM}): in every realization the number of nonzero eigenvalues of $\hat{U}_F(a)$ is $2^{N+a}$, $2^N$, and $3^{N/2+a}2^{N/2-a}$, respectively, with unit moduli to machine precision, and the disorder-averaged SFF shows the CUE ramp for every $a$ with the plateau and the Heisenberg time moving by $I_\text{GNVW}^{\pm1}$ per unit of twist.
%The late-time twisted SFF thus measures $\text{ind}\cdot a$ as $\log_d$ of the plateau, including rational indices.

\textit{\color{blue}Non-invertible symmetries.---}
The twisted rank is defined for any MPO, not only for unitaries, and the two twisted ranks carry different information.
Let $\hat{\mathcal L}$ be the MPO, in canonical form, of a non-invertible symmetry (a topological line) \cite{Bultinck2017,Lootens2023,Shao2023,SchaferNameki2024}, such as the Kramers--Wannier (KW) duality of the Ising chain \cite{SeibergSeifnashriShao2024}, a Wilson line of a non-abelian group, or the $\tau$ line of the Fibonacci chain \cite{Feiguin2007}, and let $\hat{\mathbbm{1}}$ be the unit of the MPO algebra (the identity operator, or the constraint projector of a fusion chain).
Inserting oppositely oriented unit time-translation defects extracts the right and left ranks of the blocked MPO tensor, rank $\hat{\mathcal{L}}(+1)=r_N$ and rank $\hat{\mathcal{L}}(-1)=l_N$. The antisymmetric and symmetric combinations of their logarithms probe the flow index and quantum dimension, respectively:
%We find (Sec.~\ref{sec:noninv} in \cite{SM})
\begin{align}
    \frac{1}{2}\log\frac{\rank\hat{\mathcal{L}}(1)}{\rank\hat{\mathcal{L}}(-1)}
    &=\log\text{Ind}(\hat{\mathcal L}),
    \notag\\
    \frac{1}{2}\log\frac{\rank\hat{\mathcal{L}}(1)\,\rank\hat{\mathcal{L}}(-1)}{\rank\hat{\mathbbm{1}}(1)\,\rank\hat{\mathbbm{1}}(-1)}
    &=\log d_{\hat{\mathcal L}}.
    \label{eq:noninv}
\end{align}
%The antisymmetric part is (\ref{eq:mpurelation}) at $t=1$: $\text{Ind}(\mathcal{L})$ is the GNVW index when $\mathcal{L}$ is invertible, 
These relations hold for the examples 
in End Matter,
exactly on the tensor-product chains considered and in the thermodynamic limit on the Fibonacci chain. 
%Here $d$ r is the index of Ma. Li , and Cheng on the the index of Ma, Li, and Cheng on the symmetric subalgebra for a lattice duality \cite{MaLiCheng2024}, and the Jones--Lim index on a fusion spin chain \cite{JonesLim2023}.
Here $d_{\hat{\mathcal L}}$ is the quantum dimension of the line, the Perron--Frobenius eigenvalue of its fusion matrix \cite{Chang2019}.
%; it vanishes for an MPU,  where $r_N l_N = d^{2N}$ \cite{Cirac_2017}.
The antisymmetric combination measures the directional flow of the line and
reduces to the unit-twist rank ratio (\ref{eq:mpurelation}) when $\hat{\mathcal L}$ is an MPU.
The symmetric combination is $\tfrac12\log(r_Nl_N/d^{2N})$ on a tensor-product chain,
the excess of the product of the two ranks over that of a unitary;
for an MPU it is zero, since $r_Nl_N=d^{2N}$ \cite{Cirac_2017},
whereas for a non-invertible line it is $\log d_{\hat{\mathcal L}}>0$.
%The antisymmetric response measures directional flow and extends the unit-twist MPU rank ratio. The symmetric response measures the additional dimension associated with the line; it vanishes for an MPU, where $r_Nl_N=d^{2N}$ \cite{Cirac_2017}.
%The symmetric part is $\tfrac12\log_d(r_Nl_N/d^{2N})$ in terms of the right and left ranks of the MPO tensor $\mathcal{L}_N$, since $\rank\hat{\mathcal{L}}(\pm1)=r_N\,(l_N)$ as for the MPU; it vanishes identically for an MPU, where $r_Nl_N=d^{2N}$ \cite{Cirac_2017}, whereas for a non-invertible $\mathcal{L}$ it is the quantum dimension $d_{\mathcal{L}}$ of the line, the Perron--Frobenius eigenvalue of its fusion matrix \cite{Chang2019}.
%Equation (\ref{eq:noninv}) holds exactly at finite $N$ for the $\mathbb{Z}_n$ KW dualities ($\text{Ind}=n^{-1/2}$, $d_{\mathcal{L}}=\sqrt{n}$), for the $\text{Rep}(D_8)$ operators of \cite{MaLiCheng2024}, which separate the two parts ($\text{Ind}=1$, $d_{\mathcal{L}}=2$), and for the $\text{Rep}(S_3)$ Wilson line ($d_{\mathcal{L}}=2$); for the Fibonacci line the twisted ranks are Fibonacci numbers, $\rank\hat{\mathcal{L}}_\tau(\pm1)=F_{N+3}$ against $\rank\hat{\mathbbm{1}}(\pm1)=F_{N+2}$, and their ratio converges to $d_\tau=(1+\sqrt5)/2$.
For topological injective MPOs on tensor-product chains, the corresponding algebraic definitions of the lattice quantum dimension and index in terms of left and right ranks were given in Ref.~\cite{Inamura2026}. Here the time-translation defect supplies a probe that extracts those ranks and hence the invariants. Equality of the lattice and categorical quantum dimensions requires additional structure and is established here for the explicit examples, rather than inferred from canonical form alone.
The normalization by the unit of the MPO algebra also allows us to treat constrained fusion-chain Hilbert spaces.
An untwisted closed-ring rank generally counts the sectors on which the symmetry acts nontrivially and need not determine its quantum dimension. The time-translation defect changes the contraction at the seam, exposing the left and right ranks that carry this information.

%The untwisted operator $\hat{\mathcal{L}}(0)$ on a closed ring, by contrast, contains the projector onto the sectors on which it acts invertibly (the $\mathbb{Z}_2$-even sector for the KW duality), so that its rank counts those sectors and is insensitive to $d_{\mathcal{L}}$ in general.
%In the twisted sector the bond leg of the MPO is left open and no projection occurs: the twisted rank counts the fusion paths with one more line attached, and the extra fusion matrix contributes its Perron--Frobenius eigenvalue $d_{\mathcal{L}}$.

\textit{\color{blue}Outlook.---}
Open questions are whether Hilbert-space pumping is accompanied by a transfer of quantum information between the boundaries, and what the two-dimensional counterpart of (\ref{eq:noninv}) is.
%The fermionic index $\sqrt2$ of the radical chiral Floquet phases \cite{PhysRevB.96.245116,PhysRevB.99.085115} would correspond to the pumping of a single Majorana mode between the boundaries, as for the $\pi$ Majorana mode bound to a time vortex \cite{Kishony2025}; its description requires the fermionic MPU formalism \cite{Piroli2021} and is left for future work.
The pumping extends to fermionic boundaries with the index $\sqrt2$ of the radical chiral Floquet phases \cite{PhysRevB.96.245116,PhysRevB.99.085115}.
For the Majorana translation, described by the fermionic MPU formalism \cite{Piroli2021}, a unit twist transfers a single Majorana mode between the boundaries, as for the $\pi$ Majorana mode bound to a time vortex \cite{Kishony2025}: the twisted ranks obey (\ref{eq:mpurelation}) with $\text{ind}=1/2$, and the tilted slice carries $2N+a$ Majoranas (Sec.~\ref{sec:fermion} in \cite{SM}).
The plateau of the twisted SFF then depends on the spin structure of the ring, and the dimension $\sqrt2^{\,2N+a}$ is recovered as the geometric mean over the two spin structures.
Since the twist requires only a position-dependent delay of the drive, our predictions could be tested in cold-atom \cite{Wintersperger2020,PhysRevLett.130.043201}, photonic \cite{Maczewsky2017,Mukherjee2017,Li2023}, phononic \cite{Peng2016}, and microwave \cite{PhysRevX.5.011012} realizations of the anomalous Floquet phase, and in quantum simulators \cite{Will2025,Evered2025}, where the twisted SFF is the return amplitude of the twisted circuit.

\textit{\color{blue}Acknowledgement.---}
The authors used Claude Code (Opus 5.5) for code development and debugging and manuscript revisions.
The authors critically reviewed and independently verified all outputs through analytical and numerical checks, and retain full responsibility for this work.
R.N. is supported by
JSPS KAKENHI Grant No.~JP24K06926.
S.R. is supported by the
National Science Foundation under Award No.\ DMR-2409412.

\bibliography{chiralunitarytwist.bib}

\vspace{\baselineskip}
\begin{center}\textbf{END MATTER}\end{center}
\appendix

\section{Boundary dynamics of the AFAI}
\label{em:afal}

\begin{figure}
    \centering
    \includegraphics[width=0.47\textwidth]{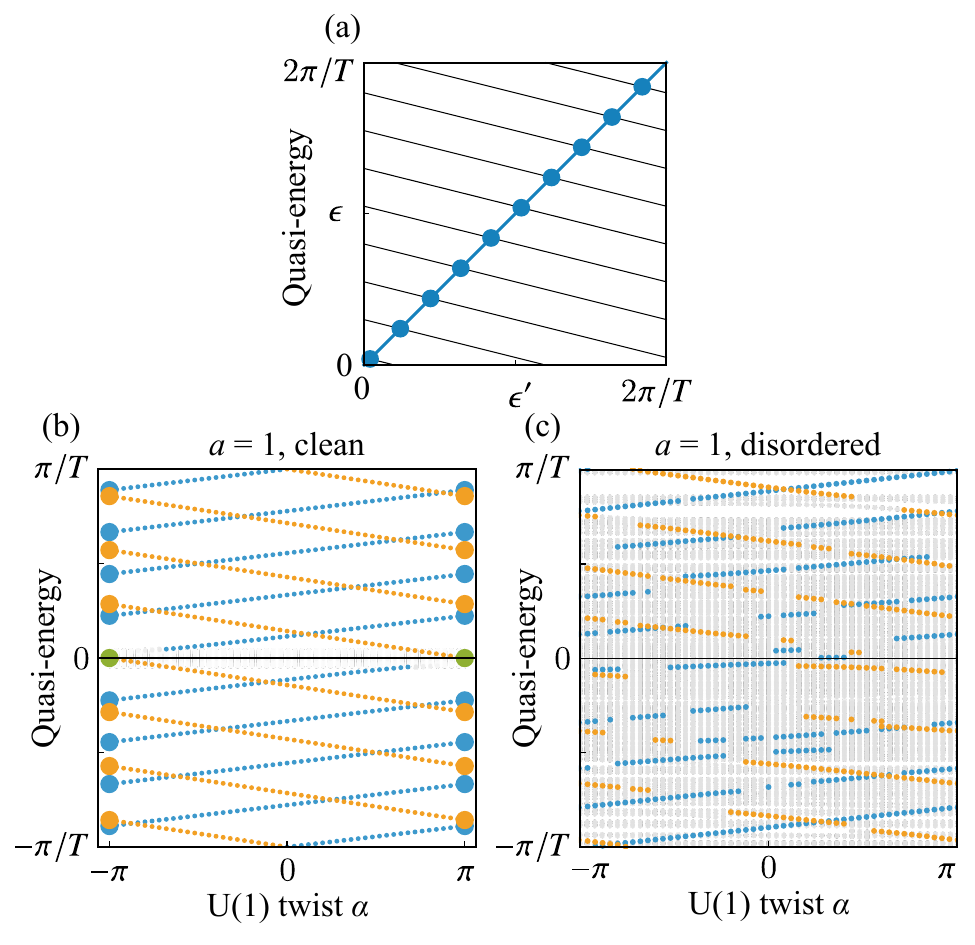}
    \caption{
    (a) The spectral flow of the quasi-energies of $U_F(e^{-i(\alpha+\theta_0-a\epsilon' T)})$ (black) as a function of $\epsilon'$ and the line $\epsilon=\epsilon'$ (blue). 
    The self-consistent solutions of $\phi U_F(e^{-i(\alpha+\theta_0-a\epsilon T)})=e^{-i\epsilon T}\phi$ are their intersections (blue points). 
    The figure corresponds to winding number $\mathcal{W}=1$ and time-translation twist $a=2$.
    (b),(c) The quasi-energies of the AFAI as a function of the U(1) twist $\alpha$ with $a=1$ (b) without and (c) with disorder set by parameters $u_r=0$, $\lambda=\pi$ and $T\delta V=4$.
        \label{fig:afai_edge}
    }
\end{figure}

Focusing on the bottom boundary of the AFAI with $J=5\pi/2T$ [Fig.~\ref{fig:afai_edge} (a)], the Floquet evolution of the boundary fermions $\bm{c}=(c_1,\cdots,c_N)$ is
\begin{align}
    \bm{c}(T)
    =
    U_F(e^{-i\theta})
    \bm{c}(0)
    \equiv
    \begin{pmatrix}
        0 & & & & e^{-i\theta}\\
        1 & 0 & & &\\
        & \ddots &\ddots & & \\
        & & 1 & 0 & \\
        & & & 1 & 0
    \end{pmatrix}
    \bm{c}(0),
    \label{eq:shifttwist_transfermatrix}
\end{align}
where $\theta=\alpha+\theta_0$, $\alpha$ is the U(1) twist phase $c_{N+1}=e^{i\alpha}c_1$, and $\theta_0=N\pi+\sum_{j=1}^Nu_jT/5$ is the disorder phase concentrated at the boundary.
Under the time-translation twist, $\theta\to\theta-a\epsilon T$, so that the Floquet operator depends on the quasi-energy $\epsilon$ and the spectrum is determined self-consistently.
For an eigenvector $\phi=(\phi_1,\cdots,\phi_N)$ with $\phi U_F=e^{-i\epsilon T}\phi$, the recursion $\phi_1=e^{i\epsilon T}\phi_2=\cdots$ gives the consistency condition $e^{i(\alpha+\theta_0)-i(N+a)\epsilon T}=1$, whose solutions are (\ref{eq:quasienergy_afai}).
For a general winding number, we regard $\epsilon'$ in $U_F(e^{-i(\alpha+\theta_0-a\epsilon'T)})$ as an independent parameter; the quasi-energies are the intersections of its spectral flow as a function of $\epsilon'$ with the line $\epsilon=\epsilon'$ [Fig.~\ref{fig:afai_edge} (a)].
Varying $\epsilon'$ from $0$ to $2\pi/T$ is equivalent to varying $\alpha$ from $0$ to $-2\pi a$, during which the spectrum is shifted to its $-\mathcal{W}a$-th neighbor by the definition of the winding number, which gives (\ref{eq:boundarymodenumber_afai}) (Sec.~\ref{sec:windingnumber} in \cite{SM}).

%The relation (\ref{eq:boundarymodenumber_afai}) is consistent with the $\pi$ Majorana mode bound to a time vortex in the Floquet Kitaev honeycomb model \cite{Kishony2025}, a time-translation twist contracted to a plaquette: without the vortex, a plaquette binds two modes at $\epsilon=\pm\pi/2T$, i.e., (\ref{eq:quasienergy_afai}) with $N=2$ and $\alpha+\theta_0=\pi$, and the vortex ($a=-1$) removes one of them, leaving the $\pi$ Majorana mode at $\epsilon=\pi/T$.

In the numerical calculation of Fig.~\ref{fig:afai_numerical} (c), the position-dependent drive (\ref{eq:afaibulk_twist}) no longer separates the five sectors cleanly, so that the bulk spectrum deviates from $0$; we recover the bulk insulation by shortening the interval $\Delta T$ during which each $H_n$ is active, keeping $J\Delta T=\pi/2$ (Sec.~\ref{sec:afai_details} in \cite{SM}).
Figures \ref{fig:afai_edge} (b),(c) show the quasi-energy as a function of $\alpha$ at $a=1$.
Boundary modes are identified by an eigenfunction weight of at least $0.8$ on $y=1$ or $y=N_y$, and the disordered case uses $u_r=0$, a staggered potential $\lambda=\pi$, and a random potential $T\delta V=4$ in (\ref{eq:disorderhamiltonian}), which is in the Anderson localized phase.
The spectral interval of the bottom boundary under the U(1) twist is smaller than that of the top.

\section{Twisted rank of a matrix product unitary}
\label{em:mpu}
Any tensor $\mathcal{U}$ generating an MPU becomes a simple tensor $\mathcal{U}_{\ge p}$ by blocking more than $p$ sites, where $p$ is at most $D^4$ \cite{Cirac_2017}.
A simple tensor $\mathcal{V}$ satisfies the isometry and separation identities \cite{Cirac_2017,PhysRevB.98.245122}
\begin{align}
    &\begin{tikzpicture}[baseline={([yshift=-0.1cm]current bounding box.center)}]
        \node[draw, fill=white, minimum size=16pt, font=\bfseries] (u1) {$\mathcal{V}$};
        \node[left = 0.1 of u1] (u1l) {};
        \node[right = 0.1 of u1] (u1r) {};
        \node[draw, fill=white, minimum size=16pt, font=\bfseries, above = 0.3 of u1] (bu1) {$\bar{\mathcal{V}}$};
        \node[left = 0.1 of bu1] (bu1l) {};
        \node[right = 0.1 of bu1] (bu1r) {};
        \draw[thin] (u1.north) -- (bu1.south);
        \draw[thin] (u1.south) -- ++(0, -0.2);
        \draw[thin] (bu1.north) -- ++(0, 0.2);
        %\node (m1r) at ($(u1r)!0.5!(bu1r)$) [draw, circle,fill=white] {};
        \draw[thin] (u1.west) -- (u1l.center) -- (bu1l.center) -- (bu1.west);
        %\draw[thin] (u1.east) -- (u1r.center) -- (m1r.south);
        %\draw[thin] (bu1.east) -- (bu1r.center) -- (m1r.north);
        \draw[thin] (u1.east) -- (u1r.center) -- node[midway, draw, fill=white, circle] (m1r) {} (bu1r.center) -- (bu1.east);
        %\draw[circle,fill=white] (m1r) {};
    \end{tikzpicture}
    \,
    =
    \,\,
    \begin{tikzpicture}[baseline={([yshift=-0.1cm]current bounding box.center)}]
        \draw[thin] (0,-1) -- ++(0,1.6);
    \end{tikzpicture}\quad, %\notag\\
    &\begin{tikzpicture}[baseline={([yshift=-0.1cm]current bounding box.center)}]
        \node[draw, fill=white, minimum size=16pt, font=\bfseries] (u1) {$\mathcal{V}$};
        \node[draw, fill=white, minimum size=16pt, font=\bfseries, above = 0.3 of u1] (bu1) {$\bar{\mathcal{V}}$};
        \node[draw, fill=white, minimum size=16pt, font=\bfseries, right = 0.4 of u1] (u2) {$\mathcal{V}$};
        \node[draw, fill=white, minimum size=16pt, font=\bfseries, above = 0.3 of u2] (bu2) {$\bar{\mathcal{V}}$};
        \draw[thin] (u1.north) -- (bu1.south);
        \draw[thin] (u2.north) -- (bu2.south);
        \draw[thin] (u1.south) -- ++(0, -0.2);
        \draw[thin] (bu1.north) -- ++(0, 0.2);
        \draw[thin] (u2.south) -- ++(0, -0.2);
        \draw[thin] (bu2.north) -- ++(0, 0.2);
        \draw[thin] (u1.west) -- ++(-0.2,0);
        \draw[thin] (bu1.west) -- ++(-0.2,0);
        \draw[thin] (u2.east) -- ++(0.2,0);
        \draw[thin] (bu2.east) -- ++(0.2,0);
        \draw[thin] (u1.east) -- (u2.west);
        \draw[thin] (bu1.east) -- (bu2.west);
    \end{tikzpicture}
    \,=\,
    \begin{tikzpicture}[baseline={([yshift=-0.1cm]current bounding box.center)}]
        \node[draw, fill=white, minimum size=16pt, font=\bfseries] (u1) {$\mathcal{V}$};
        \node[draw, fill=white, minimum size=16pt, font=\bfseries, above = 0.3 of u1] (bu1) {$\bar{\mathcal{V}}$};
        \node[draw, fill=white, minimum size=16pt, font=\bfseries, right = 0.7 of u1] (u2) {$\mathcal{V}$};
        \node[draw, fill=white, minimum size=16pt, font=\bfseries, above = 0.3 of u2] (bu2) {$\bar{\mathcal{V}}$};
        \draw[thin] (u1.north) -- (bu1.south);
        \draw[thin] (u2.north) -- (bu2.south);
        \draw[thin] (u1.south) -- ++(0, -0.2);
        \draw[thin] (bu1.north) -- ++(0, 0.2);
        \draw[thin] (u2.south) -- ++(0, -0.2);
        \draw[thin] (bu2.north) -- ++(0, 0.2);
        \draw[thin] (u1.west) -- ++(-0.2,0);
        \draw[thin] (bu1.west) -- ++(-0.2,0);
        \draw[thin] (u2.east) -- ++(0.2,0);
        \draw[thin] (bu2.east) -- ++(0.2,0);
        \node[right = 0.1 of u1] (u1r) {};
        \node[right = 0.1 of bu1] (bu1r) {};
        \node[left = 0.1 of u2] (u2l) {};
        \node[left = 0.1 of bu2] (bu2l) {};
        %\node (m1r) at ($(u1r)!0.5!(bu1r)$) [draw, circle,fill=white] {};
        %\draw[thin] (u1.east) -- (u1r.center) -- (m1r.south);
        %\draw[thin] (bu1.east) -- (bu1r.center) -- (m1r.north);
        \draw[thin] (u2.west) -- (u2l.center) -- (bu2l.center) -- (bu2.west);
        \draw[thin] (u1.east) -- (u1r.center) -- node[midway, draw, fill=white, circle] (m1r) {} (bu1r.center) -- (bu1.east);
    \end{tikzpicture},
    \label{eq:isometryseparation}
\end{align}
where $\bar{\mathcal{V}}_{jk}=\mathcal{V}_{kj}^\ast$ and the circle represents the right fixed point of the transfer matrix.
Assuming that $N$ is large enough for $\mathcal{U}_N$ to be simple, the GNVW index of $\hat{U}_F$ is evaluated by the rank-ratio index \cite{Cirac_2017,PhysRevB.98.245122}
\begin{align}
    \rank \left(
    \vcenter{\hbox{
    \begin{tikzpicture}
        \node[draw, fill=white, minimum size=16pt, font=\bfseries] (centerNode) {$\mathcal{U}_N$};
        \draw[thin] (centerNode.north) -- ++(0, 0.25);
        \draw[thin] (centerNode.south) -- ++(0, -0.25);
        \draw[thin] (centerNode.east) -- ++(0.25, 0);
        \draw[thin] (centerNode.west) -- ++(-0.25, 0);
        \draw[dotted,thick] (-0.5,0.5)  -- +(1,-1) ; % 右
    \end{tikzpicture}
    }}\right)\bigg/
    \rank \left(
    \vcenter{\hbox{
    \begin{tikzpicture}
        \node[draw, fill=white, minimum size=16pt, font=\bfseries] (centerNode) {$\mathcal{U}_N$};
        \draw[thin] (centerNode.north) -- ++(0, 0.25);
        \draw[thin] (centerNode.south) -- ++(0, -0.25);
        \draw[thin] (centerNode.east) -- ++(0.25, 0);
        \draw[thin] (centerNode.west) -- ++(-0.25, 0);
        \draw[dotted,thick] (-0.5,-0.5)  -- +(1,1) ; % 右
    \end{tikzpicture}
    }}\right)
    \equiv
    \frac{r_N}{l_N}
    %=
    %I_\text{RR}
    =
    I_\text{GNVW}^2,
    \label{eq:rankratioindex}
\end{align}
the ratio of the ranks of the singular value decompositions between the legs separated by the dotted lines, which we call the right and left ranks.
Since $r_Nl_N=d^{2N}$ \cite{Cirac_2017}, $r_N\,(l_N)=d^NI_\text{GNVW}^{\pm1}$.

The time-translation twist is implemented in the tensor network of $\hat{U}_F^t(a)$ by disconnecting the horizontal legs that cross the seam, shifting one side of them along the temporal direction, and reconnecting them,
\begin{align}
    \hat{U}_F^t(0)=
    \begin{tikzpicture}[baseline={([yshift=-0.1cm]current bounding box.center)}]
            \node[draw, fill=white, minimum size=16pt, font=\bfseries] (u1) {$\mathcal{U}_N$};
            \node[draw, fill=white, minimum size=16pt, font=\bfseries, above = 0.25 of u1] (ua) {$\mathcal{U}_N$};            
            %\draw[thin] (u1.north) -- ++(0, 0.25) node[] (u1t) {};
            \draw[thin] (u1.south) -- ++(0, -0.25) node[] (u1b) {};
            \draw[thin] (u1.west) -- ++(-0.25, 0) node[] (u1l) {};
            \draw[thin] (u1.east) -- ++(0.25, 0) node[] (u1r) {};
            \draw (u1l) arc (270:90:0.1) node[] (b12) {};
            \draw[thin] (b12.center) -- ++(0.05, 0) node[] (b13) {};
            \draw (u1r) arc (-90:90:0.1) node[] (b15) {};
            \draw[thin] (b15.center) -- ++(-0.05, 0) node[] (b16) {};
            \begin{scope}[on background layer]
                \draw[dotted] (b13.center) -- (b16.center);
            \end{scope}
            \draw[dotted] (u1b.center) -- ++(0, -0.25) {};
            \draw[thin] (ua.north) -- ++(0, 0.25) node[] (uat) {};
            %\draw[thin] (ua.south) -- ++(0, -0.25) node[] (uab) {};
            \draw[thin] (ua.west) -- ++(-0.25, 0) node[] (ual) {};
            \draw[thin] (ua.east) -- ++(0.25, 0) node[] (uar) {};
            \draw (ual) arc (270:90:0.1) node[] (ba2) {};
            \draw[thin] (ba2.center) -- ++(0.05, 0) node[] (ba3) {};
            \draw (uar) arc (-90:90:0.1) node[] (ba5) {};
            \draw[thin] (ba5.center) -- ++(-0.05, 0) node[] (ba6) {};
            \begin{scope}[on background layer]
                \draw[dotted] (ba3.center) -- (ba6.center);
            \end{scope}
            \draw[thin] (u1.north) -- (ua.south) {};
            %\draw[dotted] (u1t.center) -- (uab.center);
            \draw[dotted] (uat.center) -- ++(0, 0.25) {};
    \end{tikzpicture}\to
    \hat{U}_F^t(1)=
    \begin{tikzpicture}[baseline={([yshift=-0.1cm]current bounding box.center)}]
            \node[draw, fill=white, minimum size=16pt, font=\bfseries] (u1) {$\mathcal{U}_N$};
            \node[draw, fill=white, minimum size=16pt, font=\bfseries, above = 0.25 of u1] (ua) {$\mathcal{U}_N$};            
            %\draw[thin] (u1.north) -- ++(0, 0.25) node[] (u1t) {};
            \draw[thin] (u1.south) -- ++(0, -0.25) node[] (u1b) {};
            \draw[thin] (u1.west) -- ++(-0.25, 0) node[] (u1l) {};
            \draw[thin] (u1.east) -- ++(0.2, 0) node[] (u1r) {};
            \draw (u1l) arc (270:90:0.1) node[] (b12) {};
            \draw[thin] (b12.center) -- ++(0.05, 0) node[] (b13) {};
            \begin{scope}[on background layer]
                \draw[dotted] (b13.center) -- ++(1.5, 0) node[] (b14) {};
            \end{scope}
            \draw (b14) arc (90:-30:0.1) node[] (b15) {};
            \draw (u1r) arc (-90:-30:0.1) node[] (b16) {};
            \draw[dotted] (u1b.center) -- ++(0, -0.25) {};
            \draw[thin] (ua.north) -- ++(0, 0.25) node[] (uat) {};
            %\draw[thin] (ua.south) -- ++(0, -0.25) node[] (uab) {};
            \draw[thin] (ua.west) -- ++(-0.25, 0) node[] (ual) {};
            \draw[thin] (ua.east) -- ++(0.2, 0) node[] (uar) {};
            \draw (ual) arc (270:90:0.1) node[] (ba2) {};
            \draw[thin] (ba2.center) -- ++(0.05, 0) node[] (ba3) {};
            \begin{scope}[on background layer]
                \draw[dotted] (ba3.center) -- ++(1.5, 0) node[] (ba4) {};
            \end{scope}
            \draw (ba4) arc (90:-30:0.1) node[] (ba5) {};
            \draw (uar) arc (-90:-30:0.1) node[] (ba6) {};
            \draw[thin] (b16.center) -- (ba5.center) {};
            \draw[thin] (ba6.center) -- ++(0.15,0.45) {};
            \draw[thin] (b15.center) -- ++(-0.15,-0.45) {};
            \draw[thin] (u1.north) -- (ua.south) {};
            %\draw[dotted] (u1t.center) -- (uab.center);
            \draw[dotted] (uat.center) -- ++(0, 0.25) {};
    \end{tikzpicture}.
    \label{eq:twistnetwork}
\end{align}
Extracting a single row of the twisted network, the twist by $a$ periods is incorporated into the MPU by augmenting the physical qudits with $|a|$ auxiliary qudits of dimension $D$ and extending $\mathcal{U}_N$ by right or left translation tensors,
\begin{align}
    \hat{U}_F(a)
    =
    \left\{
    \begin{array}{ll}
        \displaystyle
        \begin{tikzpicture}[baseline={([yshift=-0.cm]current bounding box.center)}]
            \node[draw, fill=white, minimum size=16pt, font=\bfseries] (centerNode) {$\mathcal{U}_N$};
            \draw[thin] (centerNode.north) -- ++(0, 0.2);
            \draw[thin] (centerNode.south) -- ++(0, -0.2);
            \draw[thin] (centerNode.west) -- ++(-0.25, 0) node[] (b11) {};
            \draw[thin] (centerNode.east) -- ++(0.2, 0) node[] (t1l) {};
            \draw (t1l) arc (-90:0:0.2) node[] (t1a) {};
            \draw[thin] (t1a.center) -- ++(0, 0.25);
            \node[right = 0.2 of t1l] (t1r) {};
            \draw (t1r) arc (90:180:0.2) node[] (t1b) {};
            \draw[thin] (t1b.center) -- ++(0, -0.25) node[below] {};
            \draw[thin] (t1r.center) -- ++(0.2, 0) node[](t2l){};
            \node[right = 0.2 of t2l] (t2r) {};
            \draw (t2r) arc (90:180:0.2) node[] (t2b) {};
            \draw (t2l) arc (-90:0:0.2) node[] (t2a) {};
            \draw[thin] (t2a.center) -- ++(0, 0.25);
            \draw[thin] (t2b.center) -- ++(0, -0.25) node[below] {};
            \draw[thin] (t2r.center) -- ++(0.2, 0) node[](i2){};
            \node[right = 1.5 of t2l] (tnl) {};
            \draw[thin] (tnl.center) -- ++(-0.2, 0) node[](i2){};
            \draw (tnl) arc (-90:0:0.2) node[] (tna) {};
            \draw[thin] (tna.center) -- ++(0, 0.25);
            \node[right = 0.2 of tnl] (tnr) {};
            \draw (tnr) arc (90:180:0.2) node[] (tnb) {};
            \draw[thin] (tnb.center) -- ++(0, -0.25) node[below] {};
            \draw[thin] (tnr.center) -- ++(0.2, 0) node[](b14){};
            \draw[dotted] (t2r.center) -- (tnl.center);
            \draw (b11) arc (270:90:0.1) node[] (b12) {};
            \draw[thin] (b12.center) -- ++(0.05, 0) node[] (b13) {};
            \draw (b14) arc (-90:90:0.1) node[] (b15) {};
            \draw[thin] (b15.center) -- ++(-0.15, 0) node[] (b16) {};
            \begin{scope}[on background layer]
                \draw[dotted] (b13.center) -- (b16.center);
            \end{scope}
        \end{tikzpicture}
        \,&(a>0)
        \\[+10pt]
        \\
        \displaystyle
        \begin{tikzpicture}[baseline={([yshift=-0.cm]current bounding box.center)}]
            \node[draw, fill=white, minimum size=16pt, font=\bfseries] (centerNode) {$\mathcal{U}_N$};
            \draw[thin] (centerNode.north) -- ++(0, 0.2);
            \draw[thin] (centerNode.south) -- ++(0, -0.2);
            \draw[thin] (centerNode.west) -- ++(-0.25, 0) node[] (b11) {};
            \draw[thin] (centerNode.east) -- ++(0.2, 0) node[] (t1l) {};
            \draw (t1l) arc (90:0:0.2) node[] (t1b) {};
            \draw[thin] (t1b.center) -- ++(0, -0.3) node[below] {};
            \node[right = 0.2 of t1l] (t1r) {};
            \draw (t1r) arc (270:180:0.2) node[] (t1a) {};
            \draw[thin] (t1a.center) -- ++(0, 0.3);
            \draw[thin] (t1r.center) -- ++(0.2, 0) node[](t2l){};
            \draw[thin] (t2l.center) -- ++(-0.2, 0) node[](i2){};
            \draw (t2l) arc (90:0:0.2) node[] (t2b) {};
            \draw[thin] (t2b.center) -- ++(0, -0.3) node[below] {};
            \node[right = 0.2 of t2l] (t2r) {};
            \draw (t2r) arc (270:180:0.2) node[] (t2a) {};
            \draw[thin] (t2a.center) -- ++(0, 0.3);
            \draw[thin] (t2r.center) -- ++(0.2, 0) node[](i2){};
            \node[right = 1.5 of t2l] (tnl) {};
            \draw[thin] (tnl.center) -- ++(-0.2, 0) node[](i2){};
            \draw (tnl) arc (90:0:0.2) node[] (tnb) {};
            \draw[thin] (tnb.center) -- ++(0, -0.3) node[below] {};
            \node[right = 0.2 of tnl] (tnr) {};
            \draw (tnr) arc (270:180:0.2) node[] (tna) {};
            \draw[thin] (tna.center) -- ++(0, 0.3);
            \draw[thin] (tnr.center) -- ++(0.2, 0) node[](b14){};
            \draw[dotted] (t2r.center) -- (tnl.center);
            \draw (b11) arc (270:90:0.1) node[] (b12) {};
            \draw[thin] (b12.center) -- ++(0.05, 0) node[] (b13) {};
            \draw (b14) arc (-90:90:0.1) node[] (b15) {};
            \draw[thin] (b15.center) -- ++(-0.15, 0) node[] (b16) {};
            \begin{scope}[on background layer]
                \draw[dotted] (b13.center) -- (b16.center);
            \end{scope}
        \end{tikzpicture}
        \,&(a<0)
    \end{array}\right..
    \label{eq:mpu_withtwist}
\end{align}
The twisted leg carries the right (left) virtual index of $\mathcal{U}_N$ to the output, so that $\rank\hat{U}_F(\pm1)=r_N\,(l_N)$; each additional translation tensor acts as an identity operator and multiplies the rank by $D$, giving $\rank\hat{U}_F(a)=d^NI_\text{GNVW}^{\,\text{sign}[a]}D^{|a|-1}$ for $a\neq0$.
For $t>1$, we block $N/t$ sites; provided the blocked tensors are simple, the twisted legs of the $t$ rows are contracted using (\ref{eq:isometryseparation}), and the rank of $\hat{U}_F^t(a)$ is expressed through the right and left ranks of the blocked tensors.
Section~\ref{sec:shift} in \cite{SM} carries this out for $t=2$ and $3$ and yields (\ref{eq:mpu_floquetrank}).

\section{Twisted spectral form factor}
\label{em:sff}
The trace identity (\ref{eq:tilted}) relates the twisted SFF to a unitary $\hat{V}_a$ on $\mathcal D_a=\dim\mathcal H_\Sigma$ states. Its infinite-time average is
\begin{align}
    \overline K(a)=\sum_\theta m_\theta^2,
\end{align}
where $m_\theta$ is the multiplicity of each distinct eigenphase of $\hat{V}_a$. Thus $\overline K(a)=\mathcal D_a$ for nondegenerate eigenphases, and the ratio $\overline K(a)/\overline K(0)=I_\mathrm{GNVW}^a$ measures the index without requiring random-matrix statistics. Degeneracies must be accounted for separately; the derivation is given in Sec.~\ref{sec:sff} of \cite{SM}.

For an ensemble of chaotic circuits in the circular unitary ensemble (CUE) regime, the averaged SFF additionally approaches a ramp $\langle K(t,a)\rangle\simeq t$ after the Thouless time and a plateau $\mathcal D_a$ beyond the Heisenberg time $t_H=\mathcal D_a$ \cite{PhysRevX.8.041019,PhysRevLett.121.264101}. This time dependence is a statistical property of the ensemble, distinct from the exact infinite-time average above.

We test three circuits: brickwork RUCs on qubits with a shift layer ($I_\mathrm{GNVW}=2$), the corresponding circuits without the shift ($I_\mathrm{GNVW}=1$), and the alternating-dimension circuit of Ref.~\cite{Gong2022} with $(p,q)=(3,2)$ ($I_\mathrm{GNVW}=3/2$). The respective tilted-slice dimensions are
$\mathcal D_a=2^{N+a}, 
2^N, 3^{N/2+a}2^{N/2-a}$.
In every realization tested, the number of nonzero eigenvalues of $\hat U_F(a)$ agrees with $\mathcal D_a$, and their moduli equal one to machine precision. The averaged SFF approaches the CUE ramp and plateau, with one unit of twist multiplying the plateau and the Heisenberg scale by the index. The trivial-index circuit provides a control with unchanged plateau.
For the rational-index circuit the untwisted reference dimension is $6^{N/2}$, giving the plateau ratio $(3/2)^a$. Circuit ensembles, system sizes, and SFF plots are specified in Sec.~\ref{sec:sff} of \cite{SM}.

\section{Time-translation defect responses of non-invertible symmetries}
\label{em:noninv}

We summarize the examples underlying (\ref{eq:noninv}) in terms of the normalized responses
\begin{align}
    \chi_\pm=\frac{\rank\hat{\mathcal L}(\pm1)}{\rank\hat{\mathbbm{1}}(\pm1)}.
    \label{eq:noninv_responses_em}
\end{align}
The reference ranks for the two orientations are equal in these examples, so $\text{Ind}=\sqrt{\chi_+/\chi_-}$ and $d_{\hat{\mathcal L}}=\sqrt{\chi_+\chi_-}$, with the latter equality understood in the thermodynamic limit for the Fibonacci line. Table~\ref{tab:noninv_responses} collects the results; MPO tensors, rank calculations, and numerical checks are given in Sec.~\ref{sec:noninv} of \cite{SM}.
\begin{table}[tb]
   \centering
    \begin{tabular}{lcccc}
    \hline\hline
    Line & $\chi_+$ & $\chi_-$ & $\text{Ind}$ & $d_{\hat{\mathcal L}}$ \\ \hline
    $\mathbb Z_n$ KW & $1$ & $n$ & $n^{-1/2}$ & $\sqrt n$ \\
    $\mathrm{Rep}(D_8)$: $\hat{\mathcal L}_1$ & $2$ & $2$ & $1$ & $2$ \\
    $\mathrm{Rep}(D_8)$: $\hat{\mathcal L}_2$ & $1$ & $4$ & $1/2$ & $2$ \\
    $\mathrm{Rep}(S_3)$: $\hat{\mathcal L}_E$ & $2$ & $2$ & $1$ & $2$ \\
    Fibonacci: $\hat{\mathcal L}_\tau$ & $f_N$ & $f_N$ & $1$ & $\lim f_N=\phi$ \\
    \hline\hline
    \end{tabular}
       \caption{Normalized responses to unit time-translation defects. The orientation convention is $\text{Ind}=\sqrt{r_N/l_N}$. The first four rows are exact at finite size for the chains considered. In the last row, $f_N=F_{N+3}/F_{N+2}$, where $F_n$ is the Fibonacci sequence, and $\phi=(1+\sqrt5)/2$.}
    \label{tab:noninv_responses}
\end{table}

For the $\mathbb Z_n$ KW duality, one orientation leaves a projector onto the symmetric sector, whereas the other gives a unitary map with an additional boundary qudit \cite{SeibergSeifnashriShao2024}. The resulting index agrees with the symmetric-subalgebra index of Ref.~\cite{MaLiCheng2024}, with our choice of orientation. On the closed ring, the rank instead counts only the $\mathbb Z_n$-symmetric sector.
The $\mathrm{Rep}(D_8)$ lines $\hat{\mathcal L}_1=\hat{T}\,\mathrm{KW}_e\mathrm{KW}_o$ and $\hat{\mathcal L}_2=\mathrm{KW}_e\mathrm{KW}_o$ \cite{MaLiCheng2024,SeifnashriShao2024} illustrate the separation of the two responses: the translation $\hat{T}$ changes the index while leaving the quantum dimension unchanged.
For the Wilson line $\hat{\mathcal L}_E$ of the two-dimensional representation of $S_3$, the open bond carries a two-dimensional representation space in either orientation, giving equal responses.

On the Fibonacci chain, the reference operator is the constraint projector $\hat{\Pi}$. Its twisted rank is $F_{N+2}$, while that of $\hat{\mathcal L}_\tau$ is $F_{N+3}$: opening the line adds one step to the fusion path. Thus the irrational quantum dimension is recovered from a ratio of integer ranks as $N\to\infty$.
For comparison, translation on this chain has normalized responses $F_{N+3}/F_{N+2}$ and $F_{N+1}/F_{N+2}$, yielding the Jones--Lim index $\phi$ and quantum dimension one in the thermodynamic limit \cite{JonesLim2023}.

%%%%%%%%%%%%%%%%%%
%% SUPPLEMENTAL %% 
%%%%%%%%%%%%%%%%%%

\clearpage
\onecolumngrid

%\appendix
\setcounter{secnumdepth}{3} 

\makeatletter 
\@removefromreset{equation}{section} 
\makeatother 

\renewcommand{\thesection}{S\arabic{section}}
\renewcommand{\theequation}{S\arabic{equation}}
\renewcommand{\thefigure}{S\arabic{figure}}
\renewcommand{\thetable}{S\arabic{table}}
\renewcommand{\thepage}{S\arabic{page}}

\setcounter{equation}{0}
\setcounter{figure}{0}
\setcounter{table}{0}
\setcounter{page}{1}

\begin{center}
\textbf{\large Supplemental Material for ``Quantized pumping of boundary Hilbert space in chiral Floquet phases''}
\end{center}

\tableofcontents

\section{Extension to arbitrary winding numbers}
\label{sec:windingnumber}

We prove (\ref{eq:boundarymodenumber_afai}).
Let the single-particle quasi-energies of a Floquet operator $U_F(e^{-i\alpha})$ under the U(1) twisted boundary condition by $\alpha$ be given by $\epsilon_j(\alpha)$.
The winding number is defined by \cite{PhysRevX.3.031005,PhysRevX.6.021013}
\begin{align}
    \mathcal{W}
    =
    \frac{i}{2\pi}
    \int_0^{2\pi} d\alpha \text{Tr}\,U^\dagger_F(e^{-i\alpha})\frac{d}{d\alpha}U_F(e^{-i\alpha})
    =
    \frac{T}{2\pi}\sum_j\left[\epsilon_j(2\pi)-\epsilon_j(0)\right].
    \label{eq:windingnumber}
\end{align}
Specifically, the Floquet operator (\ref{eq:shifttwist_transfermatrix}) with $\theta=\alpha+\theta_0$ has the quasi-energies $\epsilon_j(\alpha)=(2\pi j+\alpha+\theta_0)/(NT)$, where $\theta_0$ is the disorder phase, and thus the winding number is 1.

\begin{figure}[b]
    \centering
    \includegraphics[width=1.0\textwidth]{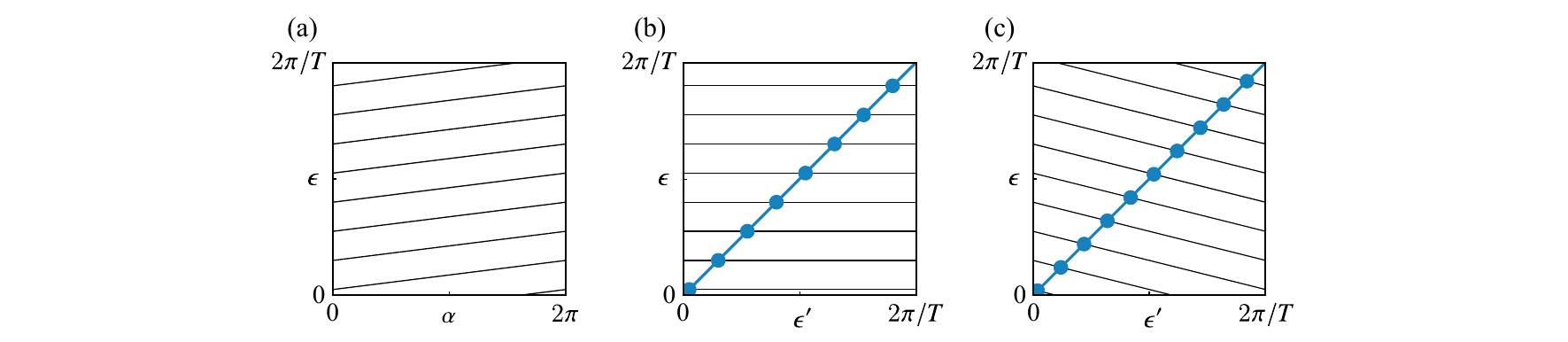}
    \caption{
        The spectral flow of the quasi-energies of the boundary Floquet operator in (\ref{eq:shifttwist_transfermatrix}) as a function 
        (a) of the U(1) phase $\alpha$ and
        (b) of the parameter $\epsilon'$ without the time-translation twist, and 
        (c) of the parameter $\epsilon'$ with the time-translation twist by $a=2$.
        \label{fig:winding}
    }
\end{figure}

The time-translation twist by $aT$ is implemented by $\alpha\to\alpha-a\epsilon T$, where $\epsilon$ is the quasi-energy of the Floquet operator.
Thus, the quasi-energies are determined self-consistently.
Instead, we can determine the quasi-energies by regarding $\epsilon'$ in the Floquet operator $U_F(e^{-i(\alpha+\theta_0-a\epsilon' T)})$ as an independent parameter and find the solution that the quasi-energy $\epsilon$ of $U_F(e^{-i(\alpha+\theta_0-a\epsilon' T)})$ coincides with $\epsilon'$.
In other words, when we plot the quasi-energies of $U_F(e^{-i(\alpha+\theta_0-a\epsilon' T)})$ as a function of $\epsilon'$, the intersections with $\epsilon=\epsilon'$ are the quasi-energies.
Fig.~\ref{fig:winding} (b) and (c) show the cases of the Floquet operator (\ref{eq:shifttwist_transfermatrix}) with $a=0$ and $2$, respectively, where the number of the intersections is increased by $a$.

The above argument can be generalized to an arbitrary winding number $\mathcal{W}$ provided that the boundary dynamics is well-separated from the bulk one and thus the boundary Floquet operator is well-defined. 
The quasi-energies are invariant as a set $\{\epsilon_j(2\pi)\}=\{\epsilon_j(0)\}$ under the U(1) twist by $2\pi$.
However, each quasi-energy is shifted to another one as $\epsilon_j(2\pi)=\epsilon_{\sigma(j)}(0)$.
Equation (\ref{eq:windingnumber}) indicates that following spectral flow the quasi-energy winds the Floquet Brillouin zone $\mathcal{W}$ times.
When $\mathcal{W}$ is negative, it means that the quasi-energy winds in the opposite direction.
Therefore the winding number determines the minimum number that the spectral flow emanates from or absorbed into the top border ($\epsilon=2\pi/T$) in Fig.~\ref{fig:winding} (a).

As can be seen from Fig.~\ref{fig:winding} (b), (c), the number of intersections is in general lower-bounded by that of the spectral flow that emanates from the top ($\epsilon=2\pi/T$) and left ($\epsilon'=0$) borders and ends at the bottom ($\epsilon=0$) and right ($\epsilon'=2\pi/T$) borders.
This lower bound is the number of topologically protected boundary modes which cannot be removed by perturbations.
The number of the quasi-energies at the left and right borders are the number of boundary modes without twist.
On the other hand, the number of spectral flows that cross the top and bottom borders is lower-bounded by the absolute value of $\mathcal{W}a$, which is positive when they emanate from the top border and negative when they are absorbed into it.
The former (latter) contributes positively (negatively) to the number of the intersections.
The above argument leads to the number of topologically protected boundary mode in (\ref{eq:boundarymodenumber_afai}).

\section{Spectra of the AFAI}
\label{sec:afai_details}

\begin{figure}
    \centering
    \includegraphics[width=1.0\textwidth]{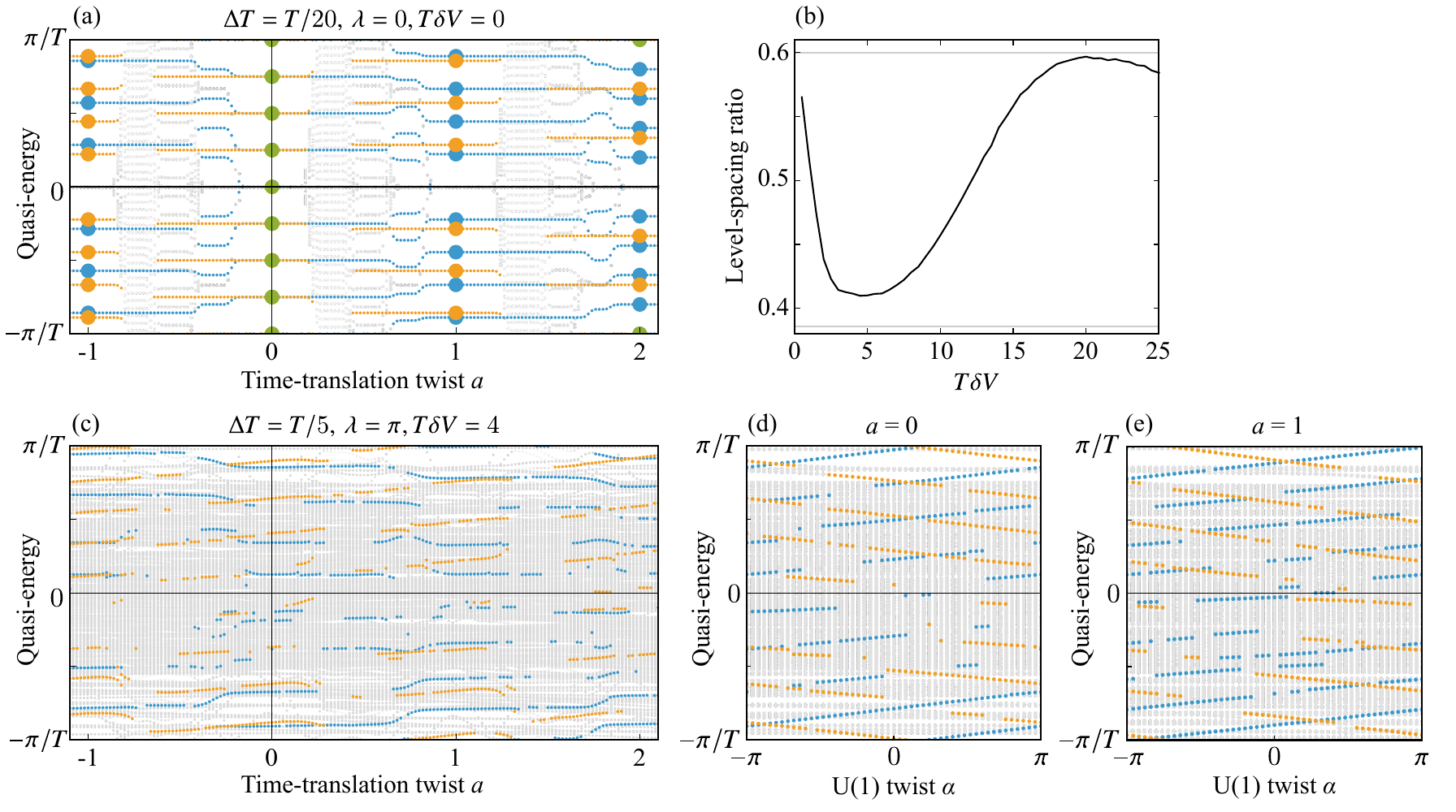}
    \caption{
        (a) The quasi-energies of the AFAI under the time-translation twist for $\Delta T=T/20$. (b) The level-spacing ratio as a function of the disorder strength. The gray lines correspond to Poissonian (bottom, 0.386) and the Gaussian unitary ensemble (top, 0.6).
        (c) The quasi-energies with potential disorder of $T\delta V=4$ as a function of the time translation twist parameter $a$.
        The quasi-energies as a function of the U(1) twist parameter $\alpha$ are shown for (d) $a=0$ and (e) $a=1$.
        \label{fig:edgetwisteigenphase}
    }
\end{figure}

Here, we show the numerical calculation of the bulk spectrum of the anomalous Floquet Anderson insulator (AFAI) under the time-translation twist.
First we make it explicit the time dependence of the AFAI by rewriting the Hamiltonian (\ref{eq:hamiltonian_afai_drive}) into the form
\begin{align}
    H(t)
    =
    \sum_{r\in A}
    \sum_{n\in[1,4]}
    H_{r,r+b_n}(t),
\end{align}
where
\begin{align}
    &H_{r,r+b_n}(t)
    =
    -J_{n}\left(t\right)
    \left(c_r^\dagger c_{r+b_n}^{\ }+\text{H.c.}\right)
    +
    \frac{I(t)}{4}\left(u_{r}n_r+u_{r+b_n}n_{r+b_n}\right)\notag\\
    &\left(
        \begin{array}{l}
            \displaystyle
            J_{n}\left(t\right)
            =
            \left\{
            \begin{array}{ll}
                J\quad &(t\in[(n-1)T/5,(n-1)T/5+\Delta T] \text{ mod }T) \\
                0\quad &(\text{otherwise})
            \end{array}
            \right.\\
            \displaystyle
            I(t)
            =
            \left\{
            \begin{array}{ll}
                1\quad &(t\in[4T/5,4T/5+\Delta T] \text{ mod }T) \\
                0\quad &(\text{otherwise})
            \end{array}
            \right.
        \end{array}
    \right)
\end{align}
is the time-dependent part of the Hamiltonian assigned to the bond between sites $r$ and $r+b_n$.
Here, we extended the definition of the coupling so that the interval during which the coupling is active can be tuned.
The original AFAI model is recovered by $\Delta T=T/5$.
We always keep $J\Delta T=\pi/2$ so that the fermions are completely swapped within each sector.
The time-independent disorder Hamiltonian is given by
\begin{align}
    H_\text{disorder}
    =
    \frac{\lambda}{2T}\sum_r(-1)^{\eta_r}n_r+\sum_r V_r n_r,
    \label{eq:disorderhamiltonian}
\end{align}
where $\eta_r=0 (1)$ for A-(B-)sublattice sites, $V_r$ is uniformly distributed in $[-\delta V,\delta V]$.
The system undergoes the Anderson transition by an infinitesimal disorder $\delta V$.
When $\lambda=\pi$, the transition from the AFAI to a trivial insulator phase occurs at $T\delta V\simeq 20$ \cite{PhysRevX.6.021013} [Fig.~\ref{fig:edgetwisteigenphase} (b)].

Following a similar method used in \cite{Kishony2025}, we incorporate the time-translation twist by a position-dependent delay of the Floquet evolution.
We consider a cylinder geometry where the twisted (open) boundary condition is imposed along the $x$ ($y$) direction.
The system size is $N_x\times N_y$.
The time-translation twist by $aT$ is implemented by
\begin{align}
    H(t,a)
    =
    \sum_{r\in A}
    \sum_{n\in[1,4]}
    H_{r,r+b_n}\left(t-\frac{(r+b_n/2)_xaT}{N_x}\right),
    \label{eq:afai_bulktwist}
\end{align}
where $r_x$ is the $x$ coordinate of the position $r$.
Notice that, unless $a$ is an integer, the shift of the Floquet evolution has a discontinuity around $x=N_x$ and $1$.
Specifically, $(r+b_1)_x=N_x-1/2$ for $r\in A$ at $x=N_x-1$, while on its neighboring bond $(r+b_3)_x=1/2$ for $r\in A$ at $x=1$.
The discontinuity line is a defect where the bulk properties of the AFAI cease to hold and where the boundary modes are transferred.
According to (\ref{eq:afai_bulktwist}), the delay of the Floquet evolution at bonds connecting $x=N_x$ and $x=1$ are either $-aT/2N_x$ or $-(N_x+1/2)aT/N_x$ depending on whether the A-sublattice sites are at $x=N_x$ or $1$.
To make the discontinuity line smooth, we redefined them to have a common delay $-aT/2N_x$, and hence the discontinuity line is close to $x=N_x$.

We numerically evaluate the quasi-energies of (\ref{eq:afai_bulktwist}) as a function of the twist parameter $a$ including non-integral case.
We use a square lattice on a cylinder of size $N_x=16$ and $N_y=16$.
Hence, the number of the boundary sites is $N=8$, and each boundary accommodates 8 modes without the twist.
The quasi-energies for $\Delta T=T/5$ and without disorder is shown in Fig.~\ref{fig:afai_numerical} (a).
The quasi-energies for $\Delta T=T/20$ without disorder is shown in Fig.~\ref{fig:edgetwisteigenphase} (a).
Unlike the case with $\Delta T=T/5$, the bulk spectrum remains at $\epsilon=0$ since the overlaps of the active time in the neighboring bonds are suppressed.
The boundary modes are highlighted in blue (bottom boundary), orange (top boundary), and green when they are overlapped.
The boundary modes are determined by the criterion that at least 0.8 of the eigenfunction weight is concentrated on $y=1$ or $N_y$.
The other modes are plotted in gray where the gray-level is determined by the inverse participation ratio.
The darker the point is, the more localized the eigenfunction is.
The number of the boundary modes agrees with (\ref{eq:boundarymodenumber_afai}).

For disordered case, we consider $\Delta T=T/5$ with disorder parameter $u_r=0$, $\lambda=\pi$, and $T\delta V=4$, which is in the Anderson localized phase [Fig.~\ref{fig:edgetwisteigenphase} (b)].
As a criterion of the boundary modes, we consider three parameters
\begin{align}
    w_x=\sum_{x=1}^{N_x}\left(\sum_{y=1}^{N_y} |\phi(r)|^2\right)^2
    \,\,\text{and}\,\,
    (w_{y,\text{bottom}},w_{y,\text{top}})=\left(\sum_{y=1}^{N_y/2}\left(\sum_{x=1}^{N_x} |\phi(r)|^2\right)^2, \sum_{y=N_y/2+1}^{N_y}\left(\sum_{x=1}^{N_x} |\phi(r)|^2\right)^2\right),
\end{align}
the former of which is used to measure how the wavefunction is extended in the $x$ direction,
while the latter two of which are used to measure how the wave function is localized in the bottom or top boundaries.
We use $w_x<0.15$ and $w_{y,\text{bottom}}$ or $w_{y,\text{top}}>0.4$ as a criterion of the boundary modes.
The quasi-energies as a function of the time-translation twist $a$ without the U(1) twist is shown in Fig.~\ref{fig:edgetwisteigenphase} (c).
While the boundary modes are less clear, it becomes relatively clear by varying the U(1) twist parameter as in Fig.~\ref{fig:edgetwisteigenphase} (d),(e), which correspond to $a=0$ and $1$, respectively.
We can see that the intervals of the quasi-energies are different for the top and bottom boundaries in Fig.~\ref{fig:edgetwisteigenphase} (e), resulting from the transfer of the boundary modes.

\section{Translation-invariant MPU}
\label{sec:shift}

\subsection{Time-translation twist in MPU}

Here, we show (\ref{eq:mpu_floquetrank}) for $t=2$ and $3$ using the properties of simple tensors.
We focus on matrix product unitaries (MPUs) that is invariant under spatial translation \cite{PhysRevB.98.245122,Cirac_2017}.
Under the periodic boundary condition, the MPU for the Floquet operator $\hat{U}_F$ is given by
\begin{align}
    &\hat{U}_F=\sum_{\bm{j},\bm{k}}
    \text{Tr}\left[\mathcal{U}_{j_1k_1}\cdots \mathcal{U}_{j_Nk_N}\right]|\bm{j}\rangle\langle \bm{k}|
    =\sum_{\bm{j},\bm{k}}\text{Tr}\left(\mathcal{U}_N\right)_{\bm{j}\bm{k}}|\bm{j}\rangle\langle \bm{k}|\notag\\
    &\Leftrightarrow\,\,
    \hat{U}_F=
    \begin{tikzpicture}[baseline={([yshift=-0.1cm]current bounding box.center)}]
        \node[draw, fill=white, minimum size=16pt] (A1) at (0.,0.){$\mathcal{U}$};
        \draw[thin] (A1.north) -- ++(0, 0.35) node[above] {$j_1$};
        \draw[thin] (A1.south) -- ++(0, -0.35) node[below] {$k_1$};
        %\draw[thin] (A1.east) -- ++(10pt, 0);
        \draw[thin] (A1.west) -- ++(-0.35, 0) node[] (b11) {};
        \draw (b11) arc (270:90:0.1) node[] (b12) {};
        \draw[thin] (b12.center) -- ++(0.15, 0) node[] (b13) {};
        \node[draw, fill=white, minimum size=16pt] (A2) at (1.,0.) {$\mathcal{U}$};
        \draw[thin] (A2.north) -- ++(0, 0.35) node[above] {$j_2$};
        \draw[thin] (A2.south) -- ++(0, -0.35) node[below] {$k_2$};
        \draw[thin] (A2.east) -- ++(0.35, 0) node[right] (i1) {};
        \draw[thin] (A2.west) -- (A1.east);
        \node[draw, fill=white, minimum size=16pt] (A3) at (3.5,0.) {$\mathcal{U}$};
        \draw[thin] (A3.north) -- ++(0, 0.35) node[above] {$j_N$};
        \draw[thin] (A3.south) -- ++(0, -0.35) node[below] {$k_N$};
        \draw[thin] (A3.east) -- ++(0.35, 0) node[] (b14) {};
        \draw[thin] (A3.west) -- ++(-0.35, 0) node[left] (i2) {};
        \draw[dotted] (i1.center) -- (i2.center);
        \draw (b14) arc (-90:90:0.1) node[] (b15) {};
        \draw[thin] (b15.center) -- ++(-0.15, 0) node[] (b16) {};
        \begin{scope}[on background layer]
            \draw[dotted] (b13.center) -- (b16.center);
        \end{scope}
    \end{tikzpicture}
    \,\,=\,\,
    \begin{tikzpicture}[baseline={([yshift=-0.1cm]current bounding box.center)}]
            \node[draw, fill=white, minimum size=16pt, font=\bfseries] (centerNode) {$\mathcal{U}_N$};
            \draw[thin] (centerNode.north) -- ++(0, 10pt) node[above] {$\bm{j}$};
            \draw[thin] (centerNode.south) -- ++(0, -10pt) node[below] {$\bm{k}$};
            \draw[thin] (centerNode.west) -- ++(-10pt, 0) node[] (b11) {};
            \draw[thin] (centerNode.east) -- ++(10pt, 0) node[] (b14) {};
            \draw (b11) arc (270:90:0.1) node[] (b12) {};
            \draw[thin] (b12.center) -- ++(0.15, 0) node[] (b13) {};
            \draw (b14) arc (-90:90:0.1) node[] (b15) {};
            \draw[thin] (b15.center) -- ++(-0.15, 0) node[] (b16) {};
            \begin{scope}[on background layer]
                \draw[dotted] (b13.center) -- (b16.center);
            \end{scope}
    \end{tikzpicture},
\end{align}
where $j,k$ specify the $d$-dimensional qudit states, Tr is taken over the $D$-dimensional bond (virtual Hilbert) space, and we block $N$ site to construct a tensor $\mathcal{U}_N$.
Here, we consider a two-dimensional tensor network of $\hat{U}_F^t$, where the horizontal and vertical axes correspond to spatial and temporal, respectively.
Under the time-translation twist by $aT$, the right-most horizontal legs are connected to the left-most ones after translating along the temporal direction by $a$ rows. 
Let the Floquet operator under the time-translation twist by $aT$ be denoted by $\hat{U}_F(a)$.
The tensor network representations of the Floquet operators with time-translation twist by $a=0,1,2$ are shown as
\begin{align}
    \hat{U}_F^t(0)=
    \begin{tikzpicture}[baseline={([yshift=-0.1cm]current bounding box.center)}]
            \node[draw, fill=white, minimum size=16pt, font=\bfseries] (u1) {$\mathcal{U}_N$};
            \node[draw, fill=white, minimum size=16pt, font=\bfseries, above = 0.25 of u1] (ua) {$\mathcal{U}_N$};            
            %\draw[thin] (u1.north) -- ++(0, 0.25) node[] (u1t) {};
            \draw[thin] (u1.south) -- ++(0, -0.25) node[] (u1b) {};
            \draw[thin] (u1.west) -- ++(-0.25, 0) node[] (u1l) {};
            \draw[thin] (u1.east) -- ++(0.25, 0) node[] (u1r) {};
            \draw (u1l) arc (270:90:0.1) node[] (b12) {};
            \draw[thin] (b12.center) -- ++(0.05, 0) node[] (b13) {};
            \draw (u1r) arc (-90:90:0.1) node[] (b15) {};
            \draw[thin] (b15.center) -- ++(-0.05, 0) node[] (b16) {};
            \begin{scope}[on background layer]
                \draw[dotted] (b13.center) -- (b16.center);
            \end{scope}
            \draw[dotted] (u1b.center) -- ++(0, -0.25) {};
            \draw[thin] (ua.north) -- ++(0, 0.25) node[] (uat) {};
            %\draw[thin] (ua.south) -- ++(0, -0.25) node[] (uab) {};
            \draw[thin] (ua.west) -- ++(-0.25, 0) node[] (ual) {};
            \draw[thin] (ua.east) -- ++(0.25, 0) node[] (uar) {};
            \draw (ual) arc (270:90:0.1) node[] (ba2) {};
            \draw[thin] (ba2.center) -- ++(0.05, 0) node[] (ba3) {};
            \draw (uar) arc (-90:90:0.1) node[] (ba5) {};
            \draw[thin] (ba5.center) -- ++(-0.05, 0) node[] (ba6) {};
            \begin{scope}[on background layer]
                \draw[dotted] (ba3.center) -- (ba6.center);
            \end{scope}
            \draw[thin] (u1.north) -- (ua.south) {};
            %\draw[dotted] (u1t.center) -- (uab.center);
            \draw[dotted] (uat.center) -- ++(0, 0.25) {};
    \end{tikzpicture},\quad
    \hat{U}_F^t(1)=
    \begin{tikzpicture}[baseline={([yshift=-0.1cm]current bounding box.center)}]
            \node[draw, fill=white, minimum size=16pt, font=\bfseries] (u1) {$\mathcal{U}_N$};
            \node[draw, fill=white, minimum size=16pt, font=\bfseries, above = 0.25 of u1] (ua) {$\mathcal{U}_N$};            
            %\draw[thin] (u1.north) -- ++(0, 0.25) node[] (u1t) {};
            \draw[thin] (u1.south) -- ++(0, -0.25) node[] (u1b) {};
            \draw[thin] (u1.west) -- ++(-0.25, 0) node[] (u1l) {};
            \draw[thin] (u1.east) -- ++(0.2, 0) node[] (u1r) {};
            \draw (u1l) arc (270:90:0.1) node[] (b12) {};
            \draw[thin] (b12.center) -- ++(0.05, 0) node[] (b13) {};
            \begin{scope}[on background layer]
                \draw[dotted] (b13.center) -- ++(1.5, 0) node[] (b14) {};
            \end{scope}
            \draw (b14) arc (90:-30:0.1) node[] (b15) {};
            \draw (u1r) arc (-90:-30:0.1) node[] (b16) {};
            \draw[dotted] (u1b.center) -- ++(0, -0.25) {};
            \draw[thin] (ua.north) -- ++(0, 0.25) node[] (uat) {};
            %\draw[thin] (ua.south) -- ++(0, -0.25) node[] (uab) {};
            \draw[thin] (ua.west) -- ++(-0.25, 0) node[] (ual) {};
            \draw[thin] (ua.east) -- ++(0.2, 0) node[] (uar) {};
            \draw (ual) arc (270:90:0.1) node[] (ba2) {};
            \draw[thin] (ba2.center) -- ++(0.05, 0) node[] (ba3) {};
            \begin{scope}[on background layer]
                \draw[dotted] (ba3.center) -- ++(1.5, 0) node[] (ba4) {};
            \end{scope}
            \draw (ba4) arc (90:-30:0.1) node[] (ba5) {};
            \draw (uar) arc (-90:-30:0.1) node[] (ba6) {};
            \draw[thin] (b16.center) -- (ba5.center) {};
            \draw[thin] (ba6.center) -- ++(0.15,0.45) {};
            \draw[thin] (b15.center) -- ++(-0.15,-0.45) {};
            \draw[thin] (u1.north) -- (ua.south) {};
            %\draw[dotted] (u1t.center) -- (uab.center);
            \draw[dotted] (uat.center) -- ++(0, 0.25) {};
    \end{tikzpicture},\quad
    \hat{U}_F^t(2)=
    \begin{tikzpicture}[baseline={([yshift=-0.1cm]current bounding box.center)}]
            \node[draw, fill=white, minimum size=16pt, font=\bfseries] (u1) {$\mathcal{U}_N$};
            \node[draw, fill=white, minimum size=16pt, font=\bfseries, above = 0.25 of u1] (u2) {$\mathcal{U}_N$};
            \node[draw, fill=white, minimum size=16pt, font=\bfseries, above = 0.25 of u2] (ua) {$\mathcal{U}_N$};            
            %\draw[thin] (u1.north) -- ++(0, 0.25) node[] (u1t) {};
            \draw[thin] (u1.south) -- ++(0, -0.25) node[] (u1b) {};
            \draw[thin] (u1.west) -- ++(-0.25, 0) node[] (u1l) {};
            \draw[thin] (u1.east) -- ++(0.2, 0) node[] (u1r) {};
            \draw (u1l) arc (270:90:0.1) node[] (b12) {};
            \draw[thin] (b12.center) -- ++(0.05, 0) node[] (b13) {};
            \begin{scope}[on background layer]
                \draw[dotted] (b13.center) -- ++(1.8, 0) node[] (b14) {};
            \end{scope}
            \draw (b14) arc (90:-30:0.1) node[] (b15) {};
            \draw (u1r) arc (-90:-30:0.1) node[] (b16) {};
            \draw[dotted] (u1b.center) -- ++(0, -0.25) {};
            \draw[thin] (u2.west) -- ++(-0.25, 0) node[] (u2l) {};
            \draw[thin] (u2.east) -- ++(0.2, 0) node[] (u2r) {};
            \draw (u2l) arc (270:90:0.1) node[] (b22) {};
            \draw[thin] (b22.center) -- ++(0.05, 0) node[] (b23) {};
            \begin{scope}[on background layer]
                \draw[dotted] (b23.center) -- ++(1.8, 0) node[] (b24) {};
            \end{scope}
            \draw (b24) arc (90:-30:0.1) node[] (b25) {};
            \draw (u2r) arc (-90:-30:0.1) node[] (b26) {};
            \draw[thin] (ua.north) -- ++(0, 0.25) node[] (uat) {};
            %\draw[thin] (ua.south) -- ++(0, -0.25) node[] (uab) {};
            \draw[thin] (ua.west) -- ++(-0.25, 0) node[] (ual) {};
            \draw[thin] (ua.east) -- ++(0.2, 0) node[] (uar) {};
            \draw (ual) arc (270:90:0.1) node[] (ba2) {};
            \draw[thin] (ba2.center) -- ++(0.05, 0) node[] (ba3) {};
            \begin{scope}[on background layer]
                \draw[dotted] (ba3.center) -- ++(1.8, 0) node[] (ba4) {};
            \end{scope}
            \draw (ba4) arc (90:-30:0.1) node[] (ba5) {};
            \draw (uar) arc (-90:-30:0.1) node[] (ba6) {};
            \draw[thin] (b16.center) -- (ba5.center) {};
            \draw[thin] (ba6.center) -- ++(0.16,0.47) {};
            \draw[thin] (b15.center) -- ++(-0.16,-0.47) {};
            \draw[thin] (b26.center) -- ++(0.47,1.3) {};
            \draw[thin] (b25.center) -- ++(-0.47,-1.3) {};
            \draw[thin] (u1.north) -- (u2.south) {};
            \draw[thin] (u2.north) -- (ua.south) {};
            %\draw[dotted] (u1t.center) -- (uab.center);
            \draw[dotted] (uat.center) -- ++(0, 0.25) {};
    \end{tikzpicture},
\end{align}
assuming $t>a$.
Notice that the number of vertical legs connecting neighboring rows is increased by $|a|$.

Equivalently, the time-translation twist is incorporated to the MPU by adding $|a|$ qudits with dimension $D$, and attaching $|a|$ right-(left-)translation operators $T$ ($\bar{T}$) when $a>0$ ($a<0$). 
The right-translation operator $T$ is defined by $[T_{jk}]_{ab} = \delta_{j,a}\delta_{k,b}$  where $j, k$ are vertical legs (physical indices) and $a,b$ are horizontal legs (virtual indices), and is given by
\begin{align}
    [T_{jk}]_{ab} 
    =
    \begin{tikzpicture}[baseline={([yshift=-.8ex]current bounding box.center)}]
    % 4方向（北・南・東・西）に足を伸ばす
    \draw (0,0.2) -- +(0, 0.2) node[above] {$j$}; % 上
    \draw (0,-0.2) -- +(0,-0.2) node[below] {$k$}; % 下
    \draw (-0.2,0)  -- +(-0.2,0) node[left]  {$a$}; % 左
    \draw (0.2,0)  -- +(0.2, 0) node[right] {$b$}; % 右
    \draw[thin] (-0.2,0.) arc [radius=0.2, start angle = 270, end angle=360];
    \draw[thin] (0.2,0.) arc [radius=0.2, start angle = 90, end angle=180];
    \end{tikzpicture},\quad
    [\bar{T}_{jk}]_{ab} 
    =
    \begin{tikzpicture}[baseline={([yshift=-.8ex]current bounding box.center)}]
    % 4方向（北・南・東・西）に足を伸ばす
    \draw (0,0.2) -- +(0, 0.2) node[above] {$j$}; % 上
    \draw (0,-0.2) -- +(0,-0.2) node[below] {$k$}; % 下
    \draw (-0.2,0)  -- +(-0.2,0) node[left]  {$a$}; % 左
    \draw (0.2,0)  -- +(0.2, 0) node[right] {$b$}; % 右
    \draw[thin] (-0.2,0.) arc [radius=0.2, start angle = 90, end angle=0];
    \draw[thin] (0.2,0.) arc [radius=0.2, start angle = 270, end angle=180];
    \end{tikzpicture}.
\end{align}
The Floquet operator with time-translation twist is given by
\begin{align}
    &\hat{U}_F(a)
    =
    \left\{
    \begin{array}{ll}
        \displaystyle
        \begin{tikzpicture}[baseline={([yshift=-0.1cm]current bounding box.center)}]
            \node[draw, fill=white, minimum size=16pt, font=\bfseries] (centerNode) {$\mathcal{U}_N$};
            \draw[thin] (centerNode.north) -- ++(0, 0.25);
            \draw[thin] (centerNode.south) -- ++(0, -0.25);
            \draw[thin] (centerNode.west) -- ++(-10pt, 0) node[] (b11) {};
            \draw[thin] (centerNode.east) -- ++(10pt, 0) node[] (t1l) {};
            \draw (t1l) arc (-90:0:0.2) node[] (t1a) {};
            \draw[thin] (t1a.center) -- ++(0, 0.3);
            \node[right = 0.2 of t1l] (t1r) {};
            \draw (t1r) arc (90:180:0.2) node[] (t1b) {};
            \draw[thin] (t1b.center) -- ++(0, -0.3) node[below] {$N+1$};
            \draw[thin] (t1r.center) -- ++(0.2, 0) node[](i1){};
            \node[right = 1.5 of t1l] (tnl) {};
            \draw[thin] (tnl.center) -- ++(-0.2, 0) node[](i2){};
            \draw (tnl) arc (-90:0:0.2) node[] (tna) {};
            \draw[thin] (tna.center) -- ++(0, 0.3);
            \node[right = 0.2 of tnl] (tnr) {};
            \draw (tnr) arc (90:180:0.2) node[] (tnb) {};
            \draw[thin] (tnb.center) -- ++(0, -0.3) node[below] {$N+a$};
            \draw[thin] (tnr.center) -- ++(0.2, 0) node[](b14){};
            \draw[dotted] (t1r.center) -- (tnl.center);
            \draw (b11) arc (270:90:0.1) node[] (b12) {};
            \draw[thin] (b12.center) -- ++(0.15, 0) node[] (b13) {};
            \draw (b14) arc (-90:90:0.1) node[] (b15) {};
            \draw[thin] (b15.center) -- ++(-0.15, 0) node[] (b16) {};
            \begin{scope}[on background layer]
                \draw[dotted] (b13.center) -- (b16.center);
            \end{scope}
        \end{tikzpicture}
        \quad&(a>0)
        \\[+30pt]
        \displaystyle
        \,\,
        \begin{tikzpicture}[baseline={([yshift=-0.1cm]current bounding box.center)}]
            \node[draw, fill=white, minimum size=16pt, font=\bfseries] (centerNode) {$\mathcal{U}_N$};
            \draw[thin] (centerNode.north) -- ++(0, 0.25);
            \draw[thin] (centerNode.south) -- ++(0, -0.25);
            \draw[thin] (centerNode.west) -- ++(-10pt, 0) node[] (b11) {};
            \draw[thin] (centerNode.east) -- ++(10pt, 0) node[] (t1l) {};
            \draw (t1l) arc (90:0:0.2) node[] (t1b) {};
            \draw[thin] (t1b.center) -- ++(0, -0.3) node[below] {$N+1$};
            \node[right = 0.2 of t1l] (t1r) {};
            \draw (t1r) arc (270:180:0.2) node[] (t1a) {};
            \draw[thin] (t1a.center) -- ++(0, 0.3);
            \draw[thin] (t1r.center) -- ++(0.2, 0) node[](i1){};
            \node[right = 1.5 of t1l] (tnl) {};
            \draw[thin] (tnl.center) -- ++(-0.2, 0) node[](i2){};
            \draw (tnl) arc (90:0:0.2) node[] (tnb) {};
            \draw[thin] (tnb.center) -- ++(0, -0.3) node[below] {$N+|a|$};
            \node[right = 0.2 of tnl] (tnr) {};
            \draw (tnr) arc (270:180:0.2) node[] (tna) {};
            \draw[thin] (tna.center) -- ++(0, 0.3);
            \draw[thin] (tnr.center) -- ++(0.2, 0) node[](b14){};
            \draw[dotted] (t1r.center) -- (tnl.center);
            \draw (b11) arc (270:90:0.1) node[] (b12) {};
            \draw[thin] (b12.center) -- ++(0.15, 0) node[] (b13) {};
            \draw (b14) arc (-90:90:0.1) node[] (b15) {};
            \draw[thin] (b15.center) -- ++(-0.15, 0) node[] (b16) {};
            \begin{scope}[on background layer]
                \draw[dotted] (b13.center) -- (b16.center);
            \end{scope}
        \end{tikzpicture}
        \quad&(a<0)
    \end{array}\right.,
    \label{eq:mpu_withtwist_sm}
\end{align}
where the legs from 1 to $N$ which are blocked into a single leg below and above $\mathcal{U}_N$ correspond to physical qudits, while those from $N+1$ to $N+|a|$ correspond to virtual qudits.

\subsection{Properties of simple tensors}

A tensor $\mathcal{V}$ is said to be \textit{simple}, when $\mathcal{V}$ satisfies the following identities for some tensors $a$ and $b$ \cite{PhysRevB.98.245122,Cirac_2017}
\begin{align}
    \begin{tikzpicture}[baseline={([yshift=-0.1cm]current bounding box.center)}]
        \node[draw, fill=white, minimum size=16pt, font=\bfseries] (u1) {$\mathcal{V}$};
        \node[left = 0.1 of u1] (u1l) {};
        \node[right = 0.1 of u1] (u1r) {};
        \node[draw, fill=white, minimum size=16pt, font=\bfseries, above = 0.3 of u1] (bu1) {$\bar{\mathcal{V}}$};
        \node[left = 0.1 of bu1] (bu1l) {};
        \node[right = 0.1 of bu1] (bu1r) {};
        \draw[thin] (u1.north) -- (bu1.south);
        \draw[thin] (u1.south) -- ++(0, -0.35);
        \draw[thin] (bu1.north) -- ++(0, 0.35);
        \node (m1r) at ($(u1r)!0.5!(bu1r)$) [draw, circle, inner sep=0pt, minimum size=10pt, fill=white] {$b$};
        \node (m1l) at ($(u1l)!0.5!(bu1l)$) [draw, circle, inner sep=0pt, minimum size=10pt, fill=white] {$a$};
        \draw[thin] (u1.west) -- (u1l.center) -- (m1l.south);
        \draw[thin] (bu1.west) -- (bu1l.center) -- (m1l.north);
        \draw[thin] (u1.east) -- (u1r.center) -- (m1r.south);
        \draw[thin] (bu1.east) -- (bu1r.center) -- (m1r.north);
    \end{tikzpicture}
    \,\,
    =
    \quad
    \begin{tikzpicture}[baseline={([yshift=-0.1cm]current bounding box.center)}]
        \draw[thin] (0,-1) -- ++(0,2);
    \end{tikzpicture}
    \quad,
    \qquad
    \begin{tikzpicture}[baseline={([yshift=-0.1cm]current bounding box.center)}]
        \node[draw, fill=white, minimum size=16pt, font=\bfseries] (u1) {$\mathcal{V}$};
        \node[draw, fill=white, minimum size=16pt, font=\bfseries, above = 0.3 of u1] (bu1) {$\bar{\mathcal{V}}$};
        \node[draw, fill=white, minimum size=16pt, font=\bfseries, right = 0.5 of u1] (u2) {$\mathcal{V}$};
        \node[draw, fill=white, minimum size=16pt, font=\bfseries, above = 0.3 of u2] (bu2) {$\bar{\mathcal{V}}$};
        \draw[thin] (u1.north) -- (bu1.south);
        \draw[thin] (u2.north) -- (bu2.south);
        \draw[thin] (u1.south) -- ++(0, -0.35);
        \draw[thin] (bu1.north) -- ++(0, 0.35);
        \draw[thin] (u2.south) -- ++(0, -0.35);
        \draw[thin] (bu2.north) -- ++(0, 0.35);
        \draw[thin] (u1.west) -- ++(-0.35,0);
        \draw[thin] (bu1.west) -- ++(-0.35,0);
        \draw[thin] (u2.east) -- ++(0.35,0);
        \draw[thin] (bu2.east) -- ++(0.35,0);
        \draw[thin] (u1.east) -- (u2.west);
        \draw[thin] (bu1.east) -- (bu2.west);
    \end{tikzpicture}
    \,\,=\,\,
    \begin{tikzpicture}[baseline={([yshift=-0.1cm]current bounding box.center)}]
        \node[draw, fill=white, minimum size=16pt, font=\bfseries] (u1) {$\mathcal{V}$};
        \node[draw, fill=white, minimum size=16pt, font=\bfseries, above = 0.3 of u1] (bu1) {$\bar{\mathcal{V}}$};
        \node[draw, fill=white, minimum size=16pt, font=\bfseries, right = 0.9 of u1] (u2) {$\mathcal{V}$};
        \node[draw, fill=white, minimum size=16pt, font=\bfseries, above = 0.3 of u2] (bu2) {$\bar{\mathcal{V}}$};
        \draw[thin] (u1.north) -- (bu1.south);
        \draw[thin] (u2.north) -- (bu2.south);
        \draw[thin] (u1.south) -- ++(0, -0.35);
        \draw[thin] (bu1.north) -- ++(0, 0.35);
        \draw[thin] (u2.south) -- ++(0, -0.35);
        \draw[thin] (bu2.north) -- ++(0, 0.35);
        \draw[thin] (u1.west) -- ++(-0.35,0);
        \draw[thin] (bu1.west) -- ++(-0.35,0);
        \draw[thin] (u2.east) -- ++(0.35,0);
        \draw[thin] (bu2.east) -- ++(0.35,0);
        \node[right = 0.1 of u1] (u1r) {};
        \node[right = 0.1 of bu1] (bu1r) {};
        \node[left = 0.1 of u2] (u2l) {};
        \node[left = 0.1 of bu2] (bu2l) {};
        \node (m1r) at ($(u1r)!0.5!(bu1r)$) [draw, circle, inner sep=0pt, minimum size=10pt, fill=white] {$b$};
        \node (m2l) at ($(u2l)!0.5!(bu2l)$) [draw, circle, inner sep=0pt, minimum size=10pt, fill=white] {$a$};
        \draw[thin] (u1.east) -- (u1r.center) -- (m1r.south);
        \draw[thin] (bu1.east) -- (bu1r.center) -- (m1r.north);
        \draw[thin] (u2.west) -- (u2l.center) -- (m2l.south);
        \draw[thin] (bu2.west) -- (bu2l.center) -- (m2l.north);
    \end{tikzpicture},
    \label{eq:isometryseparation_sm}
\end{align}
where $\bar{\mathcal{V}}_{jk}=\mathcal{V}^\ast_{kj}$ ($k$ and $j$ are indices of the physical degrees of freedom).
Let $\mathcal{U}_p$ be the tensor obtained by blocking $p$ sites of a tensor $\mathcal{U}$ generating MPU, which indicates that $\text{Tr}[\mathcal{U}_p]$ is a unitary operator.
It is known that $\mathcal{U}_p$ becomes simple by blocking more than a particular number of sites, the minimum of which is at most $D^4$, 
and that $a$ and $b$ are left and right fixed points of the transfer operator $E=(1/d)\sum_{jk}\mathcal{U}_{jk}\otimes \mathcal{U}^\ast_{jk}$ \cite{PhysRevB.98.245122,Cirac_2017}.
After a gauge transformation, the left and right fixed points can be taken as (canonical form II in \cite{Cirac_2017})
\begin{align}
    (\Phi|=\sum_{i=1}^D(i,i|,\quad
    |\rho)=\sum_{i=1}^D\rho_i|i,i)\,(\rho_i>0),
    \label{eq:phirho}
\end{align}
that is, $(\Phi|E=(\Phi|$ and $E|\rho)=|\rho)$ or equivalently,
\begin{align}
    \begin{tikzpicture}[baseline={([yshift=-0.1cm]current bounding box.center)}]
        \node[draw, fill=white, minimum size=16pt, font=\bfseries] (u1) {$\mathcal{U}$};
        \node[left = 0.1 of u1] (u1l) {};
        \node[right = 0.1 of u1] (u1r) {};
        \node[draw, fill=white, minimum size=16pt, font=\bfseries, above = 0.3 of u1] (bu1) {$\bar{\mathcal{U}}$};
        \node[left = 0.1 of bu1] (bu1l) {};
        \node[right = 0.1 of bu1] (bu1r) {};
        \draw[thin] (u1.north) -- (bu1.south);
        \draw[thin] (u1.south) -- ++(0, -0.35) node[] (b11) {};
        \draw[thin] (bu1.north) -- ++(0, 0.35) node[] (b14) {};
        %\node (m1r) at ($(u1r)!0.5!(bu1r)$) [draw, circle,fill=white] {};
        \draw[thin] (u1.west) -- (u1l.center) -- (bu1l.center) -- (bu1.west);
        \draw[thin] (u1.east) -- (u1r.center);
        \draw[thin] (bu1.east) -- (bu1r.center);
        \draw (b11) arc (180:360:0.1) node[] (b12) {};
        \draw[thin] (b12.center) -- ++(0, 0.15) node[] (b13) {};
        \draw (b14) arc (180:0:0.1) node[] (b15) {};
        \draw[thin] (b15.center) -- ++(0, -0.15) node[] (b16) {};
        \begin{scope}[on background layer]
            \draw[dotted] (b13.center) -- (b16.center);
        \end{scope}
    \end{tikzpicture}
    \,\,=\,\,
    \begin{tikzpicture}[baseline={([yshift=-0.1cm]current bounding box.center)}]
        \node[] (u1) {};
        \draw[thin] (u1.center) -- ++(0, 0.5) -- ++(0.25, 0);
        \draw[thin] (u1.center) -- ++(0, -0.5) -- ++(0.25, 0);
    \end{tikzpicture}\,\,,
    \quad
    \begin{tikzpicture}[baseline={([yshift=-0.1cm]current bounding box.center)}]
        \node[draw, fill=white, minimum size=16pt, font=\bfseries] (u1) {$\mathcal{U}$};
        \node[left = 0.1 of u1] (u1l) {};
        \node[right = 0.1 of u1] (u1r) {};
        \node[draw, fill=white, minimum size=16pt, font=\bfseries, above = 0.3 of u1] (bu1) {$\bar{\mathcal{U}}$};
        \node[left = 0.1 of bu1] (bu1l) {};
        \node[right = 0.1 of bu1] (bu1r) {};
        \draw[thin] (u1.north) -- (bu1.south);
        \draw[thin] (u1.south) -- ++(0, -0.35) node[] (b11) {};
        \draw[thin] (bu1.north) -- ++(0, 0.35) node[] (b14) {};
        %\node (m1r) at ($(u1r)!0.5!(bu1r)$) [draw, circle,fill=white] {};
        \draw[thin] (u1.west) -- (u1l.center);
        \draw[thin] (bu1l.center) -- (bu1.west);
        \node (m1r) at ($(u1r)!0.5!(bu1r)$) [draw, circle,fill=white] {};
        \draw[thin] (u1.east) -- (u1r.center) -- (m1r.south);
        \draw[thin] (bu1.east) -- (bu1r.center) -- (m1r.north);
        \draw (b11) arc (180:360:0.1) node[] (b12) {};
        \draw[thin] (b12.center) -- ++(0, 0.15) node[] (b13) {};
        \draw (b14) arc (180:0:0.1) node[] (b15) {};
        \draw[thin] (b15.center) -- ++(0, -0.15) node[] (b16) {};
        \begin{scope}[on background layer]
            \draw[dotted] (b13.center) -- (b16.center);
        \end{scope}
    \end{tikzpicture}
    =
    \begin{tikzpicture}[baseline={([yshift=-0.1cm]current bounding box.center)}]
        \node[draw, circle] (u1) {};
        \draw[thin] (u1.north) -- ++(0, 0.3) -- ++(-0.25, 0);
        \draw[thin] (u1.south) -- ++(0, -0.3) -- ++(-0.25, 0);
    \end{tikzpicture}\,\,.
\end{align}

Suppose the system size $N$ is sufficiently large and thus $\mathcal{U}_N$ is simple, the GNVW index of the unitary operator $\hat{U}_F$ is calculated by \cite{PhysRevB.98.245122,Cirac_2017}
\begin{align}
    \rank\left(
    \vcenter{\hbox{
    \begin{tikzpicture}
        \node[draw, fill=white, minimum size=16pt, font=\bfseries] (centerNode) {$\mathcal{U}_N$};
        \draw[thin] (centerNode.north) -- ++(0, 10pt);
        \draw[thin] (centerNode.south) -- ++(0, -10pt);
        \draw[thin] (centerNode.east) -- ++(10pt, 0);
        \draw[thin] (centerNode.west) -- ++(-10pt, 0);
        \draw[dotted,thick] (-0.5,0.5)  -- +(1,-1) ; % 右
    \end{tikzpicture}
    }}\right)\bigg/
    \rank\left(
    \vcenter{\hbox{
    \begin{tikzpicture}
        \node[draw, fill=white, minimum size=16pt, font=\bfseries] (centerNode) {$\mathcal{U}_N$};
        \draw[thin] (centerNode.north) -- ++(0, 10pt);
        \draw[thin] (centerNode.south) -- ++(0, -10pt);
        \draw[thin] (centerNode.east) -- ++(10pt, 0);
        \draw[thin] (centerNode.west) -- ++(-10pt, 0);
        \draw[dotted,thick] (-0.5,-0.5)  -- +(1,1) ; % 右
    \end{tikzpicture}
    }}\right)
    \equiv
    \frac{r_N}{l_N}
    =
    I_\text{GNVW}^2,
    \label{eq:rankratioindex_sm}
\end{align}
where the rank of a tensor with a diagonal dotted line indicates that of the singular value decomposition between legs separated by the line.
We refer to the numerator (denominator) of the left-hand side of (\ref{eq:rankratioindex_sm}) as the right (left) rank of $\mathcal{U}_N$ and are denoted by $r_N(l_N)$. 
When $\mathcal{U}_N$ is simple, $r_Nl_N=d^{2N}$ and thus $r_N(l_N)=I_\text{GNVW}^{\pm 1}d^{N}$ \cite{Cirac_2017}.

\subsection{$t=1$}

Assume that $\mathcal{U}_N$ is simple and thus the relation (\ref{eq:rankratioindex_sm}) holds.
For positive $a$,
\begin{align}
    \rank\hat{U}_F(1)
    &=
    \rank
    \left(

    }}\right)
    =
    r_ND,
    \label{eq:u2}
\end{align}
and thus $\rank\hat{U}_F(a>0)=r_ND^{a-1}$ since each additional virtual leg contributes to the rank by $D$.
On the other hand for negative $a$,
\begin{align}
    \rank\hat{U}_F(-1)
    &=
    \rank\left(
    %
    \right..
\end{align}

\subsection{$t=2$}
\label{sec:t2}

To discuss the $t=2$ case, we assume that the system size is twice an integer, and we block $N/2$ sites to compose a tensor $\mathcal{U}_{N/2}$.
That is, the tensors $\mathcal{U}_N$ and $\mathcal{U}_{N/2}$ are related by
\begin{align}
    %
.
    \label{eq:t2_decomposition_sm}
\end{align}
Here, we assume that 
\begin{align}
    %
\end{align}
are both simple.
Let the right and left ranks of $\mathcal{U}_{N/2}$ be denoted by $r_{N/2}$ and $l_{N/2}$ as in (\ref{eq:rankratioindex_sm}).
Comparing the rank of both sides of (\ref{eq:t2_decomposition_sm}), the rank of the RHS with a separation below is upper-bounded by $r_{N/2}d^{N/2}$.
\begin{align}
    %
\end{align}
Similarly, $l_N\le l_{N/2}d^{N/2}$ holds.
However, since $\mathcal{U}_N$ and $\mathcal{U}_{N/2}$ are both simple, $r_Nl_N=d^{2N}$ and $r_{N/2}l_{N/2}=d^N$. 
Thus, we obtain $r_N=r_{N/2}d^{N/2}$ and $l_N=l_{N/2}d^{N/2}$ \cite{Cirac_2017}.

For $a=1$,
\begin{align}
    \hat{U}_F^2(1)
    &\,\,=\,\,
    %
\,\,.
\end{align}
We use a property for a matrix $A$ and a diagonal matrix $D$ with positive entries that $\rank A^\dagger DA=\rank A$.
Using the relations (\ref{eq:isometryseparation_sm}) repeatedly and $D=I_{d^N}\otimes \rho$, where $\rho$ is the right fixed-point eigenvector in (\ref{eq:phirho}), one obtains
\begin{align}
    &[\hat{U}_F^\dagger(1)]^2\left(I_{d^N}\otimes \rho\right)\hat{U}_F^2(1)= \notag\\[+8pt]
    &
    %
 \notag\\[+8pt]
    &=
    \hat{U}_F^\dagger(1)\left(I_{d^N}\otimes \rho\right)\hat{U}_F(1).
\end{align}
Thus, we find
\begin{align}
    \rank\hat{U}_F^2(1)
    =
    \rank[\hat{U}_F^\dagger(1)]^2\left(I_{d^N}\otimes \rho\right)\hat{U}_F^2(1)
    =
    \rank\hat{U}_F^\dagger(1)\left(I_{d^N}\otimes \rho\right)\hat{U}_F(1)
    =
    \rank\hat{U}_F(1)
    =
    r_N.
\end{align}
Similarly, one can derive the $a=-1$ case by
$
    \rank\hat{U}_F^2(-1)
    =
    \rank[\hat{U}_F^\dagger(-1)]^2\hat{U}_F^2(-1)
    =
    \rank\hat{U}_F(-1)\hat{U}^\dagger(-1)
    =
    \rank\hat{U}_F(-1)=l_N
$.

For $a=2$,
\begin{align}
    \rank\hat{U}_F^2(2)
    &\,\,=\,\,
    \rank\left(
    \begin{tikzpicture}[baseline={([yshift=-0.1cm]current bounding box.center)}]
        \node[draw, fill=white, minimum size=16pt, font=\bfseries] (u11) {$\mathcal{U}_{N/2}$};
        \node[draw, fill=white, minimum size=16pt, font=\bfseries, right = 0.5 of u11] (u12) {$\mathcal{U}_{N/2}$};
        \draw[thin] (u11.east) -- (u12.west);
        \node[draw, fill=white, minimum size=16pt, font=\bfseries, above = 0.3 of u11] (u21) {$\mathcal{U}_{N/2}$};
        \node[draw, fill=white, minimum size=16pt, font=\bfseries, right = 0.5 of u21] (u22) {$\mathcal{U}_{N/2}$};
        \draw[thin] (u21.east) -- (u22.west);
        \draw[thin] (u11.north) -- (u21.south);
        \draw[thin] (u12.north) -- (u22.south);
        \draw[thin] (u11.south) -- ++(0, -0.25);
        \draw[thin] (u12.south) -- ++(0, -0.25);
        \draw[thin] (u11.west) -- ++(-10pt, 0) node[] (b11) {};
        \draw[thin] (u12.east) -- ++(10pt, 0) node[] (t11l) {};
        \draw[thin] (u21.north) -- ++(0,0.25);
        \draw[thin] (u22.north) -- ++(0,0.25);
        \draw[thin] (u21.west) -- ++(-10pt, 0) node[] (b21) {};
        \draw[thin] (u22.east) -- ++(10pt, 0) node[] (t21l) {};
        \draw (t11l) arc (-90:0:0.2) node[] (t11a) {};
        \node[right = 0.2 of t11l] (t11r) {};
        \draw (t11r) arc (90:180:0.2) node[] (t11b) {};
        \draw[thin] (t11b.center) -- ++(0, -0.3);
        \draw[thin] (t11r.center) -- ++(0.2, 0) node[](t12l){};
        \draw (t12l) arc (-90:0:0.2) node[] (t12a) {};
        \node[right = 0.2 of t12l] (t12r) {};
        \draw (t12r) arc (90:180:0.2) node[] (t12b) {};
        \draw[thin] (t12b.center) -- ++(0, -0.3);
        \draw[thin] (t12r.center) -- ++(0.2, 0) node[](b14){};
        \draw (b11) arc (270:90:0.1) node[] (b12) {};
        \draw[thin] (b12.center) -- ++(0.15, 0) node[] (b13) {};
        \draw (b14) arc (-90:90:0.1) node[] (b15) {};
        \draw[thin] (b15.center) -- ++(-0.15, 0) node[] (b16) {};
        \begin{scope}[on background layer]
            \draw[dotted] (b13.center) -- (b16.center);
        \end{scope}
        \draw (t21l) arc (-90:0:0.2) node[] (t21a) {};
        \draw[thin] (t21a.center) -- ++(0, 0.3);
        \node[right = 0.2 of t21l] (t21r) {};
        \draw (t21r) arc (90:180:0.2) node[] (t21b) {};
        \draw[thin] (t21b.center) -- (t11a.center);
        \draw[thin] (t21r.center) -- ++(0.2, 0) node[](t22l){};
        \draw (t22l) arc (-90:0:0.2) node[] (t22a) {};
        \draw[thin] (t22a.center) -- ++(0, 0.3);
        \node[right = 0.2 of t22l] (t22r) {};
        \draw (t22r) arc (90:180:0.2) node[] (t22b) {};
        \draw[thin] (t22b.center) -- (t12a.center);
        \draw[thin] (t22r.center) -- ++(0.2, 0) node[](b24){};
        \draw (b21) arc (270:90:0.1) node[] (b22) {};
        \draw[thin] (b22.center) -- ++(0.15, 0) node[] (b23) {};
        \draw (b24) arc (-90:90:0.1) node[] (b25) {};
        \draw[thin] (b25.center) -- ++(-0.15, 0) node[] (b26) {};
        \begin{scope}[on background layer]
            \draw[dotted] (b23.center) -- (b26.center);
        \end{scope}
    \end{tikzpicture}
    \right)
    \,\,=\,\,
    \rank\left(
    \begin{tikzpicture}[baseline={([yshift=-0.1cm]current bounding box.center)}]
        \node[draw, fill=white, minimum size=16pt, font=\bfseries] (u11) {$\mathcal{U}_{N/2}$};
        \node[draw, fill=white, minimum size=16pt, font=\bfseries, right = 0.5 of u11] (u12) {$\mathcal{U}_{N/2}$};
        \draw[thin] (u11.east) -- (u12.west);
        \node[draw, fill=white, minimum size=16pt, font=\bfseries, above = 0.3 of u11] (u21) {$\mathcal{U}_{N/2}$};
        \node[draw, fill=white, minimum size=16pt, font=\bfseries, right = 0.5 of u21] (u22) {$\mathcal{U}_{N/2}$};
        \draw[thin] (u21.east) -- (u22.west);
        \draw[thin] (u11.north) -- (u21.south);
        \draw[thin] (u12.north) -- (u22.south);
        \draw[thin] (u11.south) -- ++(0, -0.25);
        \draw[thin] (u12.south) -- ++(0, -0.25);
        \draw[thin] (u11.west) -- ++(-0.25, 0);
        \draw[thin] (u12.east) -- ++(0.25, 0);
        \draw[thin] (u21.north) -- ++(0,0.25);
        \draw[thin] (u22.north) -- ++(0,0.25);
        \draw[thin] (u21.west) -- ++(-0.25, 0);
        \draw[thin] (u22.east) -- ++(0.25, 0);
        \draw[dotted,thick] (-0.7,1.4) -- (2.2,-0.5);
    \end{tikzpicture}
    \right)\,\,.
\end{align}
Thus $\rank\hat{U}_F^2(\pm2)$ is the right (left) rank of the above tensor.
Since the diagonal lines intersect two $\mathcal{U}_{N/2}$ tensors, the right and left ranks are upper-bounded by $r_{N/2}^2=r_N^2d^{-N}$ and $l_{N/2}^2=l_N^2d^{-N}$, respectively, while the product of them is $d^{2N}=(r_Nl_N)^2d^{-2N}$.
Therefore, the right and left ranks are, respectively, $r_N^2d^{-N}$ and $l_N^2d^{-N}$. 
When $|a|>2$, additional virtual legs contribute as in (\ref{eq:u2}) and (\ref{eq:u-2}), and thus $\rank\hat{U}_F^2(a>2)=r_N^2d^{-N}D^{a-2}$ and $\rank\hat{U}_F^2(a<-2)=l_N^2d^{-N}D^{|a|-2}$.
In summary, 
\begin{align}
    \rank\hat{U}_F^2(a)
    =
    \left\{
\begin{array}{ll}
        l_N^2d^{-N}D^{|a|-2}\quad &(a\le -2)\\[+3pt]
        l_N\quad &(a=-1)\\[+3pt]
        d^N \quad&(a=0)\\[+3pt]
        r_N\quad &(a=1)\\[+3pt]
        r_N^2d^{-N}D^{a-2}\quad&(a\ge 2)
    \end{array}
    \right.
    =
    d^N\cdot
    \left\{
    \begin{array}{ll}
        I_\text{GNVW}^{-2}D^{|a|-2}\quad &(a\le -2)\\[+3pt]
        I_\text{GNVW}^{-1}\quad &(a=-1)\\[+3pt]
        1 \quad&(a=0)\\[+3pt]
        I_\text{GNVW}\quad &(a=1)\\[+3pt]
        I_\text{GNVW}^2D^{a-2}\quad&(a\ge 2)
    \end{array}
    \right..
\end{align}

\subsection{$t=3$}
    
We assume that the system size is three times an integer, and we block $N/3$ sites to define $\mathcal{U}_{N/3}$ as
\begin{align}
    %
.
\end{align}
We assume that the following three tensors are simple:
\begin{align}
    %
.
    \label{eq:t3simple}
\end{align}
Let the right and left ranks of $\mathcal{U}_{N/3}$ be denoted by $r_{N/3}$ and $l_{N/3}$. 
Then a similar argument as in Sec.~\ref{sec:t2} gives $r_N=r_{N/3}d^{2N/3}$ and $l_N=l_{N/3}d^{2N/3}$.

For $a=1$,
\begin{align}
    \hat{U}_F^3(1)
    &\,\,=\,\,
    %
\,\,.
\end{align}
Using the relations (\ref{eq:isometryseparation_sm}) repeatedly and $D=I_{d^N}\otimes \rho$, we obtain
\begin{align}
    [\hat{U}_F^\dagger(1)]^3\left(I_{d^N}\otimes \rho\right)\hat{U}_F^3(1)=
    %
 
    =
    \hat{U}_F^\dagger(1)\left(I_{d^N}\otimes \rho\right)\hat{U}_F(1).
\end{align}
Thus, we find $\rank\hat{U}_F^3(1)=\rank\hat{U}_F(1)=r_N$.
Similarly, one can derive $\rank\hat{U}_F^3(-1)=\rank\hat{U}_F(-1)=l_N$.

For $a=2$,
\begin{align}
    \hat{U}_F^3(2)
    &\,\,=\,\,
    %
,
\end{align}
where the vertical line is the identity operator on $N/3$ physical qubits.
Since the tensors in (\ref{eq:t3simple}) are simple, we obtain
\begin{align}
    \rank\hat{U}_F^3(2)
    \,\,=\,\,
    \rank \left(
    %
    }}\right)\,\,
    =
    r_{N/3}^2d^{N/3}
    =
    r_N^2d^{-N}.
\end{align}
Similarly, one can show that
$
    \rank\hat{U}_F^3(-2)=l_N^2d^{-N}
$.

In the case with $a=\pm 3$, there appear the right and left ranks of the first tensor in (\ref{eq:t3simple}), and their ranks are $r_N^3d^{-2N}$ and $l_N^3d^{-2N}$, respectively.
A straightforward calculation gives $\rank\hat{U}_F^3(a\ge 3)=r_N^3d^{-2N}D^{a-3}$ and $\rank\hat{U}_F^3(a\le -3)=l_N^3d^{-2N}D^{|a|-3}$.
In summary,
\begin{align}
    \rank\hat{U}_F^3(a)
    =
    \left\{
    \begin{array}{ll}
        l_N^3d^{-2N}D^{|a|-3}\quad &(a\le -3)\\[+3pt]
        l_N^2d^{-N}\quad &(a=-2)\\[+3pt]
        l_N\quad &(a=-1)\\[+3pt]
        d^N \quad&(a=0)\\[+3pt]
        r_N\quad &(a=1)\\[+3pt]
        r_N^2d^{-N}\quad &(a=2)\\[+3pt]
        r_N^3d^{-2N}D^{a-3}\quad&(a\ge 3)
    \end{array}
    \right.
    =
    d^N\cdot
    \left\{
    \begin{array}{ll}
        I_\text{GNVW}^{-3}D^{|a|-3}\quad &(a\le -3)\\[+3pt]
        I_\text{GNVW}^{-2}\quad &(a=-2)\\[+3pt]
        I_\text{GNVW}^{-1}\quad &(a=-1)\\[+3pt]
        1 \quad&(a=0)\\[+3pt]
        I_\text{GNVW}\quad &(a=1)\\[+3pt]
        I_\text{GNVW}^2\quad &(a=2)\\[+3pt]
        I_\text{GNVW}^3D^{a-3}\quad&(a\ge 3)
    \end{array}
    \right..
\end{align}

\subsection{Arbitrary $t$}

From the results obtained so far, we can generalize the rank of the Floquet operator for arbitrary $t>0$ and $a$ as
\begin{align}
    \rank\hat{U}_F^t(a)
    =
    d^N\cdot
    \left\{
    \begin{array}{ll}
        I_\text{GNVW}^{-t}D^{|a|-t}\quad &(a\le -t)\\[+3pt]
        I_\text{GNVW}^{a}\quad &(|a|\le t)\\[+3pt]
        I_\text{GNVW}^tD^{a-t}\quad&(a\ge t)
    \end{array}
    \right.,
    \label{eq:rankut}
\end{align}
Since the dimension $D$ of the virtual Hilbert space have some arbitrariness, physically relevant quantities are their ratios given by 
\begin{align}
    \sqrt{\frac{\rank\hat{U}_F^t(a)}{\rank\hat{U}_F^t(-a)}}
    =
    \left\{
    \begin{array}{ll}
        I_\text{GNVW}^{-t}\quad&(a< -t)\\[+5pt]
        I_\text{GNVW}^{a}\quad&(|a|\le t)\\[+5pt]
        I_\text{GNVW}^{t}\quad&(a>t)
    \end{array}
    \right.,
    \label{eq:twistrankratio}
\end{align}
or equivalently, 
\begin{align}
    \frac{1}{2}\log_d \frac{\rank\hat{U}_F^t(a)}{\rank\hat{U}_F^t(-a)}
    =
    \left\{
    \begin{array}{ll}
        -\text{ind}\cdot t\quad&(a< -t)\\[+5pt]
        \text{ind}\cdot a\quad&(|a|\le t)\\[+5pt]
        \text{ind}\cdot t\quad&(a>t)
    \end{array}
    \right.,
\end{align}
where $\text{ind} = \log_d I_\text{GNVW}$.
Notice that this result is not periodic with respect to $a$ modulo $t$ as in \cite{1dc6-y1x8}, since we do not trace out the physical degrees of freedom of $\hat{U}^t(a)$.

\section{Shift operator}
\label{sec:shiftrank}

As an example of the boundary Floquet operator having a nontrivial GNVW index, we consider the shift operator, which is the tensor product of the right-translation operators given as
\begin{align}
    \hat{S}
    =
    \,\,
    %
.
    \label{eq:shiftoperator}
\end{align}
The shift operator has the GNVW index $I_\text{GNVW}=d$ \cite{PhysRevX.6.041070} and $D=d$.
The shift operator under the time-translation twist is defined on $N+|a|$ qudits as
\begin{align}
    \hat{S}(a)
    =
    \left\{
    \begin{array}{ll}
        \begin{tikzpicture}[baseline={([yshift=0.5ex]current bounding box.center)}]
        % 4方向（北・南・東・西）に足を伸ばす
        \draw (0,0.2) -- +(0, 0.3); % 上
        \draw (0,-0.2) -- +(0,-0.2) node[below] {1}; % 下
        \draw (-0.2,0)  -- +(-0.2,0) ; % 左
        \draw (0.2,0)  -- +(0.4, 0) ; % 右
        \draw[thin] (-0.2,0.) arc [radius=0.2, start angle = 270, end angle=360];
        \draw[thin] (0.2,0.) arc [radius=0.2, start angle = 90, end angle=180];
        % 4方向（北・南・東・西）に足を伸ばす
        \draw (0.8,0.2) -- +(0, 0.3); % 上
        \draw (0.8,-0.2) -- +(0,-0.2) node[below] {2}; % 下
        \draw (1.0,0)  -- +(0.4, 0); % 右
        \draw[thin] (0.6,0.) arc [radius=0.2, start angle = 270, end angle=360];
        \draw[thin] (1.0,0.) arc [radius=0.2, start angle = 90, end angle=180];
        \draw (1.6,0.2) -- +(0, 0.3); % 上
        \draw (1.6,-0.2) -- +(0,-0.2) node[below] {3}; % 下
        \draw (1.8,0)  -- +(0.2, 0); % 右
        \draw[dotted] (2.2,0.)  -- +(0.4, 0);
        \draw[thin] (1.4,0.) arc [radius=0.2, start angle = 270, end angle=360];
        \draw[thin] (1.8,0.) arc [radius=0.2, start angle = 90, end angle=180];
        \draw (3.2,0.2) -- +(0, 0.3); % 上
        \draw (3.2,-0.2) -- +(0,-0.2) node[below] {$N$}; % 下
        \draw (3.,0)  -- +(-0.2,0) ; % 左
        \draw (3.4,0)  -- +(0.4, 0); % 右
        \draw[thin] (3.0,0.) arc [radius=0.2, start angle = 270, end angle=360];
        \draw[thin] (3.4,0.) arc [radius=0.2, start angle = 90, end angle=180];
        \draw (4.,0.2) -- +(0, 0.3); % 上
        \draw (4.,-0.2) -- +(0,-0.2) node[below] {$N+1$}; % 下
        \draw (4.2,0)  -- +(0.2, 0) ; % 右
        \draw[thin] (3.8,0.) arc [radius=0.2, start angle = 270, end angle=360];
        \draw[thin] (4.2,0.) arc [radius=0.2, start angle = 90, end angle=180];
        \draw[dotted] (4.6,0.)  -- +(0.4, 0);
        \draw (5.6,0.2) -- +(0, 0.3); % 上
        \draw (5.6,-0.2) -- +(0,-0.2) node[below] {$N+a$}; % 下
        \draw (5.4,0)  -- +(-0.2, 0) ; % 右
        \draw (5.8,0)  -- +(0.2, 0) ; % 右
        \draw[thin] (5.4,0.) arc [radius=0.2, start angle = 270, end angle=360];
        \draw[thin] (5.8,0.) arc [radius=0.2, start angle = 90, end angle=180];
        \draw[thin] (6.,0.) arc [radius=0.1, start angle = -90, end angle=90];
        \draw[thin] (-0.4,0.) arc [radius=0.1, start angle = 270, end angle=90];
        \draw[densely dotted] (-0.4,0.2)  -- +(6.4, 0);
        \end{tikzpicture}&\quad (a>0)  \\
\end{array}
    \right..
    \label{eq:tensornetwork_shiftandtwist}
\end{align}
The shift operator with $a\ge 0$ is still a unitary operator on $N+a$ sites, and thus $\rank\hat{S}^t(a\ge0)=d^{N+a}$.
On the other hand, the shift operator with $a<0$ is not unitary since the bottom legs at sites $N$ and $N+1$ are connected to each other.
This connection reduces the rank by $d^2$, hence $\rank\hat{S}(a<0)=d^{N+|a|-2}$.
For $\hat{S}^2(a<0)$, a similar connection occurs also between sites $N-1$ and $N+2$, reducing the rank further by $d^2$ and thus $\rank\hat{S}^2(a<0)=d^{N+|a|-4}$ when $a\le-2$.
When $t\ge -a$, no further reductions of the rank occur.
This result can be generalized to $\rank\hat{S}^t(a)=d^{N-a-2\text{min}(N,t,-a)}$.
In particular, when $N\ge\text{min}(t,-a)$, we obtain 
\begin{align}
    \rank\hat{S}^t(a)
    =
    \left\{
    \begin{array}{ll}
        d^{N+|a|-2t}&(a< -t)\\[+3pt]
        d^{N+a}\quad &(a\ge -t)\\[+3pt]
    \end{array}
    \right.,
    \label{eq:shiftrankt}
\end{align}
which agrees with (\ref{eq:mpu_floquetrank}) by substituting $d=D$ and $I_\text{GNVW}=d$.
We have numerically examined (\ref{eq:shiftrankt}) for the qubit case ($d=2$) with $N\in[1,6]$, $t\in[1,3]$, and $a\in[-5,5]$.

On the other hand, the number of nonzero eigenvalues of $\hat{S}(a)$, which is in general equal to or smaller than the rank, is confirmed numerically to be $2^{N+a}$.
This is the dimension $d^{N+a}$ of the tilted slice of Sec.~\ref{sec:tilted}, on which $\hat{S}(a)$ acts as the shift on $N+a$ sites.
The discrepancy between the rank and the number of nonzero eigenvalues for $a<0$ is resulting from Jordan blocks of size $n$,
\begin{align}
    J_n
    =
    \begin{pmatrix}
        0 & 1 & 0 & \cdots & 0 \\
        0 & 0 & 1 & \cdots & 0 \\
        \vdots & \vdots & \vdots & \ddots & \vdots \\
        0 & 0 & 0 & \cdots & 1 \\
        0 & 0 & 0 & \cdots & 0 \\
    \end{pmatrix}
    \label{eq:jordanblock}
\end{align} 
in the Jordan normal form of $\hat{S}(a)$, which contribute to the rank by $n-1$ while the eigenvalues are 0.
Since the rank of the Jordan block is reduced by a power as $\rank J_n^m = \text{max}[0,n-m]$, $\rank\hat{S}^t(a)$ is also reduced by increasing $t$.
This consideration indicates that the maximum size of the Jordan block is $-a>0$, since from (\ref{eq:shiftrankt}) $\rank\hat{S}^t(a)$ no longer decreases when $t$ is larger than $-a$.

\section{Random unitary circuits}
\label{sec:ruc}

Here, we numerically evaluate the rank of random unitary circuits (RUCs) on qubits under the time-translation twist.
We assume that the unitary operator $\hat{U}_F$ consists of a shift operator $\hat{S}$ given in (\ref{eq:shiftoperator}) and tensor products of on-site or two-site Haar random unitaries denoted by $\hat{U}_2$ and $\hat{U}_4$, respectively.
Therefore, the GNVW index of $\hat{U}_F$ is $I_\text{GNVW}=2$.
We assume that the system size $N$ is even when we consider two-site random unitaries.
%Since $\hat{U}_F^t(a)$ is unitary for positive $a$, we obtain $\text{rank}\,\hat{U}_F^t(a\ge 0)=2^{N+a}r^a$ ($r$ is the operator Schmidt rank of the two-site unitary across the boundary, thus $D=2r$) for arbitrary $t$, which agrees with (\ref{eq:mpu_floquetrank}).

We consider three types of spatially-nonuniform random unitary circuits, the shift operator following a tensor product of on-site unitary operators $\hat{U}_2=\bigotimes_{j=1}^NU_j$ where $U_j\in U(2)$,
\begin{align}
    &\hat{U}_{F,2}(a<0)=
    \hat{S}(a)\left(\hat{U}_2\otimes I_{2^{|a|}}\right)
    =
    \,\,
    \begin{tikzpicture}[baseline={([yshift=0.5ex]current bounding box.center)}]
    % 4方向（北・南・東・西）に足を伸ばす
    \draw (0,0.2) -- +(0, 0.3); % 上
    \draw (0,-0.2) -- +(0,-0.3) node[below] {}; % 下
    \draw (-0.1,-0.4) -- +(0.2, 0.2); 
    \draw (0.1,-0.4) -- +(-0.2, 0.2);
    \draw (-0.2,0)  -- +(-0.2,0) ; % 左
    \draw (0.2,0)  -- +(0.4, 0) ; % 右
    \draw[thin] (-0.2,0.) arc [radius=0.2, start angle = 270, end angle=360];
    \draw[thin] (0.2,0.) arc [radius=0.2, start angle = 90, end angle=180];
    % 4方向（北・南・東・西）に足を伸ばす
    \draw (0.8,0.2) -- +(0, 0.3); % 上
    \draw (0.8,-0.2) -- +(0,-0.3) node[below] {}; % 下
    \draw (0.7,-0.4) -- +(0.2, 0.2); 
    \draw (0.9,-0.4) -- +(-0.2, 0.2);
    \draw (1.0,0)  -- +(0.4, 0); % 右
    \draw[thin] (0.6,0.) arc [radius=0.2, start angle = 270, end angle=360];
    \draw[thin] (1.0,0.) arc [radius=0.2, start angle = 90, end angle=180];
    \draw (1.6,0.2) -- +(0, 0.3); % 上
    \draw (1.6,-0.2) -- +(0,-0.3) node[below] {}; % 下
    \draw (1.5,-0.4) -- +(0.2, 0.2); 
    \draw (1.7,-0.4) -- +(-0.2, 0.2);
    \draw (1.8,0)  -- +(0.2, 0); % 右
    \draw[dotted] (2.2,0.)  -- +(0.4, 0);
    \draw[thin] (1.4,0.) arc [radius=0.2, start angle = 270, end angle=360];
    \draw[thin] (1.8,0.) arc [radius=0.2, start angle = 90, end angle=180];
    \draw (3.2,0.2) -- +(0, 0.3); % 上
    \draw (3.2,-0.2) -- +(0,-0.3) node[below] {}; % 下
    \draw (3.1,-0.4) -- +(0.2, 0.2); 
    \draw (3.3,-0.4) -- +(-0.2, 0.2);
    \draw (3.,0)  -- +(-0.2,0) ; % 左
    \draw (3.4,0)  -- +(0.4, 0); % 右
    \draw[thin] (3.0,0.) arc [radius=0.2, start angle = 270, end angle=360];
    \draw[thin] (3.4,0.) arc [radius=0.2, start angle = 90, end angle=180];
    \draw (4.,0.2) -- +(0, 0.3); % 上
    \draw (4.,-0.2) -- +(0,-0.3) node[below] {}; % 下
    \draw (4.2,0)  -- +(0.2, 0) ; % 右
    \draw[thin] (3.8,0.) arc [radius=0.2, start angle = 90, end angle=0];
    \draw[thin] (4.2,0.) arc [radius=0.2, start angle = 270, end angle=180];
    \draw[dotted] (4.6,0.)  -- +(0.4, 0);
    \draw (5.6,0.2) -- +(0, 0.3); % 上
    \draw (5.6,-0.2) -- +(0,-0.3) node[below] {}; % 下
    \draw (5.4,0)  -- +(-0.2, 0) ; % 右
    \draw (5.8,0)  -- +(0.2, 0) ; % 右
    \draw[thin] (5.4,0.) arc [radius=0.2, start angle = 90, end angle=0];
    \draw[thin] (5.8,0.) arc [radius=0.2, start angle = 270, end angle=180];
    \draw[thin] (6.,0.) arc [radius=0.1, start angle = -90, end angle=90];
    \draw[thin] (-0.4,0.) arc [radius=0.1, start angle = 270, end angle=90];
    \draw[densely dotted] (-0.4,0.2)  -- +(6.4, 0);
    \end{tikzpicture},
\end{align}
that following a tensor product of two-site unitary operators $\hat{U}_{4e}=\bigotimes_{j=1}^{N/2}U_{2j-1,2j}$, where $U_{2j-1,2j}\in U(4)$,
\begin{align}
    \hat{U}_{F,4}(a<0)=\hat{S}(a)\left(\hat{U}_{4e}\otimes I_{2^{|a|}}\right)
    &=
    \,\,
    \begin{tikzpicture}[baseline={([yshift=0.5ex]current bounding box.center)}]
        \tikzset{
            gate/.style={
                draw,
                rounded corners=2pt,
                thick,
                fill=cyan!50!blue!20,
                minimum width=1.1cm,
                minimum height=0.2cm,
                inner sep=0pt
            }
        }
        %gates
        \node[gate] (g1) at (0.,0.5) {};
        \node[gate] (g2) at (1.6,0.5) {};
        \node[gate] (g3) at (4.0,0.5) {};
        %gate lines
        \draw ($(g1.center)+(-0.4,0.1)$) -- +(0.,0.2) node[] (t1b) {};
        \draw ($(g1.center)+(-0.4,-0.1)$) -- +(0.,-0.3);
        \draw ($(g1.center)+(0.4,0.1)$) -- +(0.,0.2) node[] (t2b) {};
        \draw ($(g1.center)+(0.4,-0.1)$) -- +(0.,-0.3);
        \draw ($(g2.center)+(-0.4,0.1)$) -- +(0.,0.2) node[] (t3b) {};
        \draw ($(g2.center)+(-0.4,-0.1)$) -- +(0.,-0.3);
        \draw ($(g2.center)+(0.4,0.1)$) -- +(0.,0.2) node[] (t4b) {};
        \draw ($(g2.center)+(0.4,-0.1)$) -- +(0.,-0.3);
        \draw ($(g3.center)+(-0.4,0.1)$) -- +(0.,0.2) node[] (t5b) {};
        \draw ($(g3.center)+(-0.4,-0.1)$) -- +(0.,-0.3);
        \draw ($(g3.center)+(0.4,0.1)$) -- +(0.,0.2) node[] (t6b) {};
        \draw ($(g3.center)+(0.4,-0.1)$) -- +(0.,-0.3);
        %shift
        \node[above = 0.15 of t1b] (t1t) {};
        \draw[thin] (t1b) arc [radius=0.2, start angle = 180, end angle=90] node[] (t1r) {};
        \draw[thin] (t1t) arc [radius=0.2, start angle = 360, end angle=270] node[] (t1l) {};
        \draw (t1t.center) -- +(0.,0.3);
        \draw (t1r.center) -- +(0.4,0);
        \node[above = 0.15 of t2b] (t2t) {};
        \draw[thin] (t2b) arc [radius=0.2, start angle = 180, end angle=90] node[] (t2r) {};
        \draw[thin] (t2t) arc [radius=0.2, start angle = 360, end angle=270] node[] (t2l) {};
        \draw (t2t.center) -- +(0.,0.3);
        \draw (t2r.center) -- +(0.4,0);
        \node[above = 0.15 of t3b] (t3t) {};
        \draw[thin] (t3b) arc [radius=0.2, start angle = 180, end angle=90] node[] (t3r) {};
        \draw[thin] (t3t) arc [radius=0.2, start angle = 360, end angle=270] node[] (t3l) {};
        \draw (t3t.center) -- +(0.,0.3);
        \draw (t3r.center) -- +(0.4,0);
        \node[above = 0.15 of t4b] (t4t) {};
        \draw[thin] (t4b) arc [radius=0.2, start angle = 180, end angle=90] node[] (t4r) {};
        \draw[thin] (t4t) arc [radius=0.2, start angle = 360, end angle=270] node[] (t4l) {};
        \draw (t4t.center) -- +(0.,0.3);
        \draw (t4r.center) -- +(0.2,0);
        \node[above = 0.15 of t5b] (t5t) {};
        \draw[thin] (t5b) arc [radius=0.2, start angle = 180, end angle=90] node[] (t5r) {};
        \draw[thin] (t5t) arc [radius=0.2, start angle = 360, end angle=270] node[] (t5l) {};
        \draw (t5t.center) -- +(0.,0.3);
        \draw (t5r.center) -- +(0.4,0);
        \draw (t5l.center) -- +(-0.2,0);
        \draw[dotted] ($(t4r.center)+(0.4,0)$) -- ($(t5l.center)+(-0.4,0)$);
        \node[above = 0.15 of t6b] (t6t) {};
        \draw[thin] (t6b) arc [radius=0.2, start angle = 180, end angle=90] node[] (t6r) {};
        \draw[thin] (t6t) arc [radius=0.2, start angle = 360, end angle=270] node[] (t6l) {};
        \draw (t6t.center) -- +(0.,0.3);
        %twist
        \node[right = 0.3 of t6r] (t7l) {};
        \draw (t6r.center) -- (t7l.center);
        \node[right = 0.15 of t7l] (t7r) {};
        \draw[thin] (t7l) arc [radius=0.2, start angle = 90, end angle=0] node[] (t7b) {};
        \draw[thin] (t7r) arc [radius=0.2, start angle = 270, end angle=180] node[] (t7t) {};
        \draw (t7b.center) -- +(0,-0.7);
        \draw (t7t.center) -- +(0,0.3);
        \draw (t7r.center) -- +(0.2,0);
        \node[right = 1.6 of t7l] (t8l) {};
        \draw (t8l.center) -- +(-0.2,0);
        \draw[dotted] ($(t7r.center)+(0.4,0)$) -- ($(t8l.center)+(-0.4,0)$);
        \node[right = 0.15 of t8l] (t8r) {};
        \draw[thin] (t8l) arc [radius=0.2, start angle = 90, end angle=0] node[] (t8b) {};
        \draw[thin] (t8r) arc [radius=0.2, start angle = 270, end angle=180] node[] (t8t) {};
        \draw (t8b.center) -- +(0,-0.7);
        \draw (t8t.center) -- +(0,0.3);
        \draw (t1l.center) -- +(-0.3,0) node[] (l0) {};
        \draw (t8r.center) -- +(0.3,0) node[] (r0) {};
        \draw (r0) arc [radius=0.11, start angle = -90, end angle=90] node[] (r1) {};
        \draw (l0) arc [radius=0.11, start angle = 270, end angle=90] node[] (l1) {};
        \draw[densely dotted] (r1.center)  -- (l1.center);
    \end{tikzpicture},
    \label{eq:brickwall_1}
\end{align}
and that following two layers of the two-site unitaries $\hat{U}_{4e}$ and $\hat{U}_{4o}=\bigotimes_{j=1}^{N/2}U_{2j,2j+1}$ in the brickwork pattern
\begin{align}
    \hat{U}_{F,(4,4)}(a<0)=\hat{S}(a)\left(\hat{U}_{4e}\otimes I_{D^{|a|}}\right)\hat{U}_{4o}(a)
    &=
    \,\,
\begin{tikzpicture}[baseline={([yshift=-0.1ex]current bounding box.center)}]
        \tikzset{
            gate/.style={
                draw,
                rounded corners=2pt,
                thick,
                fill=cyan!50!blue!20,
                minimum width=1.1cm,
                minimum height=0.2cm,
                inner sep=0pt
            }
        }
        %gates 1
        \node[gate] (gh0) at (-0.4,0.5) {};
        \fill[white] ($(gh0.center)+(0,-0.2)$) rectangle ($(gh0.west)+(-0.01,0.2)$);
        \node[gate] (g1) at (0.4,1.1) {};
        \node[gate] (g2) at (1.2,0.5) {};
        \node[gate] (g3)at (2.0,1.1) {}; 
        \node[gate] (gh1) at (2.8,0.5) {};
        \fill[white] ($(gh1.center)+(0,-0.2)$) rectangle ($(gh1.east)+(0.01,0.2)$);
        %gates 2
        \node[gate] (gh2) at (3.6,0.5) {};
        \fill[white] ($(gh2.center)+(0,-0.2)$) rectangle ($(gh2.west)+(-0.01,0.2)$);
        \node[gate] (g4) at (4.4,1.1) {};
        \node[gate] (gh3) at (5.2,0.5) {};
        \fill[white] ($(gh3.center)+(0,-0.2)$) rectangle ($(gh3.east)+(0.01,0.2)$);
        %vertical lines 1
        \draw ($(g1.center)+(-0.4,0.1)$) -- +(0.,0.3) node[] (s1b) {};
        \draw ($(g1.center)+(-0.4,-0.1)$) -- ($(gh0.center)+(0.4,0.1)$);
        \draw ($(gh0.center)+(0.4,-0.1)$) -- +(0.,-0.3);
        \draw ($(g1.center)+(0.4,0.1)$) -- +(0.,0.3) node[] (s2b) {};
        \draw ($(g1.center)+(0.4,-0.1)$) -- ($(g2.center)+(-0.4,0.1)$);
        \draw ($(g2.center)+(-0.4,-0.1)$) -- +(0.,-0.3);
        \draw ($(g3.center)+(-0.4,0.1)$) -- +(0.,0.3) node[] (s3b) {};
        \draw ($(g3.center)+(-0.4,-0.1)$) -- ($(g2.center)+(0.4,0.1)$);
        \draw ($(g2.center)+(0.4,-0.1)$) -- +(0.,-0.3);
        \draw ($(g3.center)+(0.4,0.1)$) -- +(0.,0.3) node[] (s4b) {};
        \draw ($(g3.center)+(0.4,-0.1)$) -- ($(gh1.center)+(-0.4,0.1)$);
        \draw ($(gh1.center)+(-0.4,-0.1)$) -- +(0.,-0.3);
        %dots
        \draw[dotted] ($(gh1.center)+(0.2,0)$) -- ($(gh2.center)+(-0.2,0.0)$);
        %vertical lines 2
        \draw ($(g4.center)+(-0.4,0.1)$) -- +(0.,0.3) node[] (s5b) {};
        \draw ($(g4.center)+(-0.4,-0.1)$) -- ($(gh2.center)+(0.4,0.1)$);
        \draw ($(gh2.center)+(0.4,-0.1)$) -- +(0.,-0.3);
        \draw ($(g4.center)+(0.4,0.1)$) -- +(0.,0.3) node[] (s6b) {};
        \draw ($(g4.center)+(0.4,-0.1)$) -- ($(gh3.center)+(-0.4,0.1)$);
        \draw ($(gh3.center)+(-0.4,-0.1)$) -- +(0.,-0.3);
        %twist
        \draw[thick] (gh3.center) -- +(0.2,0) node[] (t1l) {};
        \node[right = 0.15 of t1l] (t1r) {};
        \draw[thick] (t1l) arc [radius=0.2, start angle = 90, end angle=0] node[] (t1b) {};
        \draw[thick] (t1r) arc [radius=0.2, start angle = 270, end angle=180] node[] (t1t) {};
        \draw[thick] (t1b.center) -- +(0., -0.2);
        \draw[thick] (t1t.center) -- +(0., 1.5);
        \draw[thick] (t1r.center) -- +(0.2, 0.);
        \draw[dotted] ($(t1r.center)+(0.4,0)$)  -- +(0.6, 0);
        \node[right = 1.6 of t1l] (t2l) {};
        \draw[thick] (t2l.center) -- +(-0.2,0.);
        \node[right = 0.15 of t2l] (t2r) {};
        \draw[thick] (t2l) arc [radius=0.2, start angle = 90, end angle=0] node[] (t2b) {};
        \draw[thick] (t2r) arc [radius=0.2, start angle = 270, end angle=180] node[] (t2t) {};
        \draw[thick] (t2b.center) -- +(0., -0.2);
        \draw[thick] (t2t.center) -- +(0., 1.5);
        \draw[thick] (t2r.center) -- +(0.3, 0.) node[] (r0) {};
        \draw[thick] (gh0.center) -- +(-0.4,0.) node[] (l0) {}; 
        \draw[thick] (l0) arc [radius=0.11, start angle = 270, end angle=90] node[] (l1) {};
        \draw[thick] (r0) arc [radius=0.11, start angle = 270, end angle=450] node[] (r1) {};
        \draw[thick, densely dotted] (l1.center) -- (r1.center);
        %shift
        \node[above = 0.15 of s1b] (s1t) {};
        \draw[thin] (s1b) arc [radius=0.2, start angle = 180, end angle=90] node[] (s1r) {};
        \draw[thin] (s1t) arc [radius=0.2, start angle = 360, end angle=270] node[] (s1l) {};
        \draw (s1t.center) -- +(0.,0.3);
        \draw (s1r.center) -- +(0.4,0);
        \node[above = 0.15 of s2b] (s2t) {};
        \draw[thin] (s2b) arc [radius=0.2, start angle = 180, end angle=90] node[] (s2r) {};
        \draw[thin] (s2t) arc [radius=0.2, start angle = 360, end angle=270] node[] (s2l) {};
        \draw (s2t.center) -- +(0.,0.3);
        \draw (s2r.center) -- +(0.4,0);
        \node[above = 0.15 of s3b] (s3t) {};
        \draw[thin] (s3b) arc [radius=0.2, start angle = 180, end angle=90] node[] (s3r) {};
        \draw[thin] (s3t) arc [radius=0.2, start angle = 360, end angle=270] node[] (s3l) {};
        \draw (s3t.center) -- +(0.,0.3);
        \draw (s3r.center) -- +(0.4,0);
        \node[above = 0.15 of s4b] (s4t) {};
        \draw[thin] (s4b) arc [radius=0.2, start angle = 180, end angle=90] node[] (s4r) {};
        \draw[thin] (s4t) arc [radius=0.2, start angle = 360, end angle=270] node[] (s4l) {};
        \draw (s4t.center) -- +(0.,0.3);
        \draw (s4r.center) -- +(0.2,0);
        \node[above = 0.15 of s5b] (s5t) {};
        \draw[thin] (s5b) arc [radius=0.2, start angle = 180, end angle=90] node[] (s5r) {};
        \draw[thin] (s5t) arc [radius=0.2, start angle = 360, end angle=270] node[] (s5l) {};
        \draw (s5t.center) -- +(0.,0.3);
        \draw (s5r.center) -- +(0.4,0);
        \draw (s5l.center) -- +(-0.2,0);
        \draw[dotted] ($(s4r.center)+(0.4,0)$) -- ($(s5l.center)+(-0.4,0)$);
        \node[above = 0.15 of s6b] (s6t) {};
        \draw[thin] (s6b) arc [radius=0.2, start angle = 180, end angle=90] node[] (s6r) {};
        \draw[thin] (s6t) arc [radius=0.2, start angle = 360, end angle=270] node[] (s6l) {};
        \draw (s6t.center) -- +(0.,0.3);
        %twist
        \node[right = 0.2 of s6r] (s7l) {};
        \draw (s6r.center) -- (s7l.center);
        \node[right = 0.15 of s7l] (s7r) {};
        \draw[thin] (s7l) arc [radius=0.2, start angle = 90, end angle=0] node[] (s7b) {};
        \draw[thin] (s7r) arc [radius=0.2, start angle = 270, end angle=180] node[] (s7t) {};
        \draw (s7b.center) -- +(0,-1.4);
        \draw (s7t.center) -- +(0,0.3);
        \draw (s7r.center) -- +(0.2,0);
        \node[right = 1.6 of s7l] (s8l) {};
        \draw (s8l.center) -- +(-0.2,0);
        \draw[dotted] ($(s7r.center)+(0.4,0)$) -- ($(s8l.center)+(-0.4,0)$);
        \node[right = 0.15 of s8l] (s8r) {};
        \draw[thin] (s8l) arc [radius=0.2, start angle = 90, end angle=0] node[] (s8b) {};
        \draw[thin] (s8r) arc [radius=0.2, start angle = 270, end angle=180] node[] (s8t) {};
        \draw (s8b.center) -- +(0,-1.4);
        \draw (s8t.center) -- +(0,0.3);
        \draw (s1l.center) -- +(-0.6,0) node[] (l2) {};
        \draw (s8r.center) -- +(0.25,0) node[] (r2) {};
        \draw (r2) arc [radius=0.11, start angle = -90, end angle=90] node[] (r3) {};
        \draw (l2) arc [radius=0.11, start angle = 270, end angle=90] node[] (l3) {};
        \draw[densely dotted] (r3.center)  -- (l3.center);
    \end{tikzpicture},
    \label{eq:brickwall_2}
\end{align}
where the horizontal bond of the two-site unitary $U_{N,1}$ across the boundary is also twisted.
Hence, the Hilbert-space dimension of the auxiliary qudit is $D=2r$, where $r$ is the operator Schmidt rank of $U_{N,1}$.
Even though $r=4$ for two-site Haar random unitaries, we also consider the case where the operator Schmidt rank of only $U_{N,1}$ is restricted to two to consider larger twists.
In the following, we consider spatially non-uniform on-site and two-site unitaries.
We have numerically confirmed that the rank of $\hat{U}_{F,2}^t(a)$ is the same as those of the shift operator, that is, $\rank \hat{U}_{F,2}^t(a)=2^{N-a-2\text{min}(N,t,-a)}$.

On the other hand, for $\hat{U}_{F,4}$, 
we evaluated the rank up to $N=8$, $t=6$ and $|a|=5$ (Table \ref{tab:u4}).
\begin{table}[htbp]
    \caption{The rank of $\hat{U}_{F,4}(a)$ in (\ref{eq:brickwall_1}).
        Numbers in parentheses are from (\ref{eq:mpu_floquetrank}) and are shown when they are different from the numerical results.}
    \label{tab:u4}
    \setlength{\tabcolsep}{4pt} % 列間の余白を少し狭めて11列を収まりやすくする
    \begin{tabular}{cc ccccccccccc} % aの11列分を右寄せ(r)に指定
    \hline\hline
    $N$ & $t$ & $a=-5$ & $-4$ & $-3$ & $-2$ & $-1$ & $0$ & $1$ & $2$ & $3$ & $4$ & $5$ \\ \hline
    \multirow{5}{*}{4} 
    & 1 & 128 & 64 & 32 & 16 & 8 & 16 & 32 & 64 & 128 & 256 & 512 \\
    & 2 & 32 & 16 & 8 & 4 & 8 & 16 & 32 & 64 & 128 & 256 & 512 \\
    & 3 & 32\,(8) & 16\,(4) & 8\,(2) & 4 & 8 & 16 & 32 & 64 & 128 & 256 & 512 \\
    & 4 & 32\,(2) & 16\,(1) & 8\,(2) & 4 & 8 & 16 & 32 & 64 & 128 & 256 & 512 \\
    & $\ge 5$ & 32\,($\le 1$) & 16\,(1) & 8\,(2) & 4 & 8 & 16 & 32 & 64 & 128 & 256 & 512 \\\hline
    \multirow{5}{*}{6} 
    & 1 & 512 & 256 & 128 & 64 & 32 & 64 & 128 & 256 & 512 & 1024 & 2048 \\
    & 2 & 128 & 64 & 32 & 16 & 32 & 64 & 128 & 256 & 512 & 1024 & 2048 \\
    & 3 & 32 & 16 & 8 & 16 & 32 & 64 & 128 & 256 & 512 & 1024 & 2048 \\
    & 4 & 32\,(8) & 16\,(4) & 8 & 16 & 32 & 64 & 128 & 256 & 512 & 1024 & 2048 \\
    & $\ge 5$ & 32\,(2) & 16\,(4) & 8 & 16 & 32 & 64 & 128 & 256 & 512 & 1024 & 2048 \\ \hline
    \multirow{5}{*}{8} 
    & 1 & 2048 & 1024 & 512 & 256 & 128 & 256 & 512 & 1024 & 2048 & 4096 & 8192 \\
    & 2 & 512 & 256 & 128 & 64 & 128 & 256 & 512 & 1024 & 2048 & 4096 & 8192 \\
    & 3 & 128 & 64 & 32 & 64 & 128 & 256 & 512 & 1024 & 2048 & 4096 & 8192 \\
    & 4 & 32 & 16 & 32 & 64 & 128 & 256 & 512 & 1024 & 2048 & 4096 & 8192 \\
    & $\ge 5$ & 32\,(8) & 16 & 32 & 64 & 128 & 256 & 512 & 1024 & 2048 & 4096 & 8192 \\ \hline\hline
    \end{tabular}
\end{table}
When a numerical rank is different from (\ref{eq:mpu_floquetrank}), the number from (\ref{eq:mpu_floquetrank}) is shown with parenthesis.
The prediction (\ref{eq:mpu_floquetrank}) is true when $\text{min}(t,-a)\le N/2$.
Notice that the results are the same for $t\ge 5$ since we calculated up to $|a|=5$.

Numerically evaluated ranks of $\hat{U}_{F,(4,4)}$ are shown in Table \ref{tab:u44_4}.
\begin{table}[htbp]
    \caption{
        The rank of $\hat{U}_{F,(4,4)}(a)$ in (\ref{eq:brickwall_2}).
    }
    \label{tab:u44_4}
    \setlength{\tabcolsep}{4pt} % 列間の余白を少し狭めて11列を収まりやすくする
    \begin{tabular}{cc ccccccc} % aの11列分を右寄せ(r)に指定
    \hline\hline
    $N$ & $t$ & $a=-3$ & $-2$ & $-1$ & $0$ & $1$ & $2$ & $3$ \\ \hline
    \multirow{4}{*}{4} 
    & 1 & 512 & 64 & 8 & 16 & 32 & 256 & 2048 \\
    & 2 & 128\,(32) & 16\,(4) & 8 & 16 & 32 & 64 & 512 \\
    & 3 & 32\,(2) & 4 & 8 & 16 & 32 & 64 & 128 \\
    & $\ge 4$ & 8\,(2) & 4 & 8 & 16 & 32 & 64 & 128 \\ \hline
    \multirow{4}{*}{6} 
    & 1 & 2048 & 256 & 32 & 64 & 128 & 1024 & 8192\\
    & 2 & 128 & 16 & 32 & 64 & 128 & 256 & 2048 \\
    & 3 & 32\,(8) & 16 & 32 & 64 & 128 & 256 & 512 \\
    & $\ge 4$ & 8 & 16 & 32 & 64 & 128 & 256 & 512 \\\hline
    \multirow{2}{*}{8} 
    & 1 & - & 1024 & 128 & 256 & 512 & 4096 & - \\
    & $\ge 2$ & - & 64 & 128 & 256 & 512 & 1024 & - \\ \hline\hline
    \end{tabular}
\end{table}
A similar behavior can also be seen when the operator Schmidt rank of $U_{N,1}$ is two (Table \ref{tab:u44_2}).
\begin{table}[htbp]
    \caption{The rank of $\hat{U}_{F,(4,4)}(a)$ with a modification that $U_{N,1}$ has operator Schmidt rank two.}
    \label{tab:u44_2}
    \setlength{\tabcolsep}{4pt} % 列間の余白を少し狭めて11列を収まりやすくする
    \begin{tabular}{cc ccccccccccc} % aの11列分を右寄せ(r)に指定
    \hline\hline
    $N$ & $t$ & $a=-5$ & $-4$ & $-3$ & $-2$ & $-1$ & $0$ & $1$ & $2$ & $3$ & $4$ & $5$ \\ \hline
    \multirow{6}{*}{4} 
    & 1 & 2048 & 512 & 128 & 32 & 8 & 16 & 32 & 128 & 512 & 2048 & 8192 \\
    & 2 & 512\,(256) & 128\,(64) & 32\,(16) & 8\,(4) & 8 & 16 & 32 & 64 & 256 & 1024 & 4096 \\
    & 3 & 256\,(32) & 64\,(8) & 16\,(2) & 4 & 8 & 16 & 32 & 64 & 128 & 512 & 2048 \\
    & 4 & 128\,(4) & 32\,(1) & 8\,(2) & 4 & 8 & 16 & 32 & 64 & 128 & 256 & 1024 \\
    & 5 & 64\,($\le 1$) & 16\,(1) & 8\,(2) & 4 & 8 & 16 & 32 & 64 & 128 & 256 & 512 \\
    & $\ge 6$ & 32\,($\le 1$) & 16\,(1) & 8\,(2) & 4 & 8 & 16 & 32 & 64 & 128 & 256 & 512 \\ \hline
    \multirow{5}{*}{6} 
    & 1 & - & 2048 & 512 & 128 & 32 & 64 & 128 & 512 & 2048 & 8192 & - \\
    & 2 & - & 256 & 64 & 16 & 32 & 64 & 128 & 256 & 1024 & 4096 & - \\
    & 3 & - & 64\,(32) & 16\,(8) & 16 & 32 & 64 & 128 & 256 & 512 & 2048 & - \\
    & 4 & - & 32\,(4) & 8 & 16 & 32 & 64 & 128 & 256 & 512 & 1024 & - \\
    & $\ge 5$ & - & 16\,(4) & 8 & 16 & 32 & 64 & 128 & 256 & 512 & 1024 & - \\ \hline
    \multirow{3}{*}{8} 
    & 1 & - & - & 2048 & 512 & 128 & 256 & 512 & 2048 & 8192 & - & - \\
    & 2 & - & - & 256 & 64 & 128 & 256 & 512 & 1024 & 4096 & - & - \\
    & $\ge 3$ & - & - & 32 & 64 & 128 & 256 & 512 & 1024 & 2048 & - & - \\ \hline\hline
    \end{tabular}
\end{table}
The prediction (\ref{eq:mpu_floquetrank}) is true at least when $\text{min}(t,-a)\le N/2-1$.

For any case where the numerical result is different from (\ref{eq:mpu_floquetrank}), however, focusing on particular $t$ and $a$ the deviation is resolved as the system size becomes larger.
As a control, we also evaluated the rank for the brickwork circuit without the shift operator, $\hat{U}_{F,(4,4)}'=\hat{U}_{4e}\hat{U}_{4o}$, whose GNVW index is $1$ (Table \ref{tab:u44_noshift}).
The twist is implemented in the same way as in (\ref{eq:brickwall_2}) without the shift, so that $D=r$.
In this case, (\ref{eq:mpu_floquetrank}) predicts $\rank\hat{U}_F'^t(a)=2^ND^{\max(|a|-t,0)}$, which is independent of the sign of $a$ and, for $|a|\le t$, of $a$ itself.
The numerical ranks agree with this prediction for $N=6$ and $8$ and deviate only for $N=4$, $t=3$ and $|a|\ge 3$, consistent with the condition $\text{min}(t,|a|)\le N/2$ found above.
We obtained the same agreement with two-site Haar random $U_{N,1}$ ($r=4$) for $N=4$ and $6$ with $|a|\le 3$.
\begin{table}[htbp]
    \caption{The rank of $\hat{U}_{F,(4,4)}'^t(a)$, the brickwork circuit without the shift operator ($I_\text{GNVW}=1$), where $U_{N,1}$ has operator Schmidt rank two ($D=2$).
        Numbers in parentheses are from (\ref{eq:mpu_floquetrank}) and are shown when they are different from the numerical results.}
    \label{tab:u44_noshift}
    \setlength{\tabcolsep}{4pt}
    \begin{tabular}{cc ccccccccc}
    \hline\hline
    $N$ & $t$ & $a=-4$ & $-3$ & $-2$ & $-1$ & $0$ & $1$ & $2$ & $3$ & $4$ \\ \hline
    \multirow{3}{*}{4}
    & 1 & 128 & 64 & 32 & 16 & 16 & 16 & 32 & 64 & 128 \\
    & 2 & 64 & 32 & 16 & 16 & 16 & 16 & 16 & 32 & 64 \\
    & 3 & 64\,(32) & 32\,(16) & 16 & 16 & 16 & 16 & 16 & 32\,(16) & 64\,(32) \\ \hline
    \multirow{3}{*}{6}
    & 1 & 512 & 256 & 128 & 64 & 64 & 64 & 128 & 256 & 512 \\
    & 2 & 256 & 128 & 64 & 64 & 64 & 64 & 64 & 128 & 256 \\
    & 3 & 128 & 64 & 64 & 64 & 64 & 64 & 64 & 64 & 128 \\ \hline
    \multirow{3}{*}{8}
    & 1 & 2048 & 1024 & 512 & 256 & 256 & 256 & 512 & 1024 & 2048 \\
    & 2 & 1024 & 512 & 256 & 256 & 256 & 256 & 256 & 512 & 1024 \\
    & 3 & 512 & 256 & 256 & 256 & 256 & 256 & 256 & 256 & 512 \\ \hline\hline
    \end{tabular}
\end{table}

Therefore, within the circuit and the parameter range we have verified numerically, we have confirmed that Eq.~(\ref{eq:mpu_floquetrank}) is applicable to spatially non-uniform RUCs with both trivial and nontrivial GNVW index.

\section{Tilted time slice and the spectrum of the twisted Floquet operator}
\label{sec:tilted}

Here we give the argument behind (\ref{eq:tilted}) and work out the exactly solvable cases.
We consider a Floquet circuit $\hat{U}_F$ on a ring of $N$ sites with local dimensions $d_x$, $\mathcal{D}=\prod_x d_x$, composed of local unitary gates and wires; a shift is a wire that moves sideways by one site per period.
The trace $\text{Tr}\,\hat{U}_F^t(a)$ is the contraction of the spacetime network of the circuit over $t$ driving periods on a torus: in the spatial direction the network is identified by (\ref{eq:tttwist}), so that a leg that crosses the seam $x=L$ in one period re-enters the ring $a$ periods later, and in the temporal direction the top of the last period is connected to the bottom of the first.
Cutting the network along the flat slices between consecutive periods gives $\text{Tr}\,\hat{U}^t_F(a)$ with $\hat{U}_F(a)$ the twisted operator of (\ref{eq:mpu_withtwist_sm}): for $a\neq 0$ the flat slice is not closed, and $\hat{U}_F(a)$ acts on the $N$ physical legs together with the $|a|$ auxiliary legs that are in transit across the seam.

\paragraph{Tilted slice.}
Let $\Sigma$ be a curve in the network that starts at a point $\gamma_0$ on the seam, goes once around the ring crossing only wires (no gates), and ends at $\gamma_0+(L,-a)$, i.e., $a$ periods lower than it started; by the identification, its endpoints coincide and $\Sigma$ is closed (Fig.~\ref{fig:tilted}).
Such a slice exists whenever $|a|$ is small compared with $N$ (for the brickwork circuits below, $|a|\lesssim N/2$), which is the same condition under which (\ref{eq:mpu_floquetrank}) was found to hold in Sec.~\ref{sec:ruc}.
Let $\mathcal{H}_\Sigma$ be the tensor product of the wires crossing $\Sigma$.
The part of the network between $\Sigma$ and its translate $\Sigma+1$ by one period is a circuit of unitary gates from the wires of $\Sigma$ to those of $\Sigma+1$, hence a unitary $\hat{V}_a$ on $\mathcal{H}_\Sigma$, and the torus is the stack of $t$ such slabs.
Cutting the same network along $\Sigma$ instead of along the flat slices therefore gives
\begin{align}
    \text{Tr}\,\hat{U}_F^t(a)=\text{Tr}\,\hat{V}_a^t\qquad(t\ge 1).
\end{align}
Since the power sums $\text{Tr}\,A^t$, $t\ge1$, determine the multiset of nonzero eigenvalues of a finite matrix $A$, the nonzero eigenvalues of $\hat{U}_F(a)$, counted with algebraic multiplicity, are the eigenvalues of $\hat{V}_a$: there are $\dim\mathcal{H}_\Sigma$ of them and all have modulus one.
We stress that $\hat{U}_F(a)$ and $\hat{V}_a$ are different operators, the transfer matrices of the same network along two different families of slices: $\hat{U}_F(a)$ acts on the $N$ physical legs and the $|a|$ auxiliary legs that the flat slice leaves dangling at the seam, has dimension $d^ND^{|a|}$, and is not unitary, whereas $\hat{V}_a$ acts on the $\dim\mathcal{H}_\Sigma$ wires of the closed slice $\Sigma$ and is unitary.
They are not similar; only their nonzero spectra agree.
Accordingly, the trace is independent of the slicing, while the rank is not: $\rank \hat{V}_a^t=\dim\mathcal{H}_\Sigma$ for every $t$, whereas $\rank\hat{U}_F^t(a)$ exceeds $\dim\mathcal{H}_\Sigma$ by the nilpotent part for $t<|a|$ [the factors $D^{|a|-t}$ and part of $I_\text{GNVW}^{\pm t}$ in (\ref{eq:mpu_floquetrank})] and equals it for $t\ge|a|$.
In the AFAI, the position-dependent delay (\ref{eq:afaibulk_twist}) realizes the tilted slicing physically: a flat slice in laboratory time is a tilted slice in the local time of the drive.
The remaining part of $\hat{U}_F(a)$ is nilpotent; for the shift operator it consists of Jordan blocks of size up to $|a|$ (Sec.~\ref{sec:shiftrank}).
The twist on flat slices is thus the periodic boundary condition on tilted slices, and $\hat{V}_a$ is the Floquet operator of the same circuit in a boosted frame; this is the lattice version of the correspondence between the energy twist and the boost of \cite{PhysRevB.106.155128}.

\paragraph{Counting wires.}
Let $R$ be the region, for $a>0$, bounded below by $\Sigma$, above by the flat slice through the upper endpoint of $\Sigma$, and on the right by the segment of the seam of length $a$ between them [the shaded region in Fig.~\ref{fig:tilted} (a)].
The wires entering $R$ are those crossing $\Sigma$ upward (dimension $\dim\mathcal{H}_\Sigma$) and those entering through the seam from the right (dimension $\dim M_a$); the wires leaving $R$ are those crossing the flat slice (dimension $\mathcal{D}$) and those leaving through the seam to the right (dimension $\dim P_a$).
Since every gate inside $R$ is unitary, the product of the input dimensions equals the product of the output dimensions,
\begin{align}
    \dim\mathcal{H}_\Sigma\cdot\dim M_a=\mathcal{D}\cdot\dim P_a .
\end{align}
The rightward and leftward traffic through a vertical cut per period, $P$ and $M$, satisfy $\dim P/\dim M=I_\text{GNVW}$ for any cut, which is the definition of the GNVW index as the ratio of the information flowing in the two directions \cite{Gross2012,PhysRevX.6.041070}; over $a$ periods, $\dim P_a/\dim M_a=I_\text{GNVW}^a$.
Hence
\begin{align}
    \dim\mathcal{H}_\Sigma=\mathcal{D}\,I_\text{GNVW}^{\,a}=d^{\,N+\text{ind}\cdot a}
    \label{eq:dimHsigma}
\end{align}
for a uniform qudit dimension $d$.
For $a<0$ the roles of $P$ and $M$ are exchanged, and the same formula results.
In Fig.~\ref{fig:tilted}, $\hat{U}_F=\hat{S}(U_{12}\otimes U_{34})$ on $N=4$ qubits, the slice $\Sigma$ crosses $5$ wires for $a=+1$ and $3$ wires for $a=-1$, and the slabs are
\begin{align}
    \hat{V}_{+1}=\hat{S}_5\,(U_{12}\otimes I\otimes U_{34}),\qquad
    \hat{V}_{-1}=\hat{S}_3\,(U_{34})_{23}(U_{12})_{12},
\end{align}
the same circuit on $N\pm1$ sites, with one idle site for $a=+1$ (Fig.~\ref{fig:tilted}); we have checked these identities by brute-force contraction.
In relation to the twisted operator $\hat{U}_{F,4}(-1)$ in (\ref{eq:brickwall_1}), $\hat{V}_{-1}$ can be obtained by moving the rightmost gates above the shift operator and removing a cap and a cup.

\paragraph{Shift with velocity $v$.}
For the shift by $v$ sites per period on $N$ qudits of dimension $d$, the network consists of wires only and the torus contraction is $d^{\#\text{loops}}$, each closed loop contributing $\text{Tr}\,I_d=d$.
Following one wire for $v=1$, after one lap around the ring it has advanced by $N$ periods and arrives at the seam, which the twist identifies with the same site $a$ periods later: each lap costs $N+a$ periods, as if the ring had $N+a$ sites.
With time periodic in $t$, the wires close into $\gcd(N+a,t)$ loops, so that
\begin{align}
    \text{Tr}\,\hat{S}^t(a)=d^{\gcd(N+a,\,t)},
\end{align}
which is the trace of the $t$-th power of the shift on $N+a$ sites, and $\dim\mathcal{H}_\Sigma=d^{N+a}$ in agreement with (\ref{eq:dimHsigma}) with $I_\text{GNVW}=d$.
For velocity $v$ the same counting gives $\text{Tr}\,\hat{S}_v^t(a)=d^{\gcd(N+av,\,vt)}$ for $N+av\ge1$, and for a product of shifts of different qudit species $\text{Tr}=\prod_i d_i^{\gcd(N+av_i,\,v_it)}$.
This exact result complements the rank formula (\ref{eq:shiftrankt}): the rank counts the nilpotent Jordan blocks as long as $t<|a|$, whereas the trace never sees them.

\paragraph{Rational index.}
The counting does not require the index to be an integer.
In the circuit of Gong, Nahum, and Piroli \cite{Gong2022} [Eq.~(\ref{eq:swap})], the sites alternate between dimensions $p$ and $q$ and every gate maps $\mathbb{C}^p\otimes\mathbb{C}^q\to\mathbb{C}^q\otimes\mathbb{C}^p$, so that one period of two brickwork layers moves the $p$-dimensional degrees of freedom by one unit cell to the right and the $q$-dimensional ones to the left, $I_\text{GNVW}=p/q$ \cite{PhysRevX.6.041070}.
With $N/2$ unit cells, (\ref{eq:dimHsigma}) gives $\dim\mathcal{H}_\Sigma=(pq)^{N/2}(p/q)^a=p^{N/2+a}q^{N/2-a}$.
When all gates are SWAP gates, the network consists of wires only and
\begin{align}
    \text{Tr}\,\hat{U}^t(a)=p^{\gcd(N/2+a,\,t)}\,q^{\gcd(N/2-a,\,t)} .
\end{align}
We have confirmed numerically for Haar-random gates that $\hat{U}(a)$ has exactly $p^{N/2+a}q^{N/2-a}$ nonzero eigenvalues of unit modulus in every realization [for $N=6$ and $(p,q)=(2,3)$: $96,144,216,324,486$ for $a=-2,-1,0,1,2$].

\section{Twisted spectral form factor}
\label{sec:sff}

We evaluate the twisted spectral form factor (SFF)
\begin{align}
    \langle K(t,a)\rangle
    =
    \langle |\text{Tr}\,\hat{U}_F^t(a)|^2\rangle
    =
    \langle |\text{Tr}\,\hat{V}_a^t|^2\rangle,
\end{align}
where $\langle\cdots\rangle$ denotes the average over the Haar-random gates and the second equality is (\ref{eq:tilted}).
For a unitary $\hat{V}_a$ drawn from the circular unitary ensemble (CUE) on a $\mathcal{D}_a$-dimensional space, $\langle K\rangle=t$ for $t\le \mathcal{D}_a$ and $\langle K\rangle=\mathcal{D}_a$ for $t\ge\mathcal{D}_a$ \cite{PhysRevX.8.041019,PhysRevLett.121.264101}; if the twisted circuit is as chaotic as the untwisted one, the plateau and the Heisenberg time of the twisted SFF are both $\dim\mathcal{H}_\Sigma=d^{\,N+\text{ind}\cdot a}$.

We consider three circuits: the brickwork chiral circuit $\hat{U}_{F,(4,4)}$ of (\ref{eq:brickwall_2}) on qubits, whose GNVW index is $2$, with the wrap gate $U_{N,1}$ of operator Schmidt rank two decomposed and twisted as in Sec.~\ref{sec:ruc} ($D=4$); the control trivial circuit $\hat{U}'_{F,(4,4)}=\hat{U}_{4e}\hat{U}_{4o}$ without the shift, whose GNVW index is $1$ ($D=2$); and the brickwork circuit of \cite{Gong2022} that consists of unitaries $\mathbb{C}^p\otimes\mathbb{C}^q\to\mathbb{C}^q\otimes\mathbb{C}^p$ with $p=3$, $q=2$, whose GNVW index is $I_\text{GNVW}=3/2$, given by
\begin{align}
    \hat{U}(a<0)
    &=
    \,\,
    \begin{tikzpicture}[baseline={([yshift=-0.1ex]current bounding box.center)}]
        \tikzset{
            gate/.style={
                draw,
                rounded corners=2pt,
                thick,
                fill=cyan!50!blue!20,
                minimum width=1.1cm,
                minimum height=0.2cm,
                inner sep=0pt
            }
        }
        %gates 1
        \node[gate] (gh0) at (-0.4,0.5) {};
        \fill[white] ($(gh0.center)+(0,-0.2)$) rectangle ($(gh0.west)+(-0.01,0.2)$);
        \node[gate] (g1) at (0.4,1.1) {};
        \node[gate] (g2) at (1.2,0.5) {};
        \node[gate] (g3)at (2.0,1.1) {}; 
        \node[gate] (gh1) at (2.8,0.5) {};
        \fill[white] ($(gh1.center)+(0,-0.2)$) rectangle ($(gh1.east)+(0.01,0.2)$);
        %gates 2
        \node[gate] (gh2) at (3.6,0.5) {};
        \fill[white] ($(gh2.center)+(0,-0.2)$) rectangle ($(gh2.west)+(-0.01,0.2)$);
        \node[gate] (g4) at (4.4,1.1) {};
        \node[gate] (gh3) at (5.2,0.5) {};
        \fill[white] ($(gh3.center)+(0,-0.2)$) rectangle ($(gh3.east)+(0.01,0.2)$);
        %vertical lines 1
        \draw ($(g1.center)+(-0.4,0.1)$) -- +(0.,0.3);
        \draw[ultra thick] ($(g1.center)+(-0.4,-0.1)$) -- ($(gh0.center)+(0.4,0.1)$);
        \draw ($(gh0.center)+(0.4,-0.1)$) -- +(0.,-0.3);
        \draw[ultra thick] ($(g1.center)+(0.4,0.1)$) -- +(0.,0.3);
        \draw ($(g1.center)+(0.4,-0.1)$) -- ($(g2.center)+(-0.4,0.1)$);
        \draw[ultra thick] ($(g2.center)+(-0.4,-0.1)$) -- +(0.,-0.3);
        \draw ($(g3.center)+(-0.4,0.1)$) -- +(0.,0.3);
        \draw[ultra thick] ($(g3.center)+(-0.4,-0.1)$) -- ($(g2.center)+(0.4,0.1)$);
        \draw ($(g2.center)+(0.4,-0.1)$) -- +(0.,-0.3);
        \draw[ultra thick] ($(g3.center)+(0.4,0.1)$) -- +(0.,0.3);
        \draw ($(g3.center)+(0.4,-0.1)$) -- ($(gh1.center)+(-0.4,0.1)$);
        \draw[ultra thick] ($(gh1.center)+(-0.4,-0.1)$) -- +(0.,-0.3);
        %dots
        \draw[dotted] ($(gh1.center)+(0.2,0)$) -- ($(gh2.center)+(-0.2,0.0)$);
        %vertical lines 2
        \draw ($(g4.center)+(-0.4,0.1)$) -- +(0.,0.3);
        \draw[ultra thick] ($(g4.center)+(-0.4,-0.1)$) -- ($(gh2.center)+(0.4,0.1)$);
        \draw ($(gh2.center)+(0.4,-0.1)$) -- +(0.,-0.3);
        \draw[ultra thick] ($(g4.center)+(0.4,0.1)$) -- +(0.,0.3);
        \draw ($(g4.center)+(0.4,-0.1)$) -- ($(gh3.center)+(-0.4,0.1)$);
        \draw[ultra thick] ($(gh3.center)+(-0.4,-0.1)$) -- +(0.,-0.3);
        %twist
        \draw[thick] (gh3.center) -- +(0.2,0) node[] (t1l) {};
        \node[right = 0.15 of t1l] (t1r) {};
        \draw[thick] (t1l) arc [radius=0.2, start angle = 90, end angle=0] node[] (t1b) {};
        \draw[thick] (t1r) arc [radius=0.2, start angle = 270, end angle=180] node[] (t1t) {};
        \draw[thick] (t1b.center) -- +(0., -0.2);
        \draw[thick] (t1t.center) -- +(0., 0.8);
        \draw[thick] (t1r.center) -- +(0.2, 0.);
        \draw[dotted] ($(t1r.center)+(0.4,0)$)  -- +(0.6, 0);
        \node[right = 1.6 of t1l] (t2l) {};
        \draw[thick] (t2l.center) -- +(-0.2,0.);
        \node[right = 0.15 of t2l] (t2r) {};
        \draw[thick] (t2l) arc [radius=0.2, start angle = 90, end angle=0] node[] (t2b) {};
        \draw[thick] (t2r) arc [radius=0.2, start angle = 270, end angle=180] node[] (t2t) {};
        \draw[thick] (t2b.center) -- +(0., -0.2);
        \draw[thick] (t2t.center) -- +(0., 0.8);
        \draw[thick] (t2r.center) -- +(0.3, 0.) node[] (r0) {};
        \draw[thick] (gh0.center) -- +(-0.4,0.) node[] (l0) {}; 
        \draw[thick] (l0) arc [radius=0.11, start angle = 270, end angle=90] node[] (l1) {};
        \draw[thick] (r0) arc [radius=0.11, start angle = 270, end angle=450] node[] (r1) {};
        \draw[thick, densely dotted] (l1.center) -- (r1.center);
    \end{tikzpicture},
    \label{eq:swap}
\end{align}
where the thin and thick legs correspond to $\mathbb{C}^q$ and $\mathbb{C}^p$, respectively, and the wrapping gate $U_{N,1}$ is split by a singular value decomposition into a tensor on site $N$ and one on site $1$, whose bond (of dimension at most $pq$) is the leg that crosses the seam.
The gates are Haar random, independent in space and identical in every period.
%One period of the chiral qubit circuit with the two legs that cross the seam is shown in Fig.~\ref{fig:brickwork_period}; for the trivial circuit the shift layer is absent and only the bond of the split gate crosses the seam.
%\begin{figure}[t]
%    \centering
%    \includegraphics[width=0.6\textwidth]{FIG_brickwork_period.pdf}
%    \caption{
%        One period of the chiral brickwork circuit for $N=6$ (inputs at the bottom, outputs at the top).
%        The wrapping gate $U_{N,1}$ of the odd layer is split into $Y$ (site $1$) and $X$ (site $N$); its bond $k$ (red) and the shift wire $r$ (blue) are the two legs that leave through the right boundary and re-enter through the left one, displaced by $a$ periods under the twist.
%        \label{fig:brickwork_period}
%    }
%\end{figure}
For each realization we build $\hat{U}_F(a)$ as a matrix on the $N$ physical qudits and the $|a|$ auxiliary legs, diagonalize it, count the nonzero eigenvalues, and compute $K(t,a)=|\sum_k\lambda_k^t|^2$ from the eigenvalues $\lambda_k$.

\begin{figure}[t]
    \centering
    \includegraphics[width=\textwidth]{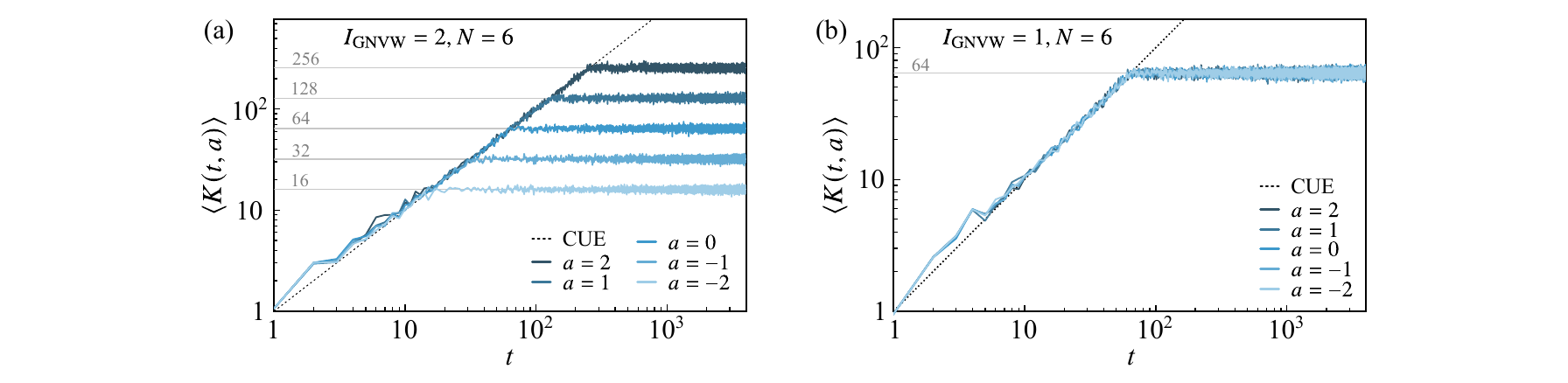}
    \caption{
        Twisted SFF $\langle K(t,a)\rangle$ of the brickwork circuits on $N=6$ qubits for $a=0,\pm1,\pm2$ (500 realizations), (a) with the shift layer ($I_\text{GNVW}=2$) and (b) without it ($I_\text{GNVW}=1$).
        Gray lines: $2^{N+a}$ in (a) and $2^N$ in (b). Dotted: the CUE ramp $\langle K\rangle=t$.
        \label{fig:sff_N6}
    }
\end{figure}
\begin{figure}[t]
    \centering
    \includegraphics[width=\textwidth]{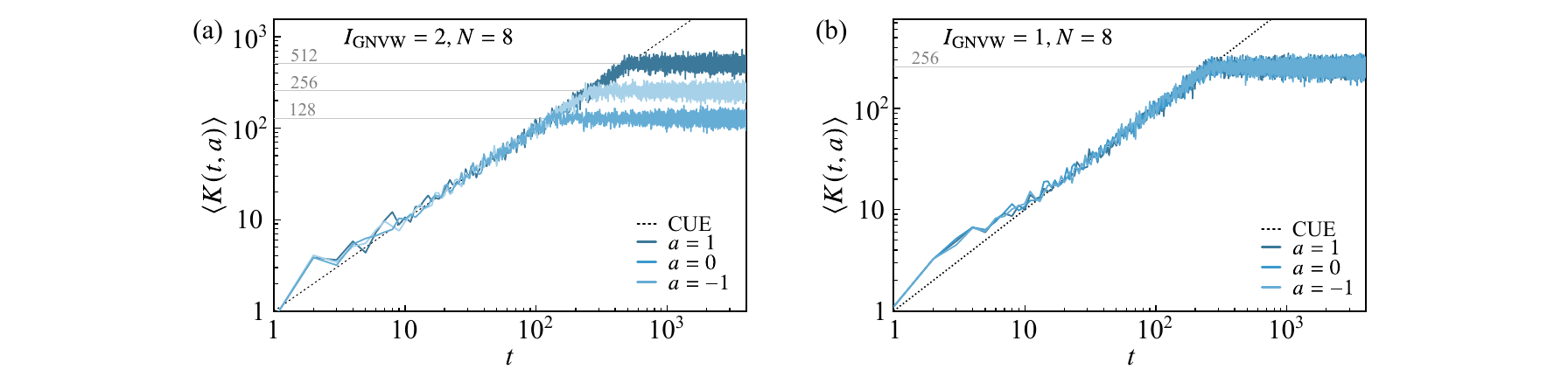}
    \caption{
        Twisted SFF of the brickwork circuits on $N=8$ qubits for $a=0,\pm1$ (100 realizations), (a) with the shift layer ($I_\text{GNVW}=2$) and (b) without it ($I_\text{GNVW}=1$).
        Gray lines: $2^{N+a}$ in (a) and $2^N$ in (b). Dotted: the CUE ramp $\langle K\rangle=t$.
        \label{fig:sff_N8}
    }
\end{figure}
For the qubit circuits we use $N=6$ with $|a|\le 2$ (500 realizations) and $N=8$ with $|a|\le1$ (100 realizations) (Figs.~\ref{fig:sff_N6} and \ref{fig:sff_N8}).
In every realization, the number of nonzero eigenvalues of $\hat{U}_{F,(4,4)}(a)$ is $2^{N+a}$ ($N=6$: $16,32,64,128,256$ for $a=-2,\ldots,2$; $N=8$: $128,256,512$ for $a=-1,0,1$), that of $\hat{U}'_{F,(4,4)}(a)$ is $2^N$ for every $a$, and all nonzero eigenvalues have modulus one within $10^{-14}$, in agreement with (\ref{eq:dimHsigma}).
The averaged SFF follows the CUE ramp $\langle K\rangle=t$ for every $a$ up to the Heisenberg time and saturates at the plateau $2^{N+a}$ for the chiral circuit and $2^N$ for the trivial one, so that the Heisenberg time moves together with the plateau.
One unit of twist thus multiplies both the plateau and the Heisenberg time by $I_\text{GNVW}^{\pm1}$ and leaves them unchanged when the index is trivial.

\begin{figure}[t]
    \centering
    \includegraphics[width=\textwidth]{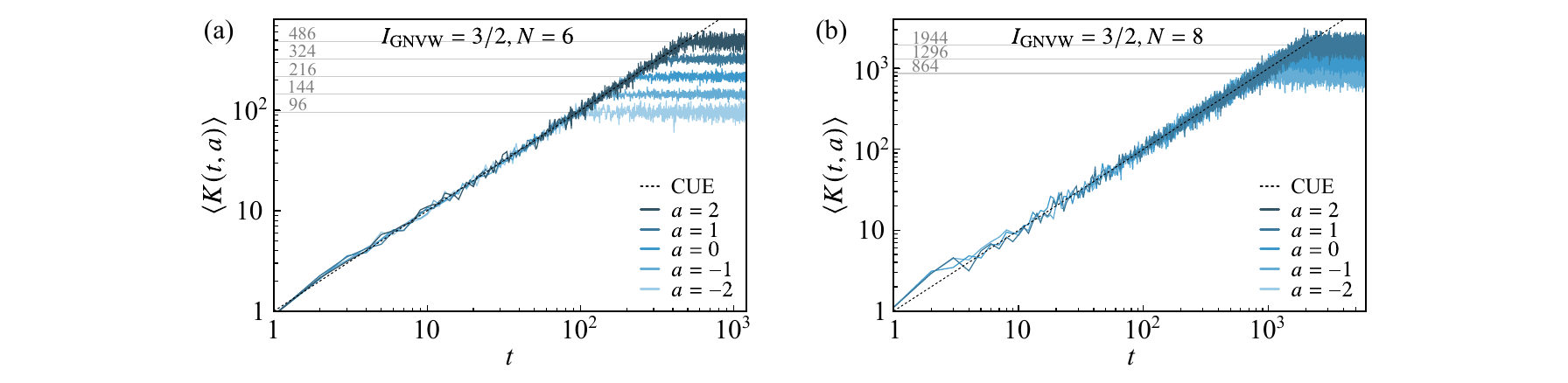}
    \caption{
        Twisted SFF of the circuit (\ref{eq:swap}) with $(p,q)=(3,2)$, $I_\text{GNVW}=3/2$, for (a) $N=6$ and $|a|\le 2$, and (b) $N=8$ and $|a|\le 1$.
        Gray lines: $3^{N/2+a}2^{N/2-a}$.
        \label{fig:sff_swap}
    }
\end{figure}

The SFF of the circuit (\ref{eq:swap}) with $(p,q)=(3,2)$ is shown in Fig.~\ref{fig:sff_swap} for $N=6$ ($|a|\le2$, 500 realizations) and $N=8$ ($|a|\le1$, 100 realizations).
The SFF converges to $3^{N/2+a}2^{N/2-a}$ after the Heisenberg time, i.e., to $96,144,216,324,486$ for $N=6$ and $864,1296,1944$ for $N=8$, which are the dimensions $p^{N/2+a}q^{N/2-a}$ of the tilted slice.
One unit of twist moves the plateau and the Heisenberg time by the rational factor $(3/2)^{\pm1}$, and exchanging $p\leftrightarrow q$ exchanges $a\leftrightarrow-a$.
The late-time twisted SFF therefore measures $\text{ind}\cdot a=\log_d(\text{plateau}/d^N)$ directly from the spectrum, without identifying boundary modes and including rational indices.

\section{Twisted ranks of non-invertible symmetry operators}
\label{sec:noninv}

The twisted operator (\ref{eq:mpu_withtwist_sm}) and its rank are defined for any matrix product operator (MPO), unitary or not.
Here we apply them to the MPOs of non-invertible symmetries on one-dimensional lattices and derive (\ref{eq:noninv}).
Throughout, $N$ is the number of sites of the open chain, $d$ the local dimension, and the MPO tensor $\hat{\mathcal L}_{jk}$ has bond dimension $\chi$.

\subsection{Setup and claims}
A non-invertible symmetry operator $\hat{\mathcal L}$ commutes with the Hamiltonian but has a kernel; its products close on a fusion algebra $\hat{\mathcal L}_a \hat{\mathcal L}_b=\sum_cN_{ab}^{\,c}\hat{\mathcal L}_c$ with non-negative integer coefficients, and the quantum dimension $d_a$ of $\hat{\hat{\mathcal L}}_a$ is the Perron--Frobenius eigenvalue of the fusion matrix $(N_a)_b^{\,c}=N_{ab}^{\,c}$ \cite{Chang2019,Shao2023,SchaferNameki2024,McGreevy2023}.
In the operator-algebra language, $\hat{\mathcal L}_a$ acts on the algebra $\mathcal{A}$ of local operators as an endomorphism $\rho_a$, and $d_a^2$ is the Jones index $[\mathcal{A}:\rho_a(\mathcal{A})]$ of the inclusion $\rho_a(\mathcal{A})\subset\mathcal{A}$ \cite{Jones1983,Longo1989}; $d_a=1$ exactly for invertible operators.
Our examples are the $\mathbb{Z}_n$ Kramers--Wannier (KW) dualities, which on the lattice satisfy $\hat{\mathcal L}^2=(\sum_{g\in\mathbb{Z}_n}g)\,\hat{T}^{-1}$ with $\hat{T}$ the one-site translation \cite{SeibergSeifnashriShao2024} and have $d_{\hat{\mathcal L}}=\sqrt{n}$; the $\text{Rep}(D_8)$ operators $\hat{\mathcal L}_1=\hat{T}\,\text{KW}_e\text{KW}_o$ and $\hat{\mathcal L}_2=\text{KW}_e\text{KW}_o$ of Ma, Li, and Cheng \cite{MaLiCheng2024,SeifnashriShao2024} ($D_1$ and $D_2$ in their notation), built from KW dualities on the even and odd sublattices, with $d_{\hat{\mathcal L}}=2$; the Wilson line $\hat{\mathcal L}_E=\chi_E(g_1\cdots g_N)$ of the two-dimensional representation of $S_3$ on the group-algebra chain, with $\hat{\mathcal L}_E^2=\hat{\mathcal L}_1+\hat{\mathcal L}_\sigma+\hat{\mathcal L}_E$ and $d_E=2$; and the $\tau$ line of the Fibonacci (golden) chain \cite{Feiguin2007,Kitaev2006}, with $\tau^2=1+\tau$ and $d_\tau=\phi=(1+\sqrt5)/2$.
All of them are MPOs of finite bond dimension \cite{Bultinck2017,Sahinoglu2021,Lootens2023,AasenMongFendley2016,Inamura2022}.

A second number attached to a symmetry operator is its flow index.
For an invertible locality-preserving operator it is the GNVW index.
For a lattice duality, which is not invertible on $\mathcal{A}$, Ma, Li, and Cheng define it by restricting $\hat{\mathcal L}$ to the symmetric subalgebra, on which $\hat{\mathcal L}$ is an automorphism; the Ising duality shifts that subalgebra by half a site and has flow index $\sqrt2$ \cite{MaLiCheng2024}.
On a fusion spin chain built from an object $X$, Jones and Lim showed that a translation has flow index $\dim X$, which is $\phi$ for the golden chain \cite{JonesLim2023}.
The flow index and the quantum dimension are independent: the Ising duality has both equal to $\sqrt2$, the operator $\hat{\mathcal L}_1$ has flow index $1$ and $d_{\hat{\mathcal L}}=2$, and the translation on the golden chain has flow index $\phi$ and $d_T=1$.

Let $\hat{\mathcal L}_N$ be the contraction of $N$ MPO tensors with the two bond legs left open, and let the right and left ranks $r_N$ and $l_N$ be defined as in (\ref{eq:rankratioindex_sm}), so that $\rank\hat{\mathcal{L}}(\pm1)=r_N\,(l_N)$.
We demand that the MPO be in canonical form, i.e., that the algebra generated by the tensors $\{\hat{\mathcal L}_{jk}\}$ has a trivial commutant (direct sums of inequivalent irreducible blocks, such as $1\oplus\eta$, are allowed); otherwise the product $r_Nl_N$ depends on the representation (the identity written with bond dimension $\chi$ gives $r_Nl_N=\chi^2d^{2N}$).
A bond gauge transformation multiplies $\hat{\mathcal{L}}(\pm1)$ by invertible matrices from the left and the right, so the twisted ranks are gauge invariant, whereas the singular values of $\hat{\mathcal{L}}(\pm1)$ are not.
Let $\hat{\mathbbm{1}}$ be the unit of the MPO algebra: the identity operator on a tensor-product chain, $\rank\hat{\mathbbm{1}}(\pm1)=d^N$, and the constraint projector $\hat{\Pi}$ on a fusion chain embedded in a tensor-product space.
Writing $r_N=\chi_+\rank\hat{\mathbbm{1}}(1)$ and $l_N=\chi_-\rank\hat{\mathbbm{1}}(-1)$, the numbers $\chi_\pm$ are the effective bond dimensions transmitted through the open operator from the right and from the left.
For an MPU, $\chi_+=\chi I_\text{GNVW}$ and $\chi_-=\chi/I_\text{GNVW}$ up to the gauge-invariant content $\chi_+/\chi_-=I_\text{GNVW}^2$, and $\chi_+\chi_-=1$ in the normalization where $r_Nl_N=d^{2N}$ \cite{Cirac_2017}.
Our claims are
\begin{align}
    \frac{\chi_+}{\chi_-}=\text{Ind}(\hat{\mathcal L})^2,
    \qquad
    \chi_+\chi_-=d_{\hat{\mathcal L}}^{\,2},
    \label{eq:chi_claims}
\end{align}
which are (\ref{eq:noninv}).
The first is the statement of the main text extended to non-invertible operators, and we take it as the definition of the flow index with a direction; the second is new and measures the non-invertibility.

\subsection{Lattice dualities}
For the $\mathbb{Z}_n$ KW MPO $W_{(s',s),(b_l,b_r)}=\delta_{b_l,s}\,\omega^{s'(s-b_r)}/\sqrt{n}$ with $\omega=e^{2\pi i/n}$, we find, exactly at finite $N$,
\begin{align}
    \hat{\mathcal{L}}(-1)^\dagger\hat{\mathcal{L}}(-1)=I_{d^{N+1}},
    \qquad
    \hat{\mathcal{L}}(+1)^\dagger\hat{\mathcal{L}}(+1)=n\,\hat{P},
    \quad
    \rank \hat{P}=d^N,
    \label{eq:KW_open}
\end{align}
with $d=n$.
The first equation says that the duality on an open chain with one extra site attached through the bond is unitary; this is the lattice duality with a boundary qudit, the operator that Seiberg, Seifnashri, and Shao describe as adding a qubit \cite{SeibergSeifnashriShao2024}.
The second says that in the other orientation the operator is $\sqrt{n}$ times a partial isometry onto a subspace of dimension $d^N$; the bond leg transmits only the $\mathbb{Z}_n$-symmetric combination.
Hence $\chi_-=n$ and $\chi_+=1$, so that $\chi_+\chi_-=n=d_{\hat{\mathcal L}}^2$ and $\chi_+/\chi_-=1/n=\text{Ind}^2$ with $\text{Ind}=n^{-1/2}$, a half-site shift to the left in our convention, the $\sqrt{n}$ of \cite{MaLiCheng2024} with a direction.
In summary,
\begin{align}
    \log\rank\hat{\mathcal{L}}(\pm1)=N\log d+\log d_{\hat{\mathcal L}}\pm\log\text{Ind},
    \qquad
    \log\rank\hat{\mathcal{L}}(0)=N\log d-2\log d_{\hat{\mathcal L}} ,
\end{align}
where $\hat{\mathcal{L}}(0)=\hat{\mathcal L}$ is the operator on the closed ring, whose rank $d^N/n$ counts only the symmetric sector; this is the statement $\hat{\mathcal L}=\hat{\mathcal L}\hat{P}_{\mathbb{Z}_n}$.
We verified (\ref{eq:KW_open}) numerically for $N=3$--$8$ and $n=2,3,4$, and for the MPO of $\hat{\mathcal L}^2$ ($t=2$), for which the ratio is $n^{-2}$ and the product $n^2$: both parts are multiplicative under composition.

For the $\text{Rep}(D_8)$ operators we use $N=8$ qubits with a two-site unit cell ($d=4$); the product tensors are irreducible.
\begin{table}[tbp]
   \begin{tabular}{lcccc}
    \hline\hline
    operator & $\log_2(2^N/\rank \hat{\mathcal L})$ & $r_N/l_N$ & $r_Nl_N/d^{2N}$ & flow index \\ \hline
    $\text{KW}_e$ & 1 & $1/2$ & 2 & $\sqrt2$ \\
    $\hat{\mathcal L}_1=\hat{T}\,\text{KW}_e\text{KW}_o$ & 2 & $1$ & 4 & 1 \\
    $\hat{\mathcal L}_2=\text{KW}_e\text{KW}_o$ & 2 & $1/4$ & 4 & 2 \\
    \hline\hline
    \end{tabular}
        \caption{Twisted ranks of the $\text{Rep}(D_8)$ operators of \cite{MaLiCheng2024} on $N=8$ qubits. The last column is the flow index of \cite{MaLiCheng2024}.}
    \label{tab:repD8}
\end{table}
Both factors of $\hat{\mathcal L}_2$ are dualities of one sublattice with $\chi_+/\chi_-=1/2$ and $\chi_+\chi_-=2$, so that $\hat{\mathcal L}_2$ has $\chi_+/\chi_-=1/4$, a shift by one full site, and $\chi_+\chi_-=4=|\mathbb{Z}_2\times\mathbb{Z}_2|$.
The operator $\hat{\mathcal L}_1$ adds a translation in the opposite direction, which multiplies $\chi_+/\chi_-$ by $d^2=4$ and leaves $\chi_+\chi_-$ unchanged (Table~\ref{tab:repD8}): the translation changes the flow index and not the quantum dimension.
This is the example in which the two parts of the twisted rank are cleanly separated; the bond dimensions of $\hat{\mathcal L}_1$ and $\hat{\mathcal L}_2$ differ ($8$ and $4$) while $d_{\hat{\mathcal L}}=2$ comes out the same, so the symmetric part does not depend on the bond dimension in canonical form.

\subsection{Wilson lines}
For the Wilson line $\hat{\mathcal L}_R=\chi_R(g_1\cdots g_N)$ on the group-algebra chain ($d=|G|$, basis $g_j\in G$), the MPO is diagonal in the group basis and its bond carries the representation matrix $R(g)$.
The open operator $\hat{\mathcal{L}}_R(\pm1)$ is block diagonal in the configuration $(g_1,\ldots,g_N)$, and each block is the invertible $\dim R\times\dim R$ matrix $R(g_1\cdots g_N)$.
Hence $\chi_+=\chi_-=\dim R=d_R$ exactly, for every $N$, and the twisted spectrum is flat.
On the closed ring, $\text{Tr}_\text{bond}R(g_1\cdots g_N)=\chi_R(g_1\cdots g_N)$ vanishes on the configurations whose holonomy lies in a conjugacy class with $\chi_R=0$; for $G=S_3$ and $R=E$ this produces the rank deficit $\log(d^N/\rank \hat{\mathcal L}_E)=\log2\neq\log d_E^{\,2}$ and the non-flat spectrum $\chi_E(h)^2\in\{4,0,1\}$ of $\hat{\mathcal L}_E^\dagger \hat{\mathcal L}_E$.
We confirmed numerically for $N=2,3,4$ that $r_N/l_N=1$ and $\tfrac12\log(r_Nl_N/d^{2N})=\log 2=\log d_E$ without any finite-size correction, and that $\hat{\mathcal L}_E^2=\hat{\mathcal L}_1+\hat{\mathcal L}_\sigma+\hat{\mathcal L}_E$ holds to machine precision.

\subsection{Topological lines on the golden chain}
We embed the golden chain into a qubit chain with the Fibonacci constraint (no two adjacent labels $1$).
The unit of the MPO algebra is the constraint projector $\hat{\Pi}$, whose twisted rank is the number of allowed label strings of length $N+1$ with one bond label attached, $\rank\hat{\Pi}(\pm1)=\sum_{b,c}(N_\tau^{N-1})_{bc}=F_{N+2}$, with $F_n$ the Fibonacci numbers.
The MPO of the $\tau$ line,
\begin{align}
    (\hat{\mathcal L}_\tau)_{x'x}=\prod_i\Big[F^{\,x_{i-1}\tau x'_i}_{\tau}\Big]_{x_ix'_{i-1}},
    \qquad
    F^{\tau\tau\tau}_\tau=\begin{pmatrix}\phi^{-1}&\phi^{-1/2}\\ \phi^{-1/2}&-\phi^{-1}\end{pmatrix},
\end{align}
is the product of the $F$-moves that carry the line across the $N$ vertices of the fusion tree; it satisfies $\hat{\mathcal L}_\tau^2=\hat{\Pi}+\hat{\mathcal L}_\tau$, $\hat{\mathcal L}_\tau=\hat{\mathcal L}_\tau^\dagger$, and $[\hat{\mathcal L}_\tau,\hat{T}]=0$ to machine precision, is irreducible, and has bond dimension $3$.
With open bonds, $\hat{\mathcal{L}}_\tau(\pm1)$ is the change of basis between the fusion tree in which the line is attached at the left end and the tree in which it is attached at the right end; both describe the space of fusion paths with fixed labels at the two ends and $N$ steps of $\tau$ in between, and the change of basis is injective on it.
Its rank is therefore the dimension of that space,
\begin{align}
    \rank\hat{\mathcal{L}}_\tau(\pm1)=\sum_{b,c}\big(N_\tau N_\tau^{\,N-1}\big)_{bc}=F_{N+3},
\end{align}
one more factor of the fusion matrix than the reference.
By the Perron--Frobenius theorem, $N_\tau^{N-1}$ projects onto the eigenvector $v_c\propto d_c$ as $N\to\infty$ and $N_\tau v=\phi v$, so that $\chi_\pm=F_{N+3}/F_{N+2}\to\phi=d_\tau$ with corrections of order $\phi^{-2N}$ (Table~\ref{tab:fib}).
\begin{table}[tbp]
   \begin{tabular}{cccc}
    \hline\hline
    $N$ & $\rank\hat{\mathcal{L}}_\tau(\pm1)$ & $\rank\hat{\Pi}(\pm1)$ & $\tfrac12\log\dfrac{\rank\hat{\mathcal{L}}_\tau(1)\rank\hat{\mathcal{L}}_\tau(-1)}{\rank\hat{\Pi}(1)\rank\hat{\Pi}(-1)}$ \\ \hline
    6 & $34=F_9$ & $21=F_8$ & 0.4818 \\
    8 & $89=F_{11}$ & $55=F_{10}$ & 0.4813 \\
    10 & $233=F_{13}$ & $144=F_{12}$ & 0.4812 \\
    \hline\hline
    \end{tabular}
        \caption{Twisted ranks of the $\tau$ line and of the constraint projector on the golden chain. The last column converges to $\log\phi=0.4812$.}
    \label{tab:fib}
\end{table}
The twisted ranks are integers, and the irrational quantum dimension emerges as the limit of ratios of Fibonacci numbers: $\rank\hat{\mathcal{L}}_\tau(\pm1)$ is the dimension of an open chain with $N+1$ sites, since the end of the twisted line carries a $\tau$.
The line is parity even, so $\chi_+=\chi_-$ and the flow index is $1$ exactly.
For the fused line $\tau^2=1\oplus\tau$ the canonical form is $\hat{\Pi}\oplus \hat{\mathcal L}_\tau$, the twisted rank is $F_{N+2}+F_{N+3}=F_{N+4}$, and $\chi_\pm\to\phi^2=d_{\tau^2}$: the symmetric part is additive under fusion, as the quantum dimension is.

The translation $\hat{T}$ on the golden chain is invertible, and (\ref{eq:chi_claims}) must return $\chi_+\chi_-=1$ and the Jones--Lim index $\text{Ind}(\hat{T})=\phi$.
Its MPO $W_{(x',x),(b_l,b_r)}=\delta_{x',b_l}\delta_{b_r,x}$, restricted to allowed pairs, gives $\rank\hat{T}(+1)=\sum_{b,c}(N_\tau^N)_{bc}=F_{N+3}$ and $\rank\hat{T}(-1)=\sum_{b,c}(N_\tau^{N-2})_{bc}=F_{N+1}$, so that
\begin{align}
    \frac{\chi_+}{\chi_-}=\frac{F_{N+3}}{F_{N+1}}\to\phi^2,
    \qquad
    \chi_+\chi_-=\frac{F_{N+3}F_{N+1}}{F_{N+2}^2}\to1 .
\end{align}
On a qudit chain the same computation gives $d^{N+1}$ and $d^{N-1}$, the ranks of the main text; the twist measures the Jones--Lim index with no modification, the qudit dimension $d$ being replaced by the Perron--Frobenius eigenvalue of the fusion matrix of the site object.

\subsection{Untwisted versus twisted ranks}

On the closed ring, i.e., for the untwisted operator $\hat{\mathcal{L}}(0)$, the bond is traced and $\hat{\mathcal L}=\hat{\mathcal L}\hat{P}$ contains its own projector $\hat{P}$ onto the sectors on which $\hat{\mathcal L}$ acts invertibly, so $\rank \hat{\mathcal L}$ counts only those sectors: the deficit $\log(d^N/\rank \hat{\mathcal L})$ is $\log n$ for the $\mathbb{Z}_n$ duality, $\log 2$ for the $S_3$ Wilson line, and $0$ for the Fibonacci line, and it is not $\log d_{\hat{\mathcal L}}^{\,2}$ in general.
The operator $\hat{\mathcal L}^\dagger \hat{\mathcal L}=\sum_xN_{\bar aa}^{\,x}\hat{\mathcal L}_x$ has eigenvalue $|S_{ai}/S_{0i}|^2$ on the sector $i$, and the quantum dimension sits in its largest eigenvalue $d_a^2$, which a rank does not weigh.
In the twisted sector the bond leg carries the sector label and no projection takes place.
The counting is then one more power of the fusion matrix, $\rank\hat{\mathcal{L}}_a(\pm1)=\sum_{b,c}(N_aN_X^{N-1})_{bc}$ for a line $a$ on a chain built from the object $X$, whose Perron--Frobenius eigenvalue is $d_a$.
Geometrically, the replica network for $\text{Tr}[(\hat{\mathcal{L}}(1)^\dagger\hat{\mathcal{L}}(1))^n]$ contains a single defect loop that winds once around the replica circle; for a topological line it can be pushed onto a boundary of the replica annulus, where it multiplies the boundary entropy by $d_a$ \cite{Cardy1989,AffleckLudwig1991,Verlinde1988,PetkovaZuber2001,FFRS2004}, for every $n$ and hence for the rank ($n\to0$).
This argument fixes only the product $\chi_+\chi_-$, since the two networks contain the same loop read in opposite directions; the ratio $\chi_+/\chi_-$ is the lattice shift of the duality, invisible to boundary entropies, and is the flow index of the main text.
A general lattice proof of (\ref{eq:chi_claims}) for an arbitrary MPO symmetry in canonical form is left for future work.

\section{Fermionic boundaries and the Majorana translation}
\label{sec:fermion}

We extend the twisted rank and the tilted slice to fermionic boundary circuits with $I_\text{GNVW}=\sqrt2$, the edge of
the radical chiral Floquet phases \cite{PhysRevB.96.245116,PhysRevB.99.085115}; with $d=2$ per mode,
$\text{ind}= \log_2{I_\text{GNVW}} = 1/2$.

\subsection{Setup}\label{sec:fsetup}
The boundary carries $N$ modes, i.e., $2N$ Majoranas $\gamma_k$, and one period is $\hat{U}_F=\hat{C}_s\hat{V}$. The
Majorana translation $\hat{C}_s$, $\gamma_k\to\gamma_{k+1}$ and $\gamma_{2N}\to s\gamma_1$ is a fermionic quantum cellular automaton (FQCA) with the fermionic GNVW index $I_\text{GNVW}=\sqrt2$
\cite{PhysRevB.99.085115, Piroli2021}; $s=\pm1$ is the boundary sign of the fermions. $\hat{V}$ is a layer of
Haar-random parity-preserving gates $U(2)\oplus U(2)$ on the modes $(1,2),(3,4),\dots$, the same in every period; acting on
four Majoranas, a generic gate contains the interaction $\gamma_{2j-1}\gamma_{2j}\gamma_{2j+1}\gamma_{2j+2}$.

As a FQCA, $\hat{C}_s$ is a graded fermionic matrix product unitary, the fermionic counterpart of the MPU of Sec.~\ref{sec:shift}, and has a bond that is one fermion mode $b=(\beta+i\beta')/2$, and its MPU is
the braid chain $\hat G=[(1-\gamma_{2N}\beta)/\sqrt2]\prod_{j=1}^{2N-1}[(1-\gamma_j\gamma_{j+1})/\sqrt2]$, which
hands $\gamma_{2N}$ to $\beta$ and $\beta$ to $\gamma_1$ \cite{Piroli2021}. On the untwisted ring,
$\hat{U}_F=\Tr_b[S_s \hat G]\hat V/\sqrt2$. The closure $S_s$ is a matrix on the closing bond: $S_s=\mathbf 1$ gives $s=-1$;
$S_s=\beta$ flips the bond state, multiplies the operator by one Majorana, $\Tr_b(\beta \hat G)=\Tr_b(\hat G)\gamma_{2N}$,
and gives the parity-odd translation with $s=+1$. Only $\beta$ flows through the bond, so the trace over its two states
returns $\sqrt2$ times a unitary, hence the weight $1/\sqrt2$.

\begin{figure}[htbp]
  \centering
  \resizebox{0.8\textwidth}{!}{%
\newcommand{\fsetupcommon}{%
  \foreach \i in {1,...,4} { \node[lab, below] at (\i,-0.3) {$\gamma_{\i}$}; \node[lab, above] at (\i,3.4) {$\gamma_{\i}$}; }
  \node[lab, below, red!70!black] at (6.0,-0.3) {$\beta$}; \node[lab, above, red!70!black] at (6.0,3.4) {$\beta$};
  \node[lab, below, gray] at (6.6,-0.3) {$\beta'$}; \node[lab, above, gray] at (6.6,3.4) {$\beta'$};
  \foreach \i in {1,...,4} { \draw[wire] (\i,-0.3) -- (\i,0.3); }
  \node[gV] at (2.5,0.5) {\scriptsize $\hat V$};
  \foreach \i in {1,...,3} { \draw[wire] (\i,0.7) -- (\i,1.1) -- (\i+1,1.9) -- (\i+1,3.4); }
  \draw[spec] (6.6,-0.3) -- (6.6,3.4);
  \draw[frame] (0.55,-0.35) rectangle (6.95,3.45);
}
\begin{tikzpicture}[
    gate/.style={draw, rounded corners=2pt, thick, fill=cyan!50!blue!20,
                 minimum width=1.75cm, minimum height=0.2cm, inner sep=0pt},
    rc/.style={rounded corners=2mm},
    aux/.style={rounded corners=2mm},   % beta is one Majorana wire, same width as the gamma wires (cf. Eq. 16)
    spec/.style={gray, densely dashed},
    Sbox/.style={draw, thick, fill=white, rounded corners=1.5pt, inner sep=1.2pt, font=\scriptsize},
    lab/.style={font=\scriptsize},
    Pbox/.style={draw, gray!70, densely dotted, rounded corners=2pt},
  ]
  % h: shift row, top: output level
  \def\h{1.2}\def\ytop{1.95}
  % common part: inputs/outputs, gates V, braid chain G on the physical wires, spectator beta'
  \def\fsixcommon{%
    \foreach \x/\k in {0.45/1, 0.9/2, 1.35/3, 1.8/4} {
      \node[lab, below] at (\x,0) {$\gamma_{\k}$};
      \node[lab, above] at (\x,\ytop) {$\gamma_{\k}$};
    }
    \node[lab, below] at (4.05,0) {$\gamma_{2N}$};  \node[lab, above] at (4.05,\ytop) {$\gamma_{2N}$};
    \node[lab, below] at (4.55,0) {$\beta$};        \node[lab, above] at (4.55,\ytop) {$\beta$};
    \node[lab, below, gray] at (5.15,0) {$\beta'$}; \node[lab, above, gray] at (5.15,\ytop) {$\beta'$};
    % braid chain: gamma_k -> gamma_{k+1}
    \foreach \x in {0.45, 0.9, 1.35, 2.7, 3.15, 3.6} {
      \draw[rc] (\x,0) -- (\x,\h) -- (\x+0.45,\h) -- (\x+0.45,\ytop);
    }
    \draw[rc] (1.8,0) -- (1.8,\h) -- (2.05,\h);
    \draw[dotted] (2.1,\h) -- (2.4,\h);
    \draw[rc] (2.45,\h) -- (2.7,\h) -- (2.7,\ytop);
    \draw (4.05,0) -- (4.05,\h-0.2);
    % ring closure: dotted line with semicircle ends, re-entering as gamma_1
    \draw (4.8,\h) arc [radius=0.11, start angle=-90, end angle=90];
    \draw[densely dotted] (4.8,\h+0.22) -- (0.15,\h+0.22);
    \draw (0.15,\h+0.22) arc [radius=0.11, start angle=90, end angle=270];
    \draw[rc] (0.15,\h) -- (0.45,\h) -- (0.45,\ytop);
    % gates V on (gamma_1..gamma_4), ..., (gamma_{2N-3}..gamma_{2N})
    \node[gate] at (1.125,0.55) {};
    \draw[dotted] (2.1,0.55) -- (2.4,0.55);
    \node[gate] at (3.375,0.55) {};
    \node[lab] at (-0.3,0.55) {$\hat V$};
    \node[lab] at (-0.3,\h+0.11) {$\hat G$};
    % spectator
    \draw[spec] (5.15,0) -- (5.15,\ytop);
  }
  % ---------------------------------------------------------------- (a) a = +1
  \begin{scope}
    \node[font=\small, anchor=west] at (-0.55,2.75) {(a) $a=+1$: $\hat U_F(+1)$};
    \fsixcommon
    % gamma_{2N} -> beta (output), with the closure S_s
    \draw[rc] (4.05,\h-0.2) -- (4.05,\h) -- (4.3,\h);
    \draw[aux] (4.3,\h) -- (4.55,\h) -- (4.55,\ytop);
    \node[Sbox] at (4.55,1.7) {$S_s$};
    % beta (input) -> closure -> gamma_1
    \draw[aux] (4.55,0) -- (4.55,\h) -- (4.8,\h);
    \node[lab, align=center] at (2.5,-0.85) {all $2N+2$ Majorana wires run through:\\ $\rank\hat U_F(+1)=\sqrt2^{\,2N+2}=2^{N+1}$};
  \end{scope}
  % ---------------------------------------------------------------- (b) a = -1
  \begin{scope}[xshift=7.3cm]
    \node[font=\small, anchor=west] at (-0.55,2.75) {(b) $a=-1$: auxiliary leg reversed, $\hat U_F(-1)$};
    \fsixcommon
    % cap: gamma_{2N} -> beta, now an output at the bottom, with S_s
    \draw[rc] (4.05,\h-0.2) -- (4.05,\h) -- (4.3,\h);
    \draw[aux] (4.3,\h) -- (4.55,\h) -- (4.55,0);
    \node[Sbox] at (4.55,0.55) {$S_s$};
    % cup: beta, now an input at the top -> closure -> gamma_1
    \draw[aux] (4.55,\ytop) -- (4.55,\h) -- (4.8,\h);
    \node[lab, align=center] at (2.5,-0.85) {cap and cup cut 2 wires, $2N$ run through:\\ $\rank\hat U_F(-1)=\sqrt2^{\,2N}=2^{N}$};
  \end{scope}

  % ================================================================ |a| = 2: two auxiliary modes b_j = (beta_j + i beta'_j)/2
  % sqrt2 U(+2) = Pi S_s G V (int_lib.jl): G couples only b_2; S_s acts on beta_2; Pi swaps b_1 <-> b_2
  % (beta_1 <-> beta_2 and beta'_1 <-> beta'_2, never beta <-> beta'). For a = -2 every auxiliary leg is reversed.
  \def\hB{1.65}\def\ytopB{3.1}
  \def\xbb{4.55}\def\xba{5.05}\def\xsb{5.6}\def\xsa{6.1}
  \def\fsixcommonB{%
    \foreach \x/\k in {0.45/1, 0.9/2, 1.35/3, 1.8/4} {
      \node[lab, below] at (\x,0) {$\gamma_{\k}$};
      \node[lab, above] at (\x,\ytopB) {$\gamma_{\k}$};
    }
    \foreach \y/\p in {0/below, \ytopB/above} {
      \node[lab, \p] at (4.05,\y) {$\gamma_{2N}$};
      \node[lab, \p] at (\xbb,\y) {$\beta_2$};
      \node[lab, \p] at (\xba,\y) {$\beta_1$};
      \node[lab, \p, gray] at (\xsb,\y) {$\beta'_2$};
      \node[lab, \p, gray] at (\xsa,\y) {$\beta'_1$};
    }
    \foreach \x in {0.45, 0.9, 1.35, 2.7, 3.15, 3.6} {
      \draw[rc] (\x,0) -- (\x,\hB) -- (\x+0.45,\hB) -- (\x+0.45,\ytopB);
    }
    \draw[rc] (1.8,0) -- (1.8,\hB) -- (2.05,\hB);
    \draw[dotted] (2.1,\hB) -- (2.4,\hB);
    \draw[rc] (2.45,\hB) -- (2.7,\hB) -- (2.7,\ytopB);
    \draw (4.05,0) -- (4.05,\hB-0.2);
    \draw[rc] (4.05,\hB-0.2) -- (4.05,\hB) -- (4.3,\hB);
    \draw (4.8,\hB) arc [radius=0.11, start angle=-90, end angle=90];
    \draw[densely dotted] (4.8,\hB+0.22) -- (0.15,\hB+0.22);
    \draw (0.15,\hB+0.22) arc [radius=0.11, start angle=90, end angle=270];
    \draw[rc] (0.15,\hB) -- (0.45,\hB) -- (0.45,\ytopB);
    \node[gate] at (1.125,0.8) {};
    \draw[dotted] (2.1,0.8) -- (2.4,0.8);
    \node[gate] at (3.375,0.8) {};
    \node[lab] at (-0.3,0.8) {$\hat V$};
    \node[lab] at (-0.3,\hB+0.11) {$\hat G$};
  }
  % swap of two columns between heights #3 (start) and #4 (end), style #1, columns #2 and #5
  \def\fswap#1#2#3#4#5{%
    \draw[#1] (#2,#3) .. controls (#2,{(#3+#4)/2}) and (#5,{(#3+#4)/2}) .. (#5,#4);
    \draw[#1] (#5,#3) .. controls (#5,{(#3+#4)/2}) and (#2,{(#3+#4)/2}) .. (#2,#4);
  }
  % ---------------------------------------------------------------- (c) a = +2
  \begin{scope}[yshift=-5.6cm]
    \node[font=\small, anchor=west] at (-0.55,3.9) {(c) $a=+2$: $\hat U_F(+2)$};
    \fsixcommonB
    % gamma_{2N} -> beta_2 -> S_s -> Pi
    \draw[aux] (4.3,\hB) -- (\xbb,\hB) -- (\xbb,2.4);
    \node[Sbox] at (\xbb,2.08) {$S_s$};
    % beta_2 (input) -> closure -> gamma_1
    \draw[aux] (\xbb,0) -- (\xbb,\hB) -- (4.8,\hB);
    % beta_1 and the spectators run up to Pi
    \draw (\xba,0) -- (\xba,2.4);
    \draw[spec] (\xsb,0) -- (\xsb,2.4);  \draw[spec] (\xsa,0) -- (\xsa,2.4);
    % Pi
    \fswap{}{\xbb}{2.4}{2.8}{\xba}
    \fswap{spec}{\xsb}{2.4}{2.8}{\xsa}
    \draw[Pbox] (4.35,2.36) rectangle (6.3,2.84);
    \node[lab, anchor=west] at (6.32,2.6) {$\hat \Pi$};
    \foreach \x in {\xbb,\xba} { \draw (\x,2.8) -- (\x,\ytopB); }
    \foreach \x in {\xsb,\xsa} { \draw[spec] (\x,2.8) -- (\x,\ytopB); }
    \node[lab, align=center] at (2.5,-0.85) {all $2N+4$ Majorana wires run through:\\ $\rank\hat U_F(+2)=\sqrt2^{\,2N+4}=2^{N+2}$};
  \end{scope}
  % ---------------------------------------------------------------- (d) a = -2
  \begin{scope}[xshift=7.3cm, yshift=-5.6cm]
    \node[font=\small, anchor=west] at (-0.55,3.9) {(d) $a=-2$: auxiliary legs reversed, $\hat U_F(-2)$};
    \fsixcommonB
    % cap: gamma_{2N} -> beta_2 -> S_s -> Pi -> beta_1, now an output at the bottom
    \draw[aux] (4.3,\hB) -- (\xbb,\hB) -- (\xbb,0.55);
    \node[Sbox] at (\xbb,1.05) {$S_s$};
    % cup: beta_2, now an input at the top -> closure -> gamma_1
    \draw[aux] (\xbb,\ytopB) -- (\xbb,\hB) -- (4.8,\hB);
    % beta_1 and the spectators, now inputs at the top, run down to Pi
    \draw (\xba,\ytopB) -- (\xba,0.55);
    \draw[spec] (\xsb,\ytopB) -- (\xsb,0.55);  \draw[spec] (\xsa,\ytopB) -- (\xsa,0.55);
    % Pi, traversed downward
    \fswap{}{\xbb}{0.55}{0.15}{\xba}
    \fswap{spec}{\xsb}{0.55}{0.15}{\xsa}
    \draw[Pbox] (4.35,0.11) rectangle (6.3,0.59);
    \node[lab, anchor=west] at (6.32,0.35) {$\hat \Pi$};
    \foreach \x in {\xbb,\xba} { \draw (\x,0.15) -- (\x,0); }
    \foreach \x in {\xsb,\xsa} { \draw[spec] (\x,0.15) -- (\x,0); }
    \node[lab, align=center] at (2.5,-0.85) {cap and cup cut 2 wires, $2N+2$ run through:\\ $\rank\hat U_F(-2)=\sqrt2^{\,2N+2}=2^{N+1}$};
  \end{scope}
\end{tikzpicture}

}
\caption{Twisted Floquet operator $\sqrt2\,\hat U_F(a)=\hat \Pi\,S_s\, \hat G\,\hat V$ of the Majorana translation
[Eq.~(\ref{eq:fX})]; inputs at the bottom. Blue: the gates $\hat V$;
$\hat G$: the braid chain $\gamma_k\to\gamma_{k+1}$, closed through the seam by the dotted line; dashed: the
spectators $\beta'_j$.
(a) $a=+1$: $\gamma_{2N}$ is stored in the Majorana $\beta$ of the auxiliary mode $b=(\beta+i\beta')/2$ and
re-enters as $\gamma_1$ one period later; $\hat \Pi=\mathbf 1$.
(b) $a=-1$: with the auxiliary leg reversed, the two seam wires become a cap and a cup.
(c) $a=+2$: two auxiliary modes $b_j=(\beta_j+i\beta'_j)/2$. $\hat G$ and $S_s$ act on $b_2$ only, and $\hat \Pi$, the
fermionic swap of $b_1$ and $b_2$, exchanges $\beta_1\leftrightarrow\beta_2$ and $\beta'_1\leftrightarrow\beta'_2$
without mixing $\beta$ and $\beta'$; the Majorana in transit passes $b_1\to b_2$ and re-enters after two periods.
(d) $a=-2$: all auxiliary legs of (c) reversed, giving one cap and one cup, while the other auxiliary leg and
the spectators run through.
The rank is $\sqrt2^{\,\#\text{through wires}}$: $2^{N+1}$, $2^{N}$, $2^{N+2}$ and $2^{N+1}$ in (a)--(d).}
  \label{fig:fsetup}
\end{figure}
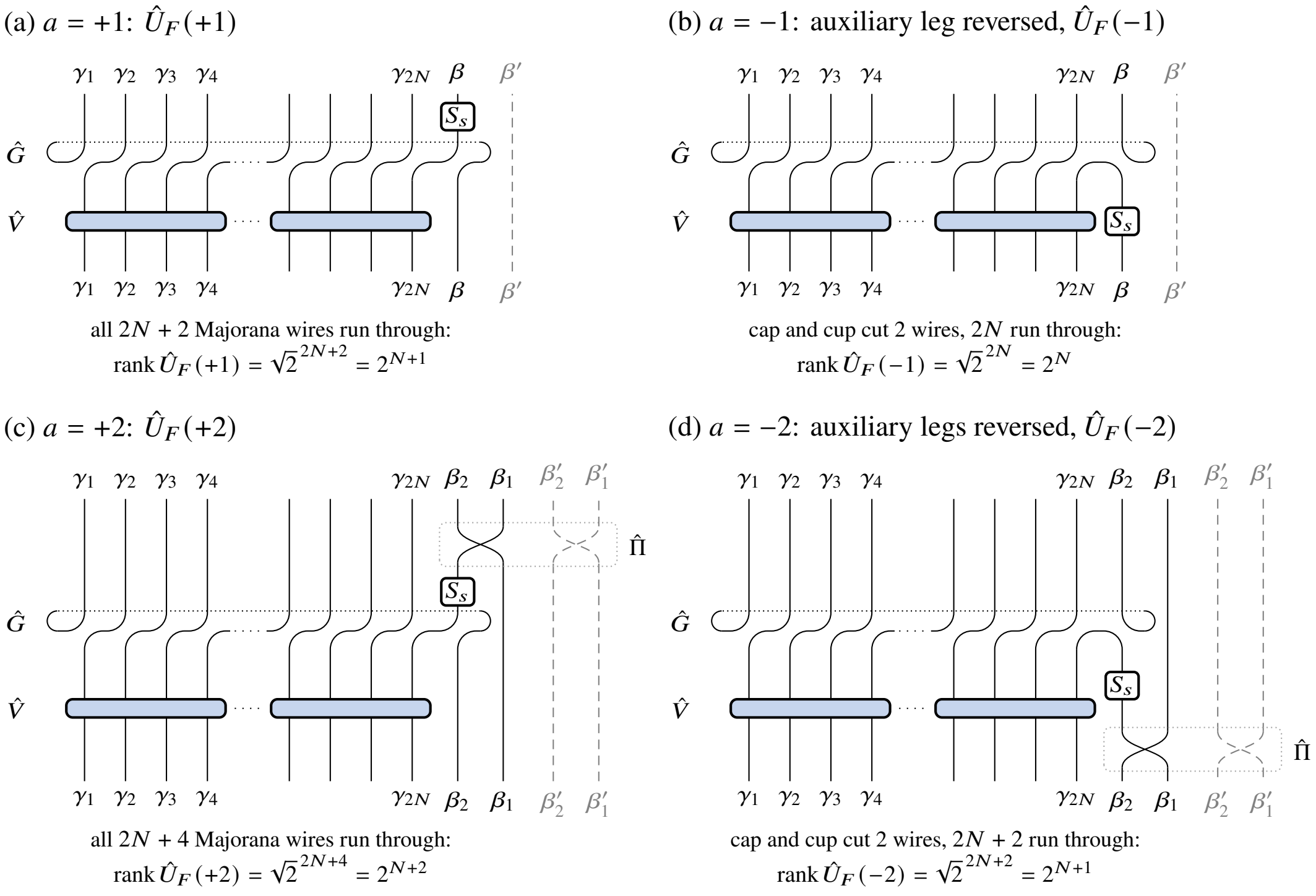

Under the twist the closing bond joins period $t$ to $t+a$; %REF
cut along flat slices it becomes $|a|$ auxiliary modes $b_j=(\beta_j+i\beta'_j)/2$, %REF
with $\beta_j$ carrying the Majorana in transit and $\beta'_j$ a spectator. For $a>0$,
\begin{align}
  \sqrt2\,\hat{U}_F(a)=\hat \Pi\,S_s\,\hat G\,\hat{V}
  \label{eq:fX}
\end{align}
on $N+a$ modes [Fig.~\ref{fig:fsetup}]. Here $\hat G$ couples the chain to the last auxiliary mode only
($\gamma_{2N}\to\beta_a$, $\beta_a\to\gamma_1$), $S_s$ acts on $\beta_a$, and
$\hat \Pi=\hat R_{1,2} \hat R_{2,3}\cdots \hat R_{a-1,a}$, with $\hat R_{k,k+1}=\tfrac12(\beta_k-\beta_{k+1})(\beta'_k-\beta'_{k+1})$ the fermionic
swap of $b_k$ and $b_{k+1}$, cyclically shifts the auxiliary modes, $b_a\to b_1$ and $b_k\to b_{k+1}$. It maps
$\beta$'s to $\beta$'s and $\beta'$'s to $\beta'$'s and never mixes them, so the Majorana stored in $b_1$ reaches $b_a$
after $a-1$ periods and re-enters as $\gamma_1$ one period later, while the spectators only circulate; $\hat \Pi=\mathbf 1$ for
$a=1$. $\sqrt2\,\hat U_F(a)$ is unitary. For $a<0$ the
auxiliary legs are reversed as in Eq.~\ref{eq:mpu_withtwist} [Fig.~\ref{fig:fsetup}(b)]; %REF
for fermions this is the fermionic partial transpose \cite{PhysRevB.95.165101} on the auxiliary modes.

\subsection{Twisted ranks}\label{sec:franks}
Each Majorana wire that runs from an input to an output contributes $\sqrt2$ to the rank (Fig.~\ref{fig:fsetup}). For
$a>0$ all $2N+2a$ wires run through. For $a<0$ the cap and the cup remove two of them, the auxiliary-to-auxiliary wires
(spectators and in-transit register) survive, and the reversed in-transit wires are capped one per period, the
nilpotent part of size $|a|$. The gates act on the physical modes only and do not change the count. Hence
\begin{align}
  \rank\hat{U}_F^t(a)=
  \begin{cases}
    2^{N+a} & (a\ge0)\\
    2^{N+\max(|a|-t,\,0)} & (a<0)
  \end{cases}
  \;=\;\sqrt2^{\,2N+a}\cdot\sqrt2^{\,|a|}\quad(t\ge|a|),
  \label{eq:frank}
\end{align}
the dimension of the tilted slice (Sec.~\ref{sec:fdim}) times $\sqrt2$ per spectator. The spectators cancel between
the two boundaries,
\begin{align}
  \frac{\rank\hat{U}_F^t(a)}{\rank\hat{U}_F^t(-a)}=2^{a}=I_\text{GNVW}^{\,2a},
  \label{eq:fratio}
\end{align}
which is Eq.~\ref{eq:mpurelation} %REF
with $\text{ind}=1/2$. Table~\ref{tab:ranktable} gives the ranks of interacting circuits for $|a|\le4$, $t\le5$,
$N=4,6,8$, both $s$. They agree with (\ref{eq:frank}) whenever $\min(t,|a|)\le N/2$, the condition found for qudit random
circuits in Sec.~\ref{sec:ruc}; %REF
beyond it the rank stops dropping, at $2^{N+|a|-\min(t,|a|,N/2)}$, while the pure Majorana translation ($\hat V=\mathbf 1$)
shows no deviation.

\begin{table}[htbp]
\centering\footnotesize
\caption{Ranks of the twisted Floquet operator $\hat U_F^t(a)$ of interacting Majorana circuits, from the
construction of Sec.~\ref{sec:fsetup} (fermionic transpose for $a<0$). Numbers in parentheses are the prediction
$\rank\hat U_F^t(a)=2^{N+a}$ ($a\ge0$), $2^{N+\max(|a|-t,0)}$ ($a<0$), shown where they differ from the numerical
rank. Both boundary signs give identical ranks. All entries, including the deviations, are
$\rank\hat U_F^t(a<0)=2^{N+|a|-\min(t,|a|,N/2)}$: the prediction holds for $\min(t,|a|)\le N/2$.}
\label{tab:ranktable}
\begin{tabular}{ccrrrrrrrrr}
\toprule
$N$ & $t$ & $a=-4$ & $-3$ & $-2$ & $-1$ & $0$ & $+1$ & $+2$ & $+3$ & $+4$\\
\midrule
\multirow{5}{*}{4} & 1 & 128 & 64 & 32 & 16 & 16 & 32 & 64 & 128 & 256\\
 & 2 & 64 & 32 & 16 & 16 & 16 & 32 & 64 & 128 & 256\\
 & 3 & 64 (32) & 32 (16) & 16 & 16 & 16 & 32 & 64 & 128 & 256\\
 & 4 & 64 (16) & 32 (16) & 16 & 16 & 16 & 32 & 64 & 128 & 256\\
 & 5 & 64 (16) & 32 (16) & 16 & 16 & 16 & 32 & 64 & 128 & 256\\
\midrule
\multirow{5}{*}{6} & 1 & 512 & 256 & 128 & 64 & 64 & 128 & 256 & 512 & 1024\\
 & 2 & 256 & 128 & 64 & 64 & 64 & 128 & 256 & 512 & 1024\\
 & 3 & 128 & 64 & 64 & 64 & 64 & 128 & 256 & 512 & 1024\\
 & 4 & 128 (64) & 64 & 64 & 64 & 64 & 128 & 256 & 512 & 1024\\
 & 5 & 128 (64) & 64 & 64 & 64 & 64 & 128 & 256 & 512 & 1024\\
\midrule
\multirow{5}{*}{8} & 1 & 2048 & 1024 & 512 & 256 & 256 & 512 & 1024 & 2048 & 4096\\
 & 2 & 1024 & 512 & 256 & 256 & 256 & 512 & 1024 & 2048 & 4096\\
 & 3 & 512 & 256 & 256 & 256 & 256 & 512 & 1024 & 2048 & 4096\\
 & 4 & 256 & 256 & 256 & 256 & 256 & 512 & 1024 & 2048 & 4096\\
 & 5 & 256 & 256 & 256 & 256 & 256 & 512 & 1024 & 2048 & 4096\\
\bottomrule
\end{tabular}
\end{table}

\subsection{Twisted dimension and spectral form factor}\label{sec:fdim}
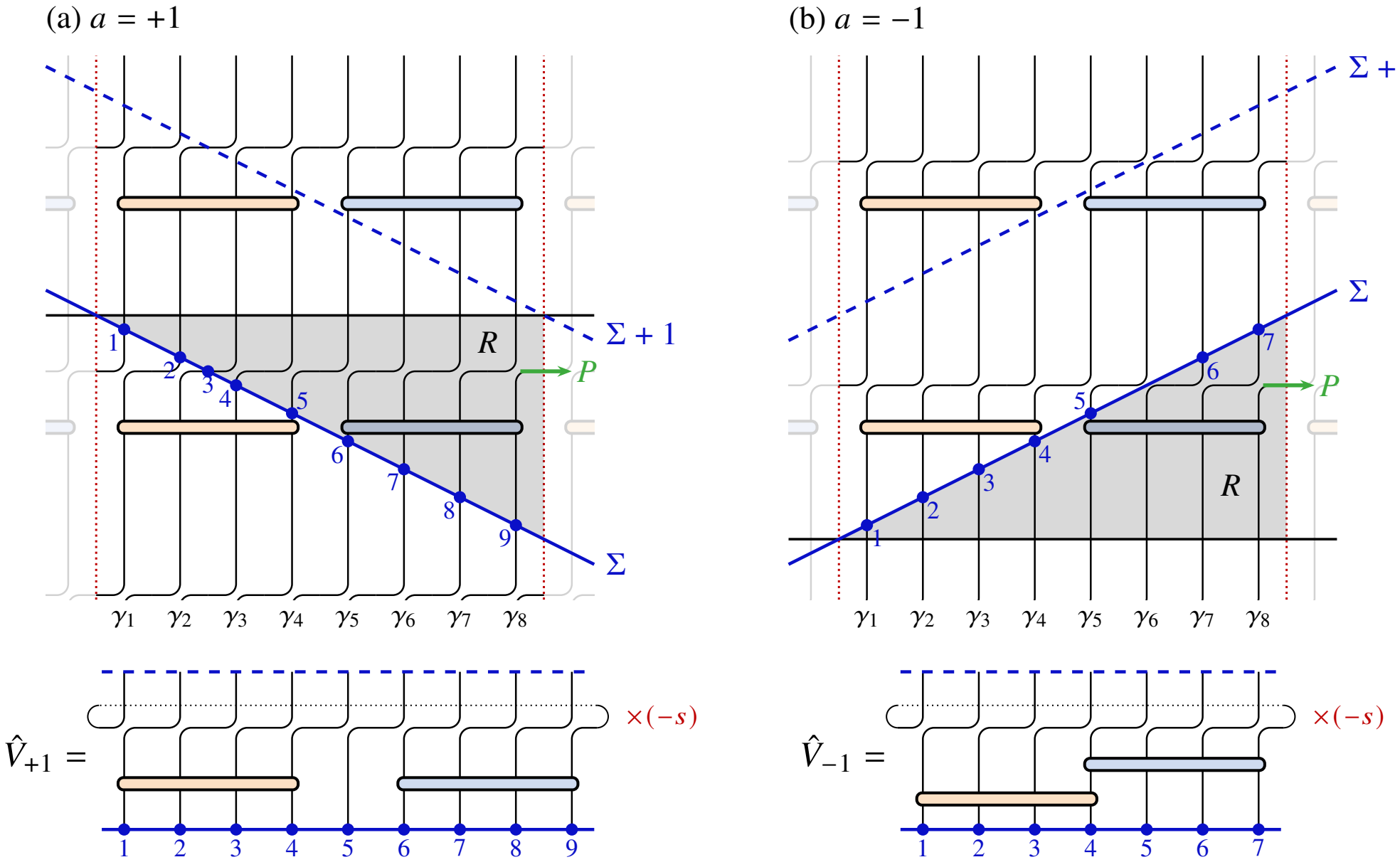
\begin{figure}[htbp]
  \centering
  \resizebox{0.75\textwidth}{!}{%
\begin{tikzpicture}[x=0.55cm, y=2.2cm,
    gA/.style={draw, rounded corners=1.5pt, line width=0.8pt, fill=fsgA, minimum width=1.77cm, minimum height=0.12cm, inner sep=0pt},
    gB/.style={draw, rounded corners=1.5pt, line width=0.8pt, fill=fsgB, minimum width=1.77cm, minimum height=0.12cm, inner sep=0pt},
    fA/.style={draw=gray!35, rounded corners=1.5pt, line width=0.8pt, fill=fsgA!30, minimum width=1.77cm, minimum height=0.12cm, inner sep=0pt},
    fB/.style={draw=gray!35, rounded corners=1.5pt, line width=0.8pt, fill=fsgB!30, minimum width=1.77cm, minimum height=0.12cm, inner sep=0pt},
    w/.style={line width=0.5pt, rounded corners=1.1mm},
    fw/.style={line width=0.5pt, rounded corners=1.1mm, gray!35},
    seam/.style={red!75!black, densely dotted, line width=0.6pt},
    lab/.style={font=\scriptsize},
    num/.style={font=\scriptsize, text=fssig, inner sep=1pt},
  ]
  % colours (defined here so the picture can be pasted into main.tex as is)
  \definecolor{fsgA}{HTML}{FCE0C4}
  \definecolor{fsgB}{HTML}{CFDBF0}
  \definecolor{fssig}{HTML}{0A10C8}
  \definecolor{fsgrn}{HTML}{3DAA3D}
  \def\yG{0.32}               % height of both gates in a period
  \def\ybot{-0.45}\def\ytopw{1.98}
  % spacetime network with jog height #1: faded copies, then the ring 0.5 < x < 8.5
  \def\stnet#1{%
    \begin{scope}
      \clip (-0.4,\ybot) rectangle (9.4,\ytopw);
      \foreach \t in {-1,0,1,2} {
        \foreach \k in {-4,...,12} { \draw[fw] (\k-0.5,\t-1+#1) -- (\k,\t-1+#1) -- (\k,\t+#1) -- (\k+0.5,\t+#1); }
        \foreach \m in {-1,0,1} { \node[fA] at (2.5+8*\m,\t+\yG) {}; \node[fB] at (6.5+8*\m,\t+\yG) {}; }
      }
    \end{scope}
    \begin{scope}
      \clip (0.5,\ybot) rectangle (8.5,\ytopw);
      \fill[white] (0.5,\ybot) rectangle (8.5,\ytopw);
      \foreach \t in {-1,0,1,2} {
        \foreach \k in {0,...,9} { \draw[w] (\k-0.5,\t-1+#1) -- (\k,\t-1+#1) -- (\k,\t+#1) -- (\k+0.5,\t+#1); }
        \node[gA] at (2.5,\t+\yG) {};
        \node[gB] at (6.5,\t+\yG) {};
      }
    \end{scope}
    \draw[seam] (0.5,\ybot) -- (0.5,\ytopw);
    \draw[seam] (8.5,\ybot) -- (8.5,\ytopw);
    \foreach \k in {1,...,8} { \node[lab, below] at (\k,\ybot) {$\gamma_{\k}$}; }
  }
  % ======================================================== (a) a = +1:  Sigma: y = 0.82 - (x-0.5)/8,  jog at 0.57
  \begin{scope}
    \def\J{0.57}
    \stnet{\J}
    \begin{scope}
      \clip (0.5,\ybot) rectangle (8.5,\ytopw);
      \fill[black, opacity=0.15] (0.5,0.82) -- (8.5,0.82) -- (8.5,-0.18) -- cycle;   % region R
    \end{scope}
    \draw[line width=0.7pt] (-0.4,0.82) -- (9.4,0.82);                               % flat slice
    \draw[fssig, line width=0.9pt] (-0.4,0.9325) -- (9.4,-0.2925);
    \draw[fssig, line width=0.9pt, dashed] (-0.4,1.9325) -- (9.4,0.7075);
    \node[fssig, anchor=west, font=\small] at (9.4,-0.29) {$\Sigma$};
    \node[fssig, anchor=west, font=\small] at (9.4,0.74) {$\Sigma+1$};
    \foreach \x/\y/\k/\p in {1/0.7575/1/below left, 2/0.6325/2/below left, 2.5/0.57/3/below, 3/0.5075/4/below left,
                             4/0.3825/5/above right, 5/0.2575/6/below left, 6/0.1325/7/below left, 7/0.0075/8/below left, 8/-0.1175/9/below left} {
      \fill[fssig] (\x,\y) circle (1.7pt);
      \node[num, \p] at (\x,\y) {\k};
    }
    \node[font=\small] at (7.5,0.70) {$R$};
    % P: the shift wire that leaves R through the seam
    \draw[fsgrn, line width=1.1pt, -{Stealth[length=4pt]}] (8.08,\J) -- (9.0,\J);
    \node[fsgrn, font=\small, anchor=west, inner sep=1pt] at (9.0,\J) {$P$};
    \node[font=\small, anchor=west] at (-0.6,2.14) {(a) $a=+1$};
  \end{scope}
  % ======================================================== (b) a = -1:  Sigma: y = 0.32 + (x-4.5)/8,  jog at 0.5075
  \begin{scope}[xshift=7.3cm]
    \def\J{0.5075}
    \stnet{\J}
    \begin{scope}
      \clip (0.5,\ybot) rectangle (8.5,\ytopw);
      \fill[black, opacity=0.15] (0.5,-0.18) -- (8.5,-0.18) -- (8.5,0.82) -- cycle;   % region R
    \end{scope}
    \draw[line width=0.7pt] (-0.4,-0.18) -- (9.4,-0.18);                             % flat slice
    \draw[fssig, line width=0.9pt] (-0.4,-0.2925) -- (9.4,0.9325);
    \draw[fssig, line width=0.9pt, dashed] (-0.4,0.7075) -- (9.4,1.9325);
    \node[fssig, anchor=west, font=\small] at (9.4,0.93) {$\Sigma$};
    \node[fssig, anchor=west, font=\small] at (9.4,1.93) {$\Sigma+1$};
    % crossings of Sigma: 2N-1 = 7 wires (Sigma passes the junction at column 6 without crossing a wire)
    \foreach \x/\y/\k/\p in {1/-0.1175/1/below right, 2/0.0075/2/below right, 3/0.1325/3/below right, 4/0.2575/4/below right,
                             5/0.3825/5/above left, 7/0.6325/6/below right, 8/0.7575/7/below right} {
      \fill[fssig] (\x,\y) circle (1.7pt);
      \node[num, \p] at (\x,\y) {\k};
    }
    \node[font=\small] at (7.5,0.06) {$R$};
    \draw[fsgrn, line width=1.1pt, -{Stealth[length=4pt]}] (8.08,\J) -- (9.0,\J);
    \node[fsgrn, font=\small, anchor=west, inner sep=1pt] at (9.0,\J) {$P$};
    \node[font=\small, anchor=west] at (-0.6,2.14) {(b) $a=-1$};
  \end{scope}
  % ======================================================== slabs
  % V_{+1}: gates on (1..4), (6..9), wire 5 idle; translation on 9 wires with the boundary sign reversed
  \begin{scope}[yshift=-3.25cm, y=1cm]
    \node[font=\normalsize, anchor=east] at (0.55,0.75) {$\hat V_{+1}=$};
    \foreach \i in {1,...,8} { \draw[w] (\i,0) -- (\i,1.0) -- (\i+1,1.0) -- (\i+1,1.55); }
    \draw[w] (9,0) -- (9,1.0) -- (9.45,1.0);
    \draw (9.45,1.0) arc [radius=0.11cm, start angle=-90, end angle=90];
    \draw[densely dotted] (9.45,1.22) -- (0.55,1.22);
    \draw (0.55,1.22) arc [radius=0.11cm, start angle=90, end angle=270];
    \draw[w] (0.55,1.0) -- (1,1.0) -- (1,1.55);
    \node[gA] at (2.5,0.45) {};
    \node[gB] at (7.5,0.45) {};
    \draw[fssig, line width=0.9pt] (0.6,0) -- (9.4,0);
    \draw[fssig, line width=0.9pt, dashed] (0.6,1.55) -- (9.4,1.55);
    \foreach \i in {1,...,9} { \fill[fssig] (\i,0) circle (1.7pt); \node[num, below=1.5pt] at (\i,0) {\i}; }
    \node[lab, red!75!black, anchor=west] at (9.75,1.1) {$\times(-s)$};
  \end{scope}
  % V_{-1}: gate on (1..4), then gate on (4..7); translation on 7 wires with the boundary sign reversed
  \begin{scope}[xshift=7.3cm, yshift=-3.25cm, y=1cm]
    \begin{scope}[xshift=0.55cm]
      \node[font=\normalsize, anchor=east] at (0.55,0.75) {$\hat V_{-1}=$};
      \foreach \i in {1,...,6} { \draw[w] (\i,0) -- (\i,1.0) -- (\i+1,1.0) -- (\i+1,1.55); }
      \draw[w] (7,0) -- (7,1.0) -- (7.45,1.0);
      \draw (7.45,1.0) arc [radius=0.11cm, start angle=-90, end angle=90];
      \draw[densely dotted] (7.45,1.22) -- (0.55,1.22);
      \draw (0.55,1.22) arc [radius=0.11cm, start angle=90, end angle=270];
      \draw[w] (0.55,1.0) -- (1,1.0) -- (1,1.55);
      \node[gA] at (2.5,0.3) {};
      \node[gB] at (5.5,0.64) {};
      \draw[fssig, line width=0.9pt] (0.6,0) -- (7.4,0);
      \draw[fssig, line width=0.9pt, dashed] (0.6,1.55) -- (7.4,1.55);
      \foreach \i in {1,...,7} { \fill[fssig] (\i,0) circle (1.7pt); \node[num, below=1.5pt] at (\i,0) {\i}; }
      \node[lab, red!75!black, anchor=west] at (7.75,1.1) {$\times(-s)$};
    \end{scope}
  \end{scope}
\end{tikzpicture}
}
\caption{Tilted time slices for $\hat U_F=\hat C_s\hat V$ on $2N=8$ Majorana wires:
the gates $\hat V$ act on $(1,\dots,4)$ and $(5,\dots,8)$, followed by the Majorana translation; time runs upward
and the dotted vertical lines are the seam. $P$ marks the shift wire that leaves the region $R$ through the seam.
(a) $a=+1$: the slice $\Sigma$ (blue) descends by one period across the ring and crosses $2N+1=9$ wires, against
$2N$ for the flat slice; the extra wire is crossed on a shift jog. The slab between $\Sigma$ and $\Sigma+1$ is
$\hat V_{+1}$ shown below: the same gates on $2N+1$ wires with wire 5 idle, followed by the translation with the
boundary sign reversed, $s\to-s$.
(b) $a=-1$: $\Sigma$ ascends by one period and passes the shift layer at a junction, where it crosses no wire, so
it crosses $2N-1=7$ wires; $\hat V_{-1}$ is the gate on $(1,\dots,4)$, then the gate on $(4,\dots,7)$, followed by
the translation on 7 wires with $s\to-s$.
In both cases $\dim\mathcal{H}_\Sigma=\sqrt2^{\,2N+a}$.}
  \label{fig:fslice}
\end{figure}

Each wire carries one Majorana, so the counting of Sec.~\ref{sec:tilted} %REF
goes through with $\sqrt2$ per wire: $\Sigma$ crosses $2N+a$ Majorana wires [Fig.~\ref{fig:fslice}(a)],
\begin{align}
  \dim\mathcal{H}_\Sigma=\sqrt2^{\,2N+a}=2^{N+a/2},
  \label{eq:fdim}
\end{align}
which is Eq.~\ref{eq:tilted_dimension} %REF
with $d=2$, $\text{ind}=1/2$. The slab is the same circuit on $2N+a$ Majoranas,
$\hat{V}_a=\hat{C}_{-s}(\gamma_1,\dots,\gamma_{2N},\beta_1,\dots,\beta_a)\hat V$ [Fig.~\ref{fig:fslice}(b)]; the boundary sign is
reversed because the Majorana that goes once around the twisted cycle also winds once in time. The spectators
$\beta'_1,\dots,\beta'_a$ have no dynamics: each period they are only moved to the next auxiliary mode, which is the
cyclic shift $\hat W_\text{s}\equiv\hat C_{+1}(\beta'_1,\dots,\beta'_a)$ of these $a$ Majoranas. For $a\ge1$ one period
factorizes exactly into the slab and this shift, $\sqrt2\,\hat U_F(a)\simeq\pm \hat{V}_a\,\hat W_\text{s}$, up to a phase and a gauge
on the auxiliary modes. Since $\hat{V}_a$ and $\hat W_\text{s}$ act on disjoint Majoranas and
products of distinct Majoranas are traceless, the trace factorizes:
\begin{align}
  \big|\Tr\hat{U}_F^t(a)\big|^2=2^{-t}\,P_{|a|}(t)\,\big|\Tr \hat{V}_a^t\big|^2,\qquad
  P_{m}(t)=\big|\Tr \hat W_\text{s}^t\big|^2=\big[1-(-1)^{m/g}\big]^g,\quad g=\gcd(m,t).
  \label{eq:fZ}
\end{align}
The factor $P_m(t)=|\operatorname{Tr}\hat W_{\mathrm{s}}^t|^2$ depends only on the $m=|a|$ spectator Majoranas. The spectator shift $\beta'_k\mapsto\beta'_{k+t}$ (indices modulo $m$)
partitions the $m$ modes into $g=\gcd(m,t)$ cycles, each of length
$\ell=m/g$. Let $O_{\mathrm{s}}$ be the matrix describing the
one-step spectator shift. By the Majorana trace identity,
$P_m(t)=\det(I+O_{\mathrm{s}}^t)$. Grouping the modes by cycle
makes $O_{\mathrm{s}}^t$ block diagonal, with $g$ identical blocks
$C_\ell$, where $C_\ell$ is the cyclic-permutation matrix for
a single cycle of length $\ell$. Thus,
$P_m(t)=[\det(I+C_\ell)]^g=[1-(-1)^{m/g}]^g$,
which equals $2^g$ when $m/g$ is odd and $0$ otherwise. Hence $P_m(t)=[1-(-1)^{m/g}]^g$. In particular, when $m$ is odd, every cycle has odd length and $P_m(t)=2^g$. $P_m$ does not depend on the circuit, and we study $K(t,a)=|\Tr \hat{V}_a^t|^2$; for
odd $2N+a$, $\Tr \hat{V}_a^t\equiv\dim\mathcal{H}_\Sigma\,\tau(\hat{V}_a^t)$ with $\tau$ the normalized trace.

The slab is parity even iff $s=(-1)^a$ ($a\neq0$), and the Clifford algebra of its $2N+a$ Majoranas fixes its level
structure (Table~\ref{tab:fblocks}). For even $a$ the plateau is $\dim\mathcal{H}_\Sigma$. For odd $a$ the parity-even
slab ($s=-1$) has exactly doubled levels (an unpaired Majorana at quasi-energy 0) and plateau $\sqrt2\dim\mathcal{H}_\Sigma$;
the parity-odd slab ($s=+1$) has pairs $(\lambda,-\lambda)$ (a Majorana at $\pi$) and plateau $\dim\mathcal{H}_\Sigma/\sqrt2$.
The two offsets are inverse, so for every $a$
\begin{align}
  \sqrt{\overline{K}_{s=-1}(a)\,\overline{K}_{s=+1}(a)}=\dim\mathcal{H}_\Sigma=\sqrt2^{\,2N+a},
  \label{eq:fprod}
\end{align}
with $\overline K$ the average over all $t$: the two spin structures of the ring together resolve the $\sqrt2$ per unit of
twist from a single twist.

\begin{table}[htbp]
  \caption{Twisted SFF of the slab $\hat{V}_a$ for CUE-random blocks, in terms of the twist $a$ and the boundary sign $s$.
  The slab has $2N+a$ Majoranas and is parity even iff $s=(-1)^a$ ($a\neq0$; for $a=0$ iff $s=-1$); a parity-odd slab
  has traceless odd powers, so $K=0$ at odd $t$. Plateau: late-time average of $K(t)$ and $K(t+1)$, equal to the average over all $t$ (for a parity-odd slab, half the even-$t$ value); $\dim\mathcal{H}_\Sigma=2^{N+a/2}=\sqrt2^{\,2N+a}$ is the
  dimension of the tilted slice.
  Last column: the parity of $n=2N+a$ and of $\hat{V}_a$.}
  \label{tab:fblocks}
  \begin{ruledtabular}
  \begin{tabular}{lllll}
    case & $\langle K(t,a)\rangle$ & plateau $\overline K$: average over $t$ & $\dim\mathcal{H}_\Sigma$ & ($n$, $\hat{V}_a$)\\ \hline
    $a$ odd, $s=-1$ & $2\min(t,2^{N+(a-1)/2})$ & $2^{N+(a+1)/2}$ & $2^{N+a/2}$ & (odd, even)\\
    $a$ odd, $s=+1$ & $0$ (odd $t$), $2\min(t,2^{N+(a-1)/2})$ (even $t$) & $2^{N+(a-1)/2}$ & $2^{N+a/2}$ & (odd, odd)\\
    $a\neq0$ even, $s=-1$ & $0$ (odd $t$), $2\min(t,2^{N+a/2})$ (even $t$) & $2^{N+a/2}$ & $2^{N+a/2}$ & (even, odd)\\
    $a\neq0$ even, $s=+1$ & $2\min(t,2^{N+a/2-1})$ & $2^{N+a/2}$ & $2^{N+a/2}$ & (even, even)\\
    $a=0$, $s=-1$ & $2\min(t,2^{N-1})$ & $2^{N}$ & $2^{N}$ & (even, even)\\
    $a=0$, $s=+1$ & $0$ (odd $t$), $2\min(t,2^{N})$ (even $t$) & $2^{N}$ & $2^{N}$ & (even, odd)
  \end{tabular}
  \end{ruledtabular}
\end{table}

\begin{figure}[htbp]
  \centering
  \includegraphics[width = 0.8\textwidth]
  {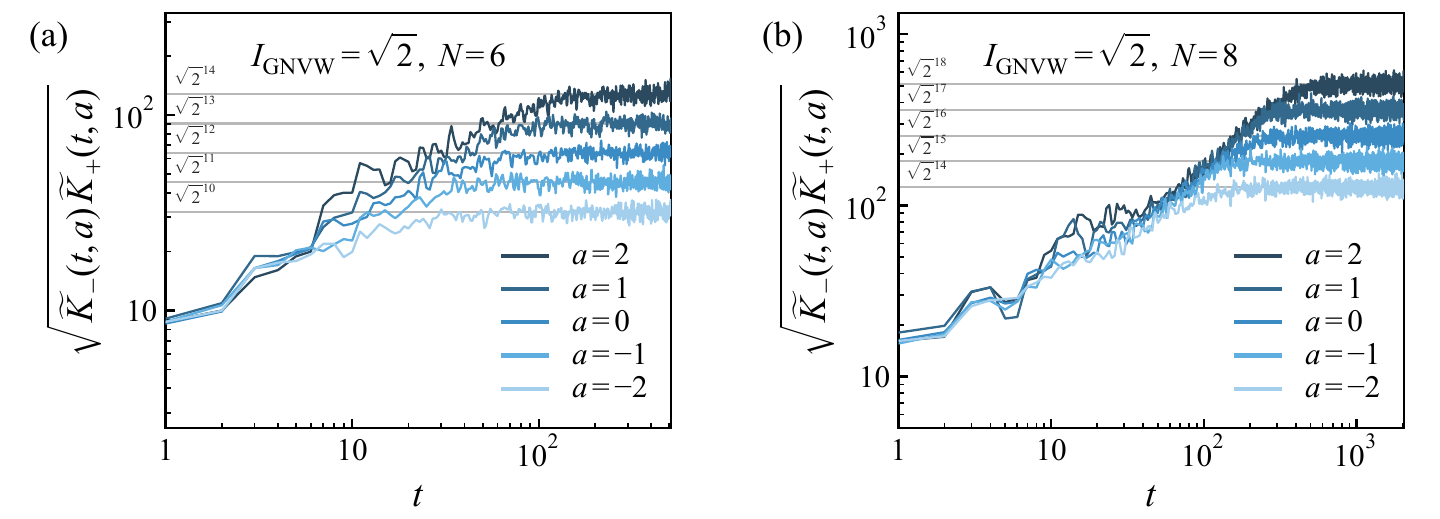}
  \caption{$\sqrt{\widetilde K_{s=-1}(t,a)\,\widetilde K_{s=+1}(t,a)}$ for $a=-2,\dots,2$, $N=6$ (left) and $N=8$ (right), with
  $\widetilde K(t)=\frac12[\langle K(t)\rangle+\langle K(t+1)\rangle]$ the average of neighbouring times (which removes the
  odd-$t$ zeros of parity-odd slabs and tends to the average over all $t$). Gray lines: the
  exact infinite-time values $\sqrt{\overline K_{s=-1}\overline K_{s=+1}}=\dim\mathcal{H}_\Sigma=\sqrt2^{\,2N+a}$, from the
  eigenvalue multiplicities.}
  \label{fig:fprod}
\end{figure}

We compute $K(t,a)$ for $a=-2,\dots,2$, both $s$, $N=6$ (100 circuits) and $N=8$ (100), the same circuits for all $a$
($a=+2$ from $\hat{V}_a$ of the factorization, $a=-2$ from the slab built from the circuit). For plotting we use the average
over neighbouring times, $\widetilde K(t,a)=\frac12[\langle K(t,a)\rangle+\langle K(t+1,a)\rangle]$. A parity-odd slab
has $K=0$ at every odd $t$, so $\widetilde K$ is half of its even-$t$ value; a parity-even slab has no odd-$t$ zeros, and
$\widetilde K=\langle K\rangle$. In both cases the late-time value of $\widetilde K$ is the average over all $t$, the plateau of
Table~\ref{tab:fblocks}; e.g., for $N=6$, $a=+1$ the even-$t$ values are $128$ for both $s$, the odd-$t$ values $128$
($s=-1$) and $0$ ($s=+1$), and $\widetilde K\to128$ and $64$. The plateaus agree with
Table~\ref{tab:fblocks} in every circuit, and (\ref{eq:fprod}) holds exactly (Fig.~\ref{fig:fprod}).

\end{document}